# Algebraic Signal Processing Theory: Cooley-Tukey Type Algorithms for DCTs and DSTs

Markus Püschel and José M. F. Moura

*Abstract*— **This paper presents a systematic methodology based on the algebraic theory of signal processing to classify and derive fast algorithms for linear transforms. Instead of manipulating the entries of transform matrices, our approach derives the algorithms by stepwise decomposition of the associated signal models, or polynomial algebras. This decomposition is based on two generic methods or algebraic principles that generalize the well-known Cooley-Tukey FFT and make the algorithms' derivations concise and transparent. Application to the 16 discrete cosine and sine transforms yields a large class of fast algorithms, many of which have not been found before.**

CONTENTS

This work was supported by NSF through awards 9988296 and 0310941.

Markus Püschel and José M. F. Moura are with the Department of Electrical and Computer Engineering, Carnegie Mellon University, Pittsburgh. E-mail: {pueschel,moura}@ece.cmu.edu .

## I. INTRODUCTION

In [1], [2], [3], we have proposed a new approach to linear signal processing (henceforth just referred to as signal processing or SP), called algebaric signal processing theory. The approach argues that the assumptions underlying SP provide structure that includes but goes beyond vector spaces and linear algebra and places SP more naturally into the context of the theory of algebras and modules, or the *representation theory* of algebras.

In recognizing this structure, we have introduced a general, axiomatic approach to SP that starts from the concept of a signal model. Given a signal model, all major SP ingredients can be derived from it, including signals, filters, convolution, associated "$z$-transform," spectrum, Fourier transform, and frequency response, among others. These concepts take different forms for different models, as shown in Table I, which is explained in detail later in Section II. For example, discrete infinite and finite (finite number of samples) 1-D time are signal models with associated $z$-transform and finite $z$-transform (defined in [1]) and the DTFT and DFT as associated Fourier transforms, respectively. Further, we developed signal models for infinite and finite 1-D space (where space is in the sense of an undirected graph versus directed graph for time, and not as 2-D versus 1-D) and showed that for the latter there are 16 reasonable alternatives corresponding to 16 finite $C$-transforms (defined in [2]) and showed that the 16 discrete cosine and sine transform (DCTs and DSTs) are the associated Fourier transforms. First results on higher-dimensional SP are also already available [4], [5], [6], [7], [8].

The algebraic theory provides a methodology for the construction of finite signal models and clarifies the role played by boundary conditions and their relation to signal extensions. In particular, we showed that any finite shift-invariant signal model is described by a polynomial algebra, which captures all the necessary information about the model. We derived and discussed in detail the polynomial algebras for the DFT, DCTs, DSTs, and most other known, as well as some new trigonometric transforms [2].

**Algebraic theory of transform algorithms.** In this paper, we apply the algebraic approach to the derivation and discovery of Fourier transform algorithms. Here, the term Fourier transform is meant in the general sense of the algebraic theory, i.e., including DFT, DFTs, DSTs, and other trigonometric like transforms. In other words, we apply the algebraic SP theory to derive fast transform algorithms. The paper extends the preliminary results shown in [9], [10].

There is a large body of literature on fast transform algorithms. With very few exceptions (for example DFT algorithms, discussed below) these algorithms are derived by clever and often lengthy manipulation of the transform coefficients. This is hard to grasp, and provides no insight into the structure or the derivation of the algorithm. Further, without an appropriate theory, it is hard to determine if all relevant classes of algorithms have been found. This is not just an academic problem as the variety of different implementation platforms and application requirements makes a thorough knowledge of the algorithm space crucial.

Our derivation of the fast algorithms is algebraic: we manipulate the signal model (or polynomial algebra) underlying a transform rather than the transform itself. We present two generic theorems for polynomial algebras that generalize the Cooley-Tukey FFT [11]. Application to the 16 DCTs and DSTs yields a large set of Cooley-Tukey type algorithms, most of which have not been found with previous methods. The algorithm derivation is concise (no tedious index manipulations) and greatly simplified as there is a clear methodology. We draw attention to the large number of algorithms in this paper. However, we do not consider *all* existing classes of algorithms. In particular, all our algorithms are non-orthogonal, i.e., they are not built from rotations. More precisely, in this paper, we will not consider orthogonal algorithms (e.g., [12]), algorithms that compute DFTs via the DCTs/DSTs [13], [14], prime-factor type algorithms [15], Rader-type algorithms [16], [17], [18], or algorithms that do not reduced the operations count [19]. The algebraic principles behind some of these algorithms will be the subject of a future paper.

**Goal of this paper.** This paper has two main goals: First, to explain how and why algorithms arise and how they can be derived in a reproducible way. Second, this paper can serve as a reference for readers whose interest is solely in the algorithms, for example, for their implementation. For this reason, all algorithms are presented in tables and in a form from which they can be easily retrieved.

**Previous work.** The approach taken in this paper to derive algorithms using polynomial algebras builds on and extends early work on DFT algorithms. The known interpretation of the $\mathrm{DFT}_n$ in terms of the polynomial algebra $\mathbb{C}[x]/(x^n-1)$ was used to derive and explain the (radix-2) Cooley-Tukey FFT by Auslander, Feig, and Winograd [20] using the Chinese remainder theorem (CRT). Equivalently, Nicholson [21] explains DFT and FFT using group theory; so does Beth [22], which generalized the approach to more general groups. Winograd's DFT algorithms [23], [24], [25], [26] and his results in complexity theory make heavy use of polynomial algebras and the CRT. So do extensions of the above work by Burrus et al. [27], [28]. Nussbaumer [29], [30], [31] uses polynomial algebras and the CRT to derive efficient 2-D FFTs that save multiplications compared to the row-column method.

For the DFT it turns out that to derive the most important FFTs, it is not necessary to work with polynomial algebras, but sufficient to work with index arithmetic modulo $n$. This approach is used in [31], [32] to provide a consistent approach to FFTs. However, this approach provides no insight into how to approach other transforms, whereas the polynomial algebra approach does, as we show in this paper. Further, this approach fits naturally with the algebraic SP theory, since polynomial algebras are a natural structure from an SP point of view as explained in [1].

The only (implicit) use of polynomial algebras for the DCTs or DSTs we found in the literature is the derivation of a DCT, type 3, algorithm by Steidl [33], [34]. These papers provided important hints for developing the work in this paper.

**Organization of the paper.** Section II provides a brief introduction to the algebraic signal processing theory. Most relevant are the signal models, or polynomial algebras, associated with the DFT and DTTs. Section III introduces notation to represent algorithms as products of structured matrices. Two algebraic methods to derive algorithms from a polynomial algebra are explained in Section IV using the DFT as an example. Then we apply these methods to derive numerous Cooley-Tukey type algorithms for the DTTs in Sections V–X. A visual organization of the most important ones can be found in Figure 2 in Section VI. Finally, we offer conclusions in Section XI.

## II. Background: Algebraic Signal Processing Theory

The algebraic signal processing theory recognizes that the structure available in *linear* signal processing (heretofore, simply signal processing or SP) goes beyond vector spaces (or linear spaces) and is actually described by *algebras* and associated *modules*, which places SP in the context of (abstract) algebra. The algebraic theory provides a consistent and

TABLE I

1-D DISCRETE INFINITE AND FINITE TIME AND SPACE SIGNAL PROCESSING AS FOUR INSTANTIATIONS OF THE GENERAL ALGEBRAIC SIGNAL PROCESSING THEORY.

| **infinite time** | **finite time** | **infinite space** | **finite space** | **generic theory** |
|---|---|---|---|---|
| series in $z^{-n}$ | polynomials in $z^{-n}$ | series in $T_n$ | polynomials in $T_n$ | $\mathcal{A}$ (algebra of filters) |
| series in $z^{-n}$ | polynomials in $z^{-n}$ | series in $C_n$ | polynomials in $C_n$ | $\mathcal{M}$ ($\mathcal{A}$-module of signals) |
| $z$-transform | finite $z$-transform(s) | $C$-transform(s) | finite $C$-transform(s) | $\Phi$ ("$z$-transform") |
| DTFT | DFTs | DSFTs | DCTs/DSTs | $\mathcal{F}$ (Fourier transform) |

generic framework for SP whose instantiations lead to many known, as well as new, ways of doing SP.

The key concept in the algebraic theory is the signal model $(\mathcal{A}, \mathcal{M}, \Phi)$. Before we define it, we introduce two algebraic terms: algebra and associated module, which model the spaces of filters and signals, respectively.

**Algebra = Space of filters.** An algebra is a vector space that is also a ring, i.e., it has a defined multiplication of elements, such that the distributivity law holds. Examples of algebras include $\mathbb{C}$ (complex numbers) and $\mathbb{C}[x]$ (set of polynomials with complex coefficients).

The crucial observation is that the set of filters in a given signal processing scenario (e.g., infinite discrete time) is usually assumed to be an algebra. Namely, the multiplication is the concatenation of filters. For example, in infinite discrete time, the set of filters (in the $z$-domain) is the algebra

$$\mathcal{A} = \{h(z^{-1}) = \sum_{n\in\mathbb{Z}} h_n z^{-n}\}, \qquad (1)$$

where, for example, the coefficient sequences $(\dots, h_{-1}, h_0, h_1, \dots) \in \ell^1(\mathbb{Z})$, i.e., are absolutely summable.

**Module = Space of signals.** Assume an algebra $\mathcal{A}$ is chosen. Then an $\mathcal{A}$-module $\mathcal{M}$ is a vector space that permits an operation of elements of $\mathcal{A}$ on $\mathcal{M}$ through linear mappings. This operation is the algebraic analogue of filtering in SP. Formally, if $h \in \mathcal{A}$, then there is an operation (written as multiplication)

$$h: \ \mathcal{M} \to \mathcal{M}, \quad m \mapsto h \cdot m,$$

which is linear, i.e., $h(s+s') = hs + hs'$, and $h(\alpha s) = \alpha(hs)$ for $s, s' \in \mathcal{M}$ and $\alpha \in \mathbb{C}$.

An example of an $\mathcal{A}$-module is $\mathcal{M} = \mathcal{A}$ itself with the operation being the multiplication in $\mathcal{A}$. When $\mathcal{M} = \mathcal{A}$, $\mathcal{M}$ is called a *regular module*.

The above properties capture exactly the structure of the signal space: every filter is a linear mapping on the signal space. The $\mathcal{A}$-module usually chosen along with $\mathcal{A}$ in (1) in infinite discrete-time signal processing is

$$\mathcal{M} = \{s = s(z^{-1}) = \sum_{n\in\mathbb{Z}} s_n z^{-n}\}, \qquad (2)$$

where the coefficient sequences are in $\ell^2(\mathbb{Z})$, i.e., of finite energy.

**Signal model.** We start with the formal definition considering infinite and finite discrete complex signals $\mathbf{s} \in \mathbb{C}^I$ over some index domain $I$. Examples include $\mathbf{s} \in \mathbb{C}^{\mathbb{Z}}$ or $\mathbf{s} \in \mathbb{C}^n$.

*Definition 1 (Signal model)* Let $V \le \mathbb{C}^I$ be a discrete vector space. A signal model for $V$ is a triple $(\mathcal{A}, \mathcal{M}, \Phi)$, where $\mathcal{A}$ is an algebra, $\mathcal{M}$ is an associated $\mathcal{A}$-module, and $\Phi$ is a bijective linear mapping

$$\Phi: \ V \to \mathcal{M}, \quad \mathbf{s} \mapsto s \in \mathcal{M}.$$

We call a signal model *regular* if $\mathcal{M} = \mathcal{A}$.

An example is the signal model commonly adopted for infinite discrete-time signal processing. Namely, $\mathcal{A}$ is defined as in (1), $\mathcal{M}$ as in (2), and $\Phi$ is the $z$-transform

$$\Phi: \ \mathbf{s} \mapsto s = \sum_{n\in\mathbb{Z}} s_n z^{-n} \in \mathcal{M}.$$

$(\mathcal{A}, \mathcal{M}, \Phi)$ is a signal model for $V = \ell^2(\mathbb{Z})$.

The purpose of the signal model is to assign a proper notion of filtering to a discrete sequence $\mathbf{s}$, which, taken by itself, does not specify how this should be done. Once a signal model is selected, all main concepts for SP can be derived: filtering or convolution (operation of $\mathcal{A}$ on $\mathcal{M}$), associated "z-transform" ($\Phi$), spectrum, frequency response, Fourier transform, and others.

The question now is, which signal models are used or make sense in SP. A partial answer was provided in [1], [2]: if shift-invariance is required, $\mathcal{A}$ has to be commutative. Further, we have shown how to derive models from basic principles, through a suitable definition of the shift operator. Using this method, we presented in [2] a theory of 1-D *space* SP.

The algebraic theory provides a comprehensive theory for *finite* signal models, i.e., models for finite sequences $\mathbf{s} \in \mathbb{C}^n$. In particular, it identifies for all trigonometric transforms $T$ the associated signal models, i.e., those that have $T$ as Fourier transform, and explain how they are obtained.

In particular, all signal models associated to the trigonometric transforms are finite, shift-invariant, and most of them are regular, i.e., $\mathcal{A} = \mathcal{M}$. The only way to obtain such models is through *polynomial algebras* as we explain next.

### A. Finite Shift-Invariant Regular 1-D Signal Models

If $(\mathcal{A}, \mathcal{M}, \Phi)$ is a shift-invariant signal model for finite 1-D sequences $\mathbf{s} = (s_0, \dots, s_{n-1})$, then, necessarily, $\mathcal{A} = \mathbb{C}[x]/p(x)$ is a *polynomial algebra* with a suitable polynomial $p(x)$. It is defined as

$$\mathbb{C}[x]/p(x) = \{q(x) \mid \deg(q) < \deg(p)\}.$$

In words, given $p(x)$, $\mathbb{C}[x]/p(x)$ is the set of all polynomials of degree smaller than $p$ with addition and multiplication modulo $p$. If $\deg(p) = n$, then $\dim(\mathbb{C}[x]/p(x)) = n$.

In this paper, we restrict ourselves to regular models, i.e., models with $\mathcal{M} = \mathcal{A} = \mathbb{C}[x]/p(x)$. With this restriction, a signal model for $V = \mathbb{C}^n$ is uniquely characterized by $p(x)$ and by a chosen basis $b = (p_0, \dots, p_{n-1})$ of $\mathcal{M}$. Namely, $\Phi$ is given by

$$\Phi : \ V \to \mathbb{C}[x]/p(x), \quad \mathbf{s} \mapsto s = \sum_{0 \le \ell < n} s_\ell p_\ell. \tag{3}$$

By construction, $\Phi$ is bijective. Conversely, every finite shift-invariant regular 1-D signal model can be expressed this way. Filtering in these models is equivalent to multiplying two polynomials (signal and filter) modulo the fixed polynomial $p(x)$. Note that (3) clarifies the role of the $z$-transform in SP: the equation shows that the bijective map $\Phi$, which generalizes the $z$-transform, is simply an artifact to fix the basis of the signal module $\mathcal{M}$.

**Example: finite z-transform.** As an example consider the model $\mathcal{A} = \mathcal{M} = \mathbb{C}[x]/(x^n - 1)$ with basis $b = (x^0, \dots, x^{n-1})$ in $\mathcal{M}$ and thus, for $\mathbf{s} = (s_0, \dots, s_{n-1})^T \in \mathbb{C}^n$,

$$\Phi : \ \mathbf{s} \mapsto s = s(x) = \sum_{0 \le k < n} s_k x^k \in \mathbb{C}[x]/(x^n - 1) \tag{4}$$

is the *finite $z$-transform*. After applying the model, filtering is defined, for $h = h(x) \in \mathcal{A}$ and $s = s(x) \in \mathcal{M}$ as

$$h(x)s(x) \bmod (x^n - 1),$$

which is equivalent to computing the circular convolution of the coefficient sequences $\mathbf{h}$ and $\mathbf{s}$ (e.g., [31]).

**Fourier transform.** The Fourier transform for signal models of the form $(\mathbb{C}[x]/p(x), \mathbb{C}[x]/p(x), \Phi)$ is obtained from the well-known Chinese remainder theorem (or CRT, see Appendix I). We assume that the zeros of $p(x)$ are pairwise distinct, given by $\alpha = (\alpha_0, \dots, \alpha_{n-1})$. Then the CRT provides the decomposition

$$\begin{array}{rcl} \mathcal{F} : \ \mathbb{C}[x]/p(x) & \to & \bigoplus_{0 \le k < n} \mathbb{C}[x]/(x - \alpha_k), \\ s(x) & \mapsto & (s(\alpha_0), \dots, s(\alpha_{n-1})). \end{array} \tag{5}$$

The mapping $\mathcal{F}$ is the *Fourier transform* for the signal model, and $\mathbb{C}[x]/(x - \alpha_k)$, $0 \le k < n$, are the spectral components of $\mathcal{M} = \mathbb{C}[x]/p(x)$.

To obtain a matrix representation of $\mathcal{F}$, we choose bases. The basis $b = (p_0, \dots, p_{n-1})$ of $\mathcal{M}$ is provided by the model (namely by $\Phi$). In each spectral component, which has dimension 1, we choose the basis $(x^0)$. The standard procedure to derive the matrix representation for $\mathcal{F}$ is apply $\mathcal{F}$ to the base vectors $p_\ell$, determine the coordinate vectors of the images, and place them in the columns of a matrix. By abuse of notation, we denote this matrix also by $\mathcal{F}$. Because

$$p_\ell(x) \equiv p_\ell(\alpha_k) \bmod (x - \alpha_k),$$

we obtain

$$\mathcal{F} = \mathcal{P}_{b,\alpha} = [p_\ell(\alpha_k)]_{0 \le k, \ell < n}.$$

We call $\mathcal{P}_{b,\alpha}$ a *polynomial transform*. It is uniquely determined by the signal model. This definition is different from Nussbaumer's in [29], [31].

Other Fourier transforms for the same model arise through the degrees of freedom in choosing the bases in the spectral components $\mathbb{C}[x]/(x - \alpha_k)$. In the most general case, we choose a basis $(\beta_k x^0)$ in each component, which yields the generic Fourier transform.

$$\mathcal{F} = \operatorname{diag}(1/\beta_0, \dots, 1/\beta_{n-1}) \mathcal{P}_{b,\alpha}. \tag{6}$$

Returning to our previous example $\mathcal{A} = \mathcal{M} = \mathbb{C}[x]/(x^n - 1)$ and $\Phi$ given in (4), we compute

$$\mathbb{C}[x]/(x^n - 1) \to \bigoplus_{0 \le k < n} \mathbb{C}[x]/(x - \omega_n^k),$$

where $\omega_n = e^{-2\pi j/n}$, and thus

$$\mathcal{P}_{b,\alpha} = [\omega_n^{k\ell}]_{0 \le k, \ell < n} = \mathrm{DFT}_n$$

is the discrete Fourier transform, which also motivates the name finite $z$-transform for (4).

### B. Signal Models for DFTs and DTTs

In this section we provide the signal models for 4 types of DFTs, the 16 DCTs and DSTs. We refer to the DCTs and DSTs collectively as DTTs (discrete trigonometric transforms) even though this class is actually larger (e.g., including discrete Hartley transform and real discrete Fourier transforms). Further, we define 4 types of skew DTTs, which were introduced in [2], and which are necessary to derive a complete set of algorithms.

Each of these transforms is a Fourier transform for a finite shift-invariant regular 1-D signal model. As said before, these models are uniquely determined by $p(x)$ (defining $\mathcal{A} = \mathcal{M} = \mathbb{C}[x]/p(x)$) and the basis $b$ (defining $\Phi$). The model in turn uniquely determines the associated *polynomial* Fourier transform $\mathcal{P}_{b,\alpha}$. To characterize an *arbitrary* Fourier transform, we need to specify in addition the diagonal matrix in (6). We do this in the following by providing a function $f$ such that the diagonal matrix is given by

$$D_f = \operatorname{diag}_{0 \le \ell < n}(f(\alpha_\ell)),$$

where $\alpha_\ell$ are, as before, the zeros of $p(x)$.

Due to lack of space, we will not provide in the paper detailed derivations of the signal models; we refer the reader to [1], [2] for details.

**DFTs.** The DFTs are Fourier transforms for finite time models. We distinguish 4 types, DFT type 1–4. Type 1 and 3 are special cases of a $\mathrm{DFT}(a)$, $a \in \mathbb{C} \backslash \{0\}$, all of which are polynomial transforms.

For example, the signal model associated to $\mathrm{DFT}(a)$ is given by $\mathcal{A} = \mathcal{M} = \mathbb{C}[x]/(x^n - a)$ and $\Phi : \ \mathbf{s} \mapsto \sum_{0 \le k < n} s_k x^k$. The zeros of $x^n - a$ are the $n$ $n$th roots of $a$ and thus straightforward computation yields

$$\mathrm{DFT}(a) = \mathcal{P}_{b,\alpha} = \mathrm{DFT}_n \operatorname{diag}_{0 \le \ell < n}(\sqrt[n]{a}^{\,\ell}), \tag{7}$$

where $\sqrt[n]{a} = |a|^{1/n} e^{\nu j/n}$ for $a = |a| e^{\nu j}$.

**DTTs.** The 16 DTTs are Fourier transforms for finite space models, which are defined in Table III. In contrast to the time models, the basis polynomials are now Chebyshev polynomials of the first ($T_k$), second ($U_k$), third ($V_k$), or fourth ($W_k$) kind.

TABLE II

SIGNAL MODELS ASSOCIATED TO THE DFTS.

| $\mathcal{F}$ | $p(x)$ | $b$ | $f = f(\ell)$ |
|---|---|---|---|
| DFT = DFT-1 | $x^n - 1$ | $x^k$ | 1 |
| DFT-2 | $x^n - 1$ | $x^k$ | $\alpha_\ell^{1/2}$ |
| DFT-3 | $x^n + 1$ | $x^k$ | 1 |
| DFT-4 | $x^n + 1$ | $x^k$ | $\alpha_\ell^{1/2}$ |
| DFT($a$) | $x^n - a$ | $x^k$ | 1 |

TABLE III

SIGNAL MODELS ASSOCIATED TO THE 16 DTTS (DCTS AND DSTS).

| | $\mathcal{F}$ | $p = p(x)$ | $b$ | $f = f(\theta),\ \cos\theta = \alpha_\ell$ |
|---|---|---|---|---|
| $T$-group | DCT-3 | $T_n$ | $T_k$ | 1 |
| | DST-3 | $T_n$ | $U_k$ | $\sin(\theta)$ |
| | DCT-4 | $T_n$ | $V_k$ | $\cos(\theta/2)$ |
| | DST-4 | $T_n$ | $W_k$ | $\sin(\theta/2)$ |
| $U$-group | DCT-1 | $(x^2-1)U_{n-2}$ | $T_k$ | 1 |
| | DST-1 | $U_n$ | $U_k$ | $\sin(\theta)$ |
| | DCT-2 | $(x-1)U_{n-1}$ | $V_k$ | $\cos(\theta/2)$ |
| | DST-2 | $(x+1)U_{n-1}$ | $W_k$ | $\sin(\theta/2)$ |
| $V$-group | DCT-7 | $(x+1)V_{n-1}$ | $T_k$ | 1 |
| | DST-7 | $V_n$ | $U_k$ | $\sin(\theta)$ |
| | DCT-8 | $V_n$ | $V_k$ | $\cos(\theta/2)$ |
| | DST-8 | $(x+1)V_{n-1}$ | $W_k$ | $\sin(\theta/2)$ |
| $W$-group | DCT-5 | $(x-1)W_{n-1}$ | $T_k$ | 1 |
| | DST-5 | $W_n$ | $U_k$ | $\sin(\theta)$ |
| | DCT-6 | $(x-1)W_{n-1}$ | $V_k$ | $\cos(\theta/2)$ |
| | DST-6 | $W_n$ | $W_k$ | $\sin(\theta/2)$ |

See Appendix II for their definition and properties that we will use in this paper.

As an example consider the most commonly used $\mathrm{DCT\text{-}2}_n$. The associated model is given from Table III by $\mathcal{A} = \mathcal{M} = \mathbb{C}[x]/(x-1)U_{n-1}$. The zeros of $(x-1)U_{n-1}$ are given by $\alpha_k = \cos(k\pi/n)$, $0 \le k < n$ (see Table XXIII in Appendix II). Thus the unique polynomial Fourier transform for the model is given by

$$\mathcal{P}_{b,\alpha} = [V_\ell(\alpha_k)]_{0\le k,\ell<n} = \left[\frac{\cos\frac{k(\ell+1/2)\pi}{n}}{\cos\frac{(k+1/2)\pi}{2n}}\right]_{0\le k,\ell<n}. \quad (8)$$

Multiplying $\mathcal{P}_{b,\alpha}$ from the left by the scaling diagonal

$$\mathrm{diag}_{0\le k<n}(\cos(\mathrm{acos}(\alpha_k)/2))$$

cancels the denominator to yield

$$\mathrm{DCT\text{-}2}_n = [\cos\tfrac{k(\ell+1/2)\pi}{n}]_{0\le k,\ell<n},$$

which identifies DCT-2 as a Fourier transform for the specified signal model.

The definitions of all 16 DTTs are given in Table IV. Types 1, 4, 5, 8 are symmetric; types 2, 3 and 6, 7 are transposes of each other, respectively.

Every DTT has a polynomial transform counterpart, which we write as $\overline{\mathrm{DTT}}$. For example $\overline{\mathrm{DCT\text{-}2}}_n$ is the matrix in (8).

TABLE IV

8 TYPES OF DCTS AND DSTS OF SIZE $n$. THE ENTRY AT ROW $k$ AND COLUMN $\ell$ IS GIVEN FOR $0 \le k, \ell < n$.

| type | DCTs | DSTs |
|---|---|---|
| 1 | $\cos k\ell\frac{\pi}{n-1}$ | $\sin(k+1)(\ell+1)\frac{\pi}{n+1}$ |
| 2 | $\cos k(\ell+\frac{1}{2})\frac{\pi}{n}$ | $\sin(k+1)(\ell+\frac{1}{2})\frac{\pi}{n}$ |
| 3 | $\cos(k+\frac{1}{2})\ell\frac{\pi}{n}$ | $\sin(k+\frac{1}{2})(\ell+1)\frac{\pi}{n}$ |
| 4 | $\cos(k+\frac{1}{2})(\ell+\frac{1}{2})\frac{\pi}{n}$ | $\sin(k+\frac{1}{2})(\ell+\frac{1}{2})\frac{\pi}{n}$ |
| 5 | $\cos k\ell\frac{\pi}{n-\frac{1}{2}}$ | $\sin(k+1)(\ell+1)\frac{\pi}{n+\frac{1}{2}}$ |
| 6 | $\cos k(\ell+\frac{1}{2})\frac{\pi}{n-\frac{1}{2}}$ | $\sin(k+1)(\ell+\frac{1}{2})\frac{\pi}{n+\frac{1}{2}}$ |
| 7 | $\cos(k+\frac{1}{2})\ell\frac{\pi}{n-\frac{1}{2}}$ | $\sin(k+\frac{1}{2})(\ell+1)\frac{\pi}{n+\frac{1}{2}}$ |
| 8 | $\cos(k+\frac{1}{2})(\ell+\frac{1}{2})\frac{\pi}{n+\frac{1}{2}}$ | $\sin(k+\frac{1}{2})(\ell+\frac{1}{2})\frac{\pi}{n-\frac{1}{2}}$ |

TABLE V

4 TYPES OF SKEW DTTS AND ASSOCIATED SIGNAL MODELS. THE PARAMETER $r$ IS IN $0 \le r \le 1$. FOR $r = 1/2$ THEY REDUCE TO THE $T$-GROUP DTTS.

| $\mathcal{F}$ | $p = p(x)$ | $b$ | $f = f(\theta),\ \cos\theta = \alpha_\ell$ |
|---|---|---|---|
| DCT-3($r$) | $T_n - \cos r\pi$ | $T_k$ | 1 |
| DST-3($r$) | $T_n - \cos r\pi$ | $U_k$ | $\sin(\theta)$ |
| DCT-4($r$) | $T_n - \cos r\pi$ | $V_k$ | $\cos(\theta/2)$ |
| DST-4($r$) | $T_n - \cos r\pi$ | $W_k$ | $\sin(\theta/2)$ |

For the DCTs of types 1,3,5,7, the scaling function is 1 (Table III) and thus they are equal to their polynomial counterpart. We will later see that in some cases, the polynomial DTTs have a lower arithmetic cost than the corresponding DTTs, which makes them suitable choices in application, where the transform output is scaled.

We divide the DTTs into 4 groups, called $T$-, $U$-, $V$-, and $W$-group depending on $p$ as shown in Table III. Within each group, the algebra and module are (almost) the same. This leads to sparse relationships between DTTs in one group as we have shown in [2]; examples we will use are in Appendix III.

Further, within a group, DTTs are pairwise *dual* (they have flipped associated boundary conditions [2]), which means that they can be translated into each other without additional arithmetic operations (see (105) in Appendix III).

**Skew DTTs.** We introduced the skew DTTs in [3] since their associated signal models are also reasonable space models, but, more importantly, because they are important building blocks of Cooley-Tukey type algorithms as we will show in this paper. There are 4 types of skew DTTs, each parameterized by $0 \le r \le 1$. They generalize the four $T$-group DTTs (DCT/DST of type 3/4) and have the same scaling functions as these do. The models that define these transforms are shown in Table V. The corresponding polynomial versions are again denoted using a bar as in $\overline{\mathrm{DCT\text{-}3}}_n(r)$.

To obtain the exact form of these transforms, we need the zeros of the polynomial $T_n - \cos r\pi$ and choose an order of these zeros. This is done in the following lemma.

*Lemma 1* Let $0 \le r \le 1$. We have the factorization

$$T_n - \cos r\pi = 2^{n-1} \prod_{0\le i<n} (x - \cos \tfrac{r+2i}{n}\pi), \tag{9}$$

which determines the zeros of $T_n - \cos r\pi$. We order the zeros as $\alpha = (\cos r_0\pi, \dots, \cos r_{n-1}\pi)$, such that $0 \le r_i \le 1$, and $r_i < r_j$ for $i < j$. The list of the $r_\ell$ is given by the concatenation

$$(r_\ell)_{0\le\ell<n} = \bigcup_{0\le i<n/2} (\tfrac{r+2i}{n}, \tfrac{2-r+2i}{n})$$

for $n$ even, and by

$$(r_\ell)_{0\le\ell<n} = \Big( \bigcup_{0\le i<\frac{n-1}{2}} (\tfrac{r+2i}{n}, \tfrac{2-r+2i}{n})\Big) \cup (\tfrac{r+n-1}{n})$$

for $n$ odd. In the particular case of $r = 1/2$ or $\cos r\pi = 0$, we thus have $\alpha = (\cos(\ell + 1/2)\pi/n)_{0\le\ell<n}$ as in Table XXIII in Appendix II.

For example, the $\text{DCT-3}_n(r)$ is given by the matrix

$$\text{DCT-3}_n(r) = [\cos k r_\ell \pi]_{0\le k,\ell<n},$$

where the $r_\ell$ are provided by Lemma 1.

Relationships between the skew DTTs and skew and non-skew DTTs are shown in Appendix III.

## III. Background: Fast Transform Algorithms

In this section, we explain the notation that we use to represent and manipulate transform algorithms followed by a brief discussion on the quality of algorithms.

### A. Representation of Algorithms

We discuss two representations for transforms[1] and their algorithms. Traditionally, transforms in SP are written as summation like

$$y_k = \sum_{0\le\ell<n} t_{k,\ell} s_\ell, \tag{10}$$

where $\mathbf{s} = (s_0, \dots, s_{n-1})^T$ is the input signal, $\mathbf{y} = (y_0, \dots, y_{n-1})^T$ the output signal, and $t_{k,\ell}$ the transform coefficients. This representation is usually adopted because these transforms are thought of as truncated versions of infinite series expansions. Correspondingly, algorithms are written as sequences of such summations, cleverly organized to reduce the operations count.

A different approach, equivalent in content, represents transforms as matrix-vector products

$$\mathbf{y} = T\mathbf{s}, \quad \text{where } T = [t_{k,\ell}]_{0\le k,\ell<n}. \tag{11}$$

The transform matrix is $T$, and transform algorithms correspond to factorizations of $T$ into a product of sparse structured matrices. This approach was adopted for the DFT in [35], [32], but also for other transforms in various research papers on fast transform algorithms.

[1] By "transforms," we mean here those computing some sort of spectrum of finite length discrete signals like the DFT or DTTs.

In the algebraic signal processing theory, we adopt the second approach for two reasons. First, transforms (in a very general sense) arise in the theory as matrices, namely as decompositions of signal models (which includes a chosen basis) into its spectral components by base changes. More importantly, transform algorithms are derived in the algebraic theory through a decomposition of the model in steps, where the steps correspond to sparse base changes or sparse matrices.

Second, we will argue below that there are many advantages of the matrix representation from an algorithmic and implementation point of view.

**Notation.** As mentioned above, we represent transform algorithms as sparse structured matrix factorizations. These are built from basic matrices and operators.

As basic matrices, we use the $n \times n$ identity matrix $\mathrm{I}_n$, the opposite identity matrix $\mathrm{J}_n$ ($\mathrm{I}_n$ with the columns in reversed order), and the butterfly matrix

$$\mathrm{F}_2 = \begin{bmatrix} 1 & 1 \\ 1 & -1 \end{bmatrix}.$$

Further, we use permutation matrices; most importantly the $n \times n$ stride permutation matrix, which can be defined for $m|n$ by

$$\mathrm{L}^n_m : \ i_2 \tfrac{n}{m} + i_1 \mapsto i_1 m + i_2, \quad 0 \le i_1 < \tfrac{n}{m},\ 0 \le i_2 < m. \tag{12}$$

This definition shows that $\mathrm{L}^n_m$ transposes a $\frac{n}{m} \times m$ matrix stored in row-major order. Alternatively, we can write

$$\begin{aligned} \mathrm{L}^n_m : \quad & i \mapsto im \bmod n-1, \quad \text{for } 0 \le i < n-1, \\ & n-1 \mapsto n-1. \end{aligned}$$

Since the last point $n-1$ is fixed, we can define an *odd* stride permutation $\widehat{\mathrm{L}}$ for $m \mid n+1$ as the restriction of $\mathrm{L}^{n+1}_m$ to the first $n$ points,

$$\widehat{\mathrm{L}}^n_m : \ i \mapsto im \bmod n. \tag{13}$$

Analogous to the stride permutation, $(\widehat{\mathrm{L}}^n_m)^{-1} = \widehat{\mathrm{L}}^n_{(n+1)/m}$, and

$$\mathrm{L}^n_m = \widehat{\mathrm{L}}^{n-1}_m \oplus \mathrm{I}_1 .$$

Other permutation matrices may be defined by their corresponding permutation

$$P : \ i \mapsto f(i), \quad 0 \le i < n,$$

which means that the matrix $P$ has in row $i$ the entry 1 at position $f(i)$ and 0 else. In this paper, matrix indices start with 0.

Diagonal matrices are written as $\mathrm{diag}(\alpha_0, \dots, \alpha_{n-1})$.

Other matrices that serve as building blocks will be defined as needed.

Further we use matrix operators, like the product of matrices, the direct sum

$$A \oplus B = \begin{bmatrix} A & \\ & B \end{bmatrix},$$

and the Kronecker or tensor product

$$A \otimes B = [a_{k,\ell} B]_{k,\ell}, \quad \text{for } A = [a_{k,\ell}].$$

Often, we will also construct a larger matrix as a matrix of matrices, e.g.,

$$\begin{bmatrix} A & B \\ B & A \end{bmatrix}.$$

**Transposition and Inversion.** If an algorithm for a transform is given as a product of sparse matrices built from the constructs above, then an algorithm for the transposed or inverse of the transform can be readily derived using mathematical properties including

$$\begin{aligned} (AB)^T &= B^T A^T, & (AB)^{-1} &= B^{-1}A^{-1}, \\ (A \oplus B)^T &= A^T \oplus B^T, & (A \oplus B)^{-1} &= A^{-1} \oplus B^{-1}, \\ (A \otimes B)^T &= A^T \otimes B^T, & (A \otimes B)^{-1} &= A^{-1} \otimes B^{-1}. \end{aligned}$$

Permutation matrices are orthogonal, i.e., $P^T = P^{-1}$. The transposition or inversion of diagonal matrices is obvious. Note that in general the inverse of a sparse matrix is not sparse.

**Advantages of representation.** We believe the structured representation of matrices to be advantageous because of the following reasons: 1) The representation is more visual and thus easier to grasp by human readers than nested sums with many indices. 2) Algorithms can be easier manipulated, e.g., transposed or inverted, using mathematical properties. 3) Properties of the algorithm, e.g., orthogonality, or which parts can be computed in parallel, can be readily identified. 4) Finally, the sparse matrix product representation can be automatically translated into programs using SPIRAL [36], [37].

Even though our approach simplifies the derivation of algorithms, the sheer number of matrices and cost formulas in the remainder of the paper makes it hard to assure correctness. We solved this problem using two computer algebra systems. Firstly, we used SPIRAL, which includes a modified version of GAP/AREP [38], [39], and provides infrastructure for working with structured matrices as shown here. This way, all formulas in the paper were formally verified (for several problem sizes). Secondly, we used Maple [40] to solve the numerous recurrences in our cost analysis.

**Quality of algorithms.** There are many different factors that determine the quality of a given transform algorithm; the relative importance of these factors is determined by the chosen implementation platform and the specific requirements of the application context. While traditionally the arithmetic cost of transform algorithms is the focus of analysis, the characteristics of modern platforms make the algorithmic structure an equally important feature. Further, the numerical stability of an algorithm is important to ensure the accuracy of the output, in particular for fixed-point implementation. Because of this, the knowledge of the entire algorithm space for a transform is not just of academic interest.

**Arithmetic cost.** We will analyze the number of operations of the algorithms presented below; we will use the notation of a triple $(a, m, m_2)$, where $a$ is the number of additions or subtractions, $m_2$ the number of multiplications by a 2-power not equal to 1, and $m$ the number of remaining multiplications by constants not equal to $-1$. The total operations count is then given by $f = a + m + m_2$.

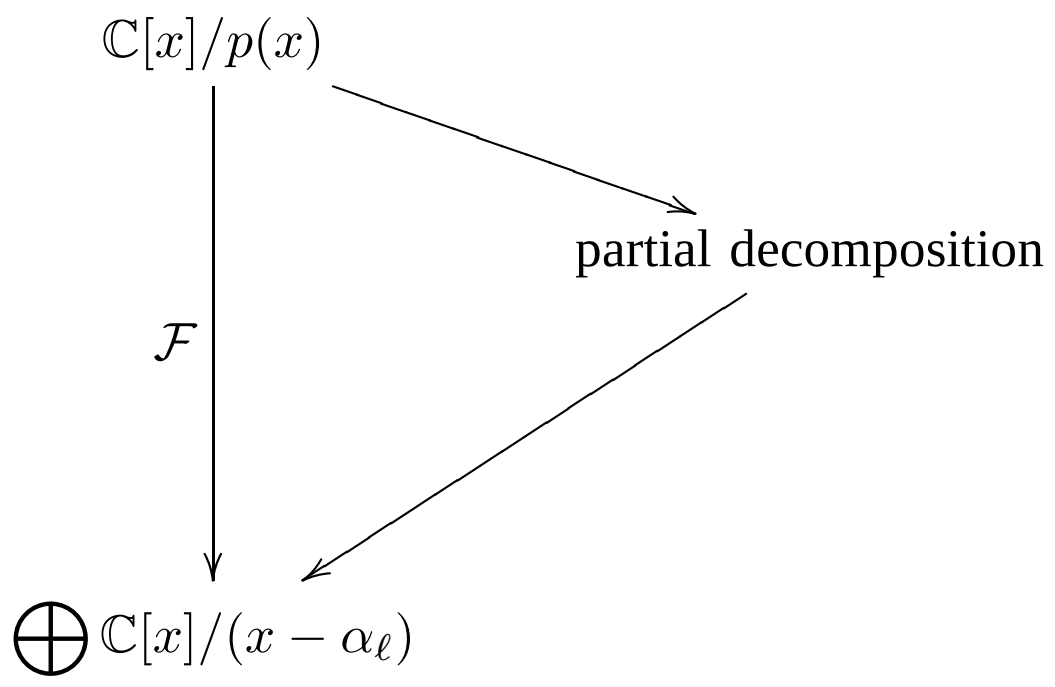

Fig. 1. Basic idea behind the algebraic derivation of Cooley-Tukey type algorithms for a Fourier transform $\mathcal{F}$.

In many SP publications the term complexity is used for the operations count or arithmetic cost . In a strict sense this is not correct, since complexity is a property of a problem (like computing a DFT), not of an algorithm (like a specific FFT). Thus we will use the term cost.

## IV. Algebraic Derivation of Fast Transform Algorithms for 1-D Polynomial Algebras

In this section, we start with our algebraic theory of Fourier transform algorithms, where the term "Fourier transform" is meant in the general sense of the algebraic signal processing theory (e.g., including the DCTs, DSTs, and other trigonometric transforms). As mentioned before, we consider only finite shift-invariant regular signal models, i.e., models of the form $\mathcal{A} = \mathcal{M} = \mathbb{C}[x]/p(x)$ and

$$\Phi : \mathbb{C}^n \to \mathcal{M}, \quad \mathbf{s} \mapsto \sum_{0 \le \ell < n} s_\ell p_\ell,$$

where $b = (p_0, \ldots, p_{n-1})$ is a basis for $\mathcal{M}$. Further, we assume that $p$ has pairwise different zeros, which causes the spectrum to consist of distinct one-dimensional submodules. The Fourier transform in these cases is given by the CRT (5) and is viewed as a matrix (6).

Assume a transform $\mathcal{F}$ is given as a matrix. The algebraic approach derives algorithms by manipulating the associated signal model $(\mathcal{A}, \mathcal{M}, \Phi)$, not by manipulating the matrix entries of $\mathcal{F}$. Fig. 1 visualizes this approach for $\mathcal{A} = \mathcal{M} = \mathbb{C}[x]/p(x)$. We saw in (6) that $\mathcal{F}$ decomposes $\mathbb{C}[x]/p(x)$ into one-dimensional polynomial algebras: its spectrum. Fast algorithms arise, as we will show, by performing this decomposition in steps using an intermediate submodule and associated subalgebra. This technique naturally leads to recursive algorithms, i.e., algorithms that decompose transforms into a product of sparse matrices including smaller transforms of the same or a different type. The advantage of the algebraic derivation is that it identifies a few general principles that account for many different algorithms when instantiated for different transforms. Further, the derivation is often greatly simplified, as we hope it will become clear, since the only task required is reading of base change matrices.

In this paper, we focus on explaining and deriving, as we will call them, "Cooley-Tukey type" algorithms. As the name suggests, these algorithms will include, as well as generalize, the equally named algorithms for the DFT. The latter will serve as examples in this section. Our main focus in the remainder of this paper will then be the derivation of analogous algorithms for the DCTs and DSTs, most of which have not been reported in the literature. All these new algorithms are non-orthogonal, i.e., are not constructed exclusively from butterflies and $2 \times 2$ rotations. Orthogonal algorithms do exist and will be captured algebraically in a future paper. Also "Rader" type algorithms, which apply when the above decomposition methods fail (for the DFT in the case of a prime size), will be explained in a future paper.

The existence and usefulness of algorithms for one of the above signal models relies on both $p(x)$ and $b$. Specifically, algorithms may arise from two different basic principles, which manifest themselves as a property of $p$:

1) *Cooley-Tukey type (factorization)*: $p(x) = q(x) \cdot r(x)$ factorizes; and
2) *Cooley-Tukey type (decomposition)*: $p(x) = q(r(x))$ decomposes.

Clearly, 1) is always possible (if we consider the basefield $\mathbb{C}$), but 2) is a special property of $p$. In each of these cases, as we will show, we obtain a matrix factorization of $\mathcal{F}$; its usefulness as a fast algorithm, however, depends on $b$.

In the remainder of this section, we derive the general form of two types of recursive algorithms based on the above. In each case the algorithm is derived by a stepwise decomposition of $\mathcal{M} = \mathbb{C}[x]/p(x)$ with basis $b$. We focus on Fourier transforms that are polynomial transforms $\mathcal{F} = \mathcal{P}_{b,\alpha}$. Since general Fourier transforms have the form $\mathcal{F} = D \operatorname{diag} \mathcal{P}_{b,\alpha}$, $D$ a diagonal matrix, the results can be readily extended.

### A. Cooley-Tukey Type Algorithms: Factorization

A simple way to decompose $\mathbb{C}[x]/p(x)$ in steps is to use a factorization $p(x) = q(x) \cdot r(x)$ of $p$. Formally, let $k = \deg(q)$ and $m = \deg(r)$, then

$$\begin{aligned} & \mathbb{C}[x]/p(x) \\ \to\ & \mathbb{C}[x]/q(x) \oplus \mathbb{C}[x]/r(x) && (14) \\ \to\ & \bigoplus_{0\le i<k} \mathbb{C}[x]/(x-\beta_i) \oplus \bigoplus_{0\le i<m} \mathbb{C}[x]/(x-\gamma_j) && (15) \\ \to\ & \bigoplus_{0\le \ell<n} \mathbb{C}[x]/(x-\alpha_\ell). && (16) \end{aligned}$$

Here the $\beta_i$ are the zeros of $q$ and the $\gamma_j$ are the zeros of $r$, i.e., both are a subset of the zeros $\alpha_k$ of $p$. Both steps (14) and (15) use the Chinese remainder theorem, whereas (16) is just a reordering of the spectrum. The corresponding factorization of the Fourier transform is provided in the following theorem.

*Theorem 1 (Cooley-Tukey Type Algorithm by Factorization)* Let $p(x) = q(x) \cdot r(x)$, and $c$ and $d$ be a basis of $\mathbb{C}[x]/q(x)$ and $\mathbb{C}[x]/r(x)$, respectively. Further, denote by $\beta$ and $\gamma$ the lists of zeros of $q$ and $r$, respectively. Then

$$\mathcal{P}_{b,\alpha} = P(\mathcal{P}_{c,\beta} \oplus \mathcal{P}_{d,\gamma})B,$$

In particular, the matrix $B$ corresponds to the base change in (14) mapping the basis $b$ to the concatenation $(c, d)$ of the bases $c$ and $d$, and $P$ is the permutation matrix mapping the concatenation $(\beta, \gamma)$ to the list of zeros $\alpha$ in (16).

Note that the factorization of $\mathcal{P}_{b,\alpha}$ in Theorem 1 is useful as a fast algorithm, i.e., reduces the arithmetic cost, only if $B$ is sparse or can be multiplied with efficiently. Referring to Fig. 1, the "partial decomposition" is step (14).

We consider next two examples: the DFT, which will justify why we refer to algorithms based on Theorem 1 as "Cooley-Tukey type," and then the more general case of a Vandermonde matrix, which is a (polynomial) Fourier transform for the generic finite time model.

**Example: DFT.** The DFT is a (polynomial) Fourier transform for the regular signal model given by $\mathcal{A} = \mathcal{M} = \mathbb{C}[x]/(x^n - 1)$ with basis $b = (1, x, \dots, x^{n-1})$. For the example, we assume $n = 2m$ and use the decomposition $x^n - 1 = (x^m - 1)(x^m + 1)$. Applying Theorem 1 yields the following decomposition steps:

$$\begin{aligned} & \mathbb{C}[x]/(x^n - 1) \\ \to\ & \mathbb{C}[x]/(x^m - 1) \oplus \mathbb{C}[x]/(x^m + 1) && (17) \\ \to\ & \bigoplus_{0\le i<m} \mathbb{C}[x]/(x-\omega_n^{2i}) \oplus \bigoplus_{0\le i<m} \mathbb{C}[x]/(x-\omega_n^{2i+1}) && (18) \\ \to\ & \bigoplus_{0\le \ell<n} \mathbb{C}[x]/(x-\omega_n^{\ell}). && (19) \end{aligned}$$

As bases in the smaller modules $\mathbb{C}[x]/(x^m - 1)$ and $\mathbb{C}[x]/(x^m + 1)$, we choose $c = d = (1, x, \dots, x^{m-1})$. We note that from this point on the derivation of the algorithm is entirely mechanical.

First, we derive the base change matrix $b$ corresponding to (17). To do so, we have to express the base elements $x^k \in b$ in the basis $(c, d)$ (concatenation); the coordinate vectors are the columns of $B$. For $0 \le k < m$, $x^k$ is actually contained in $c$ and $d$, so the first $m$ columns of $B$ are

$$B = \begin{bmatrix} \mathrm{I}_m & * \\ \mathrm{I}_m & * \end{bmatrix},$$

where the entries $*$ are determined next. For the base elements $x^{m+k}$, $0 \le k < m$, we have

$$\begin{aligned} x^{m+k} &\equiv x^k \bmod (x^m - 1), \\ x^{m+k} &\equiv -x^k \bmod (x^m + 1), \end{aligned}$$

which yields the final result

$$B = \begin{bmatrix} \mathrm{I}_m & \mathrm{I}_m \\ \mathrm{I}_m & -\mathrm{I}_m \end{bmatrix} = \mathrm{DFT}_2 \otimes \mathrm{I}_m .$$

Next, we consider step (18). $\mathbb{C}[x]/(x^m - 1)$ is decomposed by a $\mathrm{DFT}_m$ and $\mathbb{C}[x]/(x^m + 1)$ by a DFT-$3_m$ (Table III). Finally, the permutation in step (19) is the perfect shuffle $\mathrm{L}^n_m$, which interleaves the even and odd spectral components (even and odd exponents of $\omega_n$). The algorithm obtained is

$$\mathrm{DFT}_n = \mathrm{L}^n_m(\mathrm{DFT}_m \oplus \text{DFT-}3_m)(\mathrm{DFT}_2 \otimes \mathrm{I}_m).$$

To obtain a better known form, we apply the fact that DFT-$3_m = \mathrm{DFT}_m \cdot D_m$, where $D_m = \operatorname{diag}_{0\le i<m}(\omega_n^i)$ to get

$$\begin{aligned} \mathrm{DFT}_n &= \mathrm{L}^n_m(\mathrm{DFT}_m \oplus \mathrm{DFT}_m D_m)(\mathrm{DFT}_2 \otimes \mathrm{I}_m) \\ &= \mathrm{L}^n_m(\mathrm{I}_2 \otimes \mathrm{DFT}_m)(\mathrm{I}_m \oplus D_m)(\mathrm{DFT}_2 \otimes \mathrm{I}_m) \end{aligned}$$

The last expression is a radix-2 decimation-in-frequency Cooley-Tukey FFT; the corresponding decimation-in-time version is obtained by transposition using that the DFT is symmetric. The entries of the diagonal matrix $\mathrm{I}_m \oplus D_m$ are commonly called *twiddle factors*.

**Example: Vandermonde matrix.** As a second example, we consider now the general case of a separable polynomial $p(x)$ with zeros $\alpha_k$, $0 \leq k < n$, but keep the basis $b = (1, x, \dots, x^{n-1})$ of $\mathbb{C}[x]/p(x)$. The associated regular signal model is the generic case of a finite time model, the "timeness" being due to the chosen basis, see [1]. The corresponding (polynomial) Fourier transform is given by the Vandermonde matrix

$$\mathcal{F} = [\alpha_k^\ell]_{0 \leq k,\ell < n}.$$

To derive a fast algorithm for $\mathcal{F}$, we assume that $n = 2m$. We choose an arbitrary factorization $p(x) = q(x) \cdot r(x)$ with $\deg(q) = \deg(r) = m$ and use Theorem 1 to obtain a factorization of the form

$$\mathcal{F} = P(\mathcal{F}_1 \oplus \mathcal{F}_2)B, \tag{20}$$

where $\mathcal{F}_1, \mathcal{F}_2$ are Vandermonde matrices for $\mathbb{C}[x]/q(x)$ and $\mathbb{C}[x]/r(x)$, respectively, and $B$ has the form

$$B = \begin{bmatrix} \mathrm{I}_m & A \\ \mathrm{I}_m & A' \end{bmatrix}.$$

It can be shown that $A$ and $A'$ are both a product of two Toeplitz matrices [41] and can thus be multiplied with using $O(n \log(n))$ operations. If $n$ is a 2-power, then recursive application of (20) hence yields an $O(n \log^2(n))$ algorithm for $\mathcal{F}$.

**Remarks.** Theorem 1 is well-known, as it is the CRT for polynomials expressed in matrix form. The above DFT example is equivalent to the derivation in [20]. Theorem 1 is also used as the first step in the derivation of Winograd DFT algorithms [26]. There, the polynomial $x^n - 1$ is completely factored over the rational numbers, and the DFT decomposed accordingly.

The algorithm derivation method in Theorem 1 is always applicable if the basefield is $\mathbb{C}$, but in general the base change matrix $B$ will be dense and without useful structure. Otherwise, every polynomial transform would have a fast algorithm, which by the current state of knowledge is not the case. The subsequent methods are different in that respect, they require a special property of $p(x)$, and only this property leads to the typical Cooley-Tukey FFT structure for general radices.

### *B. Cooley-Tukey Type Algorithms: Decomposition*

A more interesting factorization of $\mathcal{F} = \mathcal{P}_{b,\alpha}$ can be derived if $p(x)$ *decomposes* into two polynomials, $p(x) = q(r(x))$. If $\deg(q) = k$ and $\deg(r) = m$, then $\deg(p) = n = km$, i.e., the degree of $p$ is necessarily composite. In this case, the polynomial $r(x)$ generates a subalgebra $\mathcal{B}$ of $\mathcal{A} = \mathbb{C}[x]/p(x)$ consisting of all polynomials in $r(x)$. Setting $y = r(x)$ makes the structure of $\mathcal{B}$ evident: $\mathcal{B} = \mathbb{C}[y]/q(y)$.

Let $\beta = (\beta_0, \dots, \beta_{k-1})$ be the zeros of $q$ and let $\alpha'_i = (\alpha'_{i,0}, \dots, \alpha'_{i,m-1})$ be the zeros of $r(x) - \beta_i$, $0 \leq i < k$. Then

$$\begin{aligned} p(x) &= \prod_{0 \leq i < k} (r(x) - \beta_i) \\ &= \prod_{0 \leq i < k} \prod_{0 \leq j < m} (x - \alpha'_{i,j}). \end{aligned}$$

In particular, each $\alpha'_{i,j}$ is a zero $\alpha_\ell$ of $p$. Now we decompose $\mathbb{C}[x]/p(x)$ in the following steps:

$$\begin{aligned} \mathbb{C}[x]/p(x) &\to \mathbb{C}[x]/q(r(x)) && (21) \\ &\to \bigoplus_{0 \leq i < k} \mathbb{C}[x]/(r(x) - \beta_i) && (22) \\ &\to \bigoplus_{0 \leq i < k} \bigoplus_{0 \leq j < m} \mathbb{C}[x]/(x - \alpha'_{i,j}) && (23) \\ &\to \bigoplus_{0 \leq \ell < n} \mathbb{C}[x]/(x - \alpha_\ell). && (24) \end{aligned}$$

Steps (22) and (23) use the Chinese remainder theorem. To derive the corresponding factorization of $\mathcal{P}_{b,\alpha}$ into four factors, we choose a basis $c = (q_0, \dots, q_{k-1})$ for $\mathbb{C}[y]/q(y)$, and for each $\mathbb{C}[x]/(r(x) - \beta_i)$ in (22) the *same basis* $d = (r_0, \dots, r_{m-1})$ . Then, in the first step (21), we do not change $\mathcal{A}$ but only make a base change in $\mathcal{A}$ from the given basis $b$ to the new basis

$$\begin{aligned} b' = \; & (r_0 q_0(r), \dots, r_{m-1} q_0(r), \\ & \dots \\ & r_0 q_{k-1}(r), \dots, r_{m-1} q_{k-1}(r)), \end{aligned} \tag{25}$$

which is a product of the "coarse" basis of the subalgebra $\mathcal{B} \leq \mathcal{A}$ with the "fine" common basis of the $\mathbb{C}[x]/(r(x) - \beta_i)$. We call $B$ the base change matrix for $b \to b'$.

Next, we compute the base change matrix $M$ corresponding to the coarse decomposition (22) with respect to the basis $b'$ in $\mathbb{C}[x]/p(x)$ and the basis $d$ in each summand on the right hand side. Let $r_\ell(x) q_j(r(x)) \in b'$. Then

$$r_\ell(x) q_j(r(x)) \equiv r_\ell(x) q_j(\beta_i) \bmod (r(x) - \beta_i),$$

which is $q_j(\beta_i)$ times the $\ell$th base vector $r_\ell(x)$ in $d$. Thus we get

$$M = [q_j(\beta_i) \cdot \mathrm{I}_m]_{0 \leq i,j < k} = \mathcal{P}_{c,\beta} \otimes \mathrm{I}_m.$$

The third step (23) decomposes the summands in (22) by Fourier transforms $\mathcal{P}_{d,\alpha'_i}$, respectively. The final step (24) reorders the one-dimensional summands by a suitable permutation $P$. We summarize the resulting factorization in the following theorem.

*Theorem 2 (Cooley-Tukey Type Algorithms by Decomposition)* Let $p(x) = q(r(x))$. Using previous notation,

$$\mathcal{P}_{b,\alpha} = P\Big(\bigoplus_{0 \leq i < k} \mathcal{P}_{d,\alpha'_i}\Big)(\mathcal{P}_{c,\beta} \otimes \mathrm{I}_m)B,$$

where $B$ is the base change matrix mapping $b$ to $b'$, and $P$ is the permutation matrix mapping the concatenation of the $\alpha'_i$ onto $\alpha$ in (24).

As in Theorem 1, the usefulness of the factorization as fast algorithm depends on the base change matrix $B$. Referring to Fig. 1, the "partial decomposition" is step (22).

Note that the intermediate decomposition step in (22) has $k$ summands, whereas the intermediate step in (14) has only 2 summands. However, this difference is not the point, as Theorem 1 could be easily extended to more than 2 summands. It is the decomposition property of $p(x)$ that creates a subalgebra generated by $r(x)$, which ensures that the conquer step is sparse and has the Kronecker product structure $\mathcal{P}_{c,\beta} \otimes \mathrm{I}_m$, which intuitively is a "coarse" polynomial transform for $\mathbb{C}[x]/p(x)$.

As an example we consider again the DFT.

**Example: DFT.** Let $\mathcal{A} = \mathcal{M} = \mathbb{C}[x]/(x^n - 1)$ with basis $b = (1, x, \ldots, x^{n-1})$ be the regular signal model associated to the $\mathrm{DFT}_n$. Further, assume that $n = km$, which is necessary for decomposition.

The polynomial $p(x) = x^n - 1$ then decomposes

$$x^n - 1 = (x^m)^k - 1, \qquad (26)$$

i.e., $p(x) = q(r(x))$ with $q(x) = x^k - 1$ and $r(x) = x^m$. Thus Theorem 2 is applicable. The zeros of $q(x)$ are $\beta_i = \omega_k^i$, $0 \leq i < k$. Using this theorem's notation, we choose $c = (1, x, \ldots, x^{k-1})$ as basis in $\mathbb{C}[x]/q(x)$, $d = (1, x, \ldots, x^{m-1})$ as basis in the modules $\mathbb{C}[x]/(x^m - \omega_k^i)$. We find that $b' = b$ in (25), which implies $B = \mathrm{I}_n$.

Thus, the matrix $\mathrm{DFT}_k \otimes \mathrm{I}_m$ performs the following coarse decomposition corresponding to (22):

$$\mathbb{C}[x]/(x^n - 1) \rightarrow \bigoplus_{0 \leq i < k} \mathbb{C}[x]/(x^m - \omega_k^i).$$

The modules $\mathbb{C}[x]/(x^m - \omega_k^i)$ are decomposed, respectively, by (7), which takes the form

$$\mathrm{DFT}_m(\omega_k^i) = \mathrm{DFT}_m \cdot \mathrm{diag}_{j=0}^{m-1}(\omega_n^{ij}),$$

namely as

$$\mathbb{C}[x]/(x^m - \omega_k^i) \rightarrow \bigoplus_{0 \leq j < m} \mathbb{C}[x]/(x - \omega_n^{jk+i}).$$

At this point, corresponding to (23), $\mathbb{C}[x]/p(x)$ is completely decomposed, but the spectrum is ordered according to $jk + i$, $0 \leq i < m$, $0 \leq j < k$ ($j$ runs faster). The desired order is $im + j$. Thus, we need to apply the permutation

$$jk + i \mapsto im + j,$$

which is exactly the stride permutation $\mathrm{L}_m^n$ in (12).

In summary, we obtain the Cooley-Tukey decimation-in-frequency FFT with arbitrary radix:

$$\begin{aligned} & \mathrm{L}_m^n \Big( \bigoplus_{0 \leq i < k} \mathrm{DFT}_m \cdot \mathrm{diag}_{j=0}^{m-1}(\omega_n^{ij}) \Big) (\mathrm{DFT}_k \otimes \mathrm{I}_m) \\ = \; & \mathrm{L}_m^n (\mathrm{I}_k \otimes \mathrm{DFT}_m)\, \mathrm{T}_m^n (\mathrm{DFT}_k \otimes \mathrm{I}_m), \end{aligned} \qquad (27)$$

where the matrix $\mathrm{T}_m^n$ is diagonal and usually called the *twiddle matrix*. Transposition of (27) yields the corresponding decimation-in-time version.

Again, we note that after recognizing the decomposition property (26), the derivation is completely mechanical.

**Remarks.** Theorem 2 makes use of the CRT (in (22) and (23)), but it is the decomposition property of $x^n - 1$ that produced the typical structure. The previous work on the algebraic derivation of this FFT did not make use of decompositions. As we briefly discuss next, the decomposition is a special case of a more general algebraic principle.

### C. *Remarks on Algebraic Principles*

The algorithms derived in this section are based on the factorization or decomposition of the polynomial $p(x)$ in the signal model provided by $\mathbb{C}[x]/p(x)$ (and basis $b$). This is pleasantly simple, but it is also of interest to identify the (more general) principle from the representation theory of algebras that lies behind that. This is important, as other signal models may not be regular or represented by a polynomial algebra in one variable, but the algebraic principle may still apply.

We focus on the decomposition property of $p(x)$ and be brief, assuming some familiarity with representation theory. The key concept underlying Theorem 2 is *induction* as implicit in step (21). Namely, $r(x)$ generates a subalgebra $\mathcal{B} = \langle r(x) \rangle \leq \mathcal{A}$, which is isomorphic (setting $y = r(x)$) to $\mathbb{C}[y]/q(y)$. Further, $d = (r_0, \ldots, r_{m-1})$ is a transversal of $\mathcal{B}$ in $\mathcal{A}$, which means $\mathcal{A}$ is a direct sum of the vector spaces $r_i\mathcal{B}$:

$$\mathcal{A} = r_0\mathcal{B} \oplus \ldots \oplus r_{m-1}\mathcal{B}. \qquad (28)$$

This shows that the regular $\mathcal{A}$-module is an induction of the regular $\mathcal{B}$-module with transversal $d$: $\mathcal{A} = \mathcal{B} \uparrow_d \mathcal{A}$. The natural basis of this induction is $b'$ in (25), which has a structure corresponding to (28). The purpose of step (21) is to make this induction explicit, and Theorem 2 is a decomposition theorem for inductions of (regular modules of) polynomial algebras.

This is a satisfying insight since in prior work [42], [43] we derived the corresponding theorem for inductions of (modules of) group algebras, which has a very similar form [42, Th. 2 in the appendix]. Further, we have shown (also in [42]) that at least some of the *orthogonal* DTT algorithms are based on it. Further, we have used already a different generalization of Theorem 2, namely to polynomial algebras in *two* variables (which provide two-dimensional signal models) to derive a Cooley-Tukey type algorithm in [5] for the new transform introduced in [4].

## V. Cooley-Tukey Type DTT Algorithms (Factorization)

In this section, we derive recursive DTT algorithms by applying Theorem 1, i.e., by factorizing the polynomial $p$ in the module $\mathbb{C}[x]/p(x)$ associated to a given $\mathrm{DTT}_n$. To do so, we will use the following *rational* factorizations of Chebyshev polynomials.

*Lemma 2* The following factorizations hold for the Chebyshev polynomials $T, U, V, W$:

i) $T_3 = x(4x^2 - 3)$
ii) $U_{2n-1} = 2U_{n-1}T_n$.
iii) $U_{2n} = V_nW_n$.
iv) $V_{3n+1} = 2V_n(T_{2n+1} - 1/2)$.

v) $W_{3n+1} = 2W_n(T_{2n+1} + 1/2)$.

*Proof:* Follows from the closed form of the polynomials given in Table XXIII and trigonometric identities. ■

The factorizations in Lemma 2 give rise to size 3 algorithms for DTTs in the $T$-group and recursive algorithms for DTTs in the $U$-, $V$-, and $W$-groups. These are derived in the following. We will not provide a cost analysis in this section, since most of the following algorithms are special cases of more general Cooley-Tukey algorithms to be introduced starting from Section VI.

### A. $T$-Group DTT Algorithms for Size 3

We derive algorithms based on Lemma 2, i), i.e., for DTTs in the $T$-group (DTTs of type 3 and 4) of size 3. As an example, we consider a DCT-4$_3$. We start with the polynomial version $\overline{\text{DCT-4}}_3$, which is a polynomial transform for $\mathbb{C}[x]/T_3$ with $V$-basis $(V_0, V_1, V_2) = (1, 2x-1, 4x^2-2x-1)$. The zeros of $T_3$ are $(\sqrt{3}/2, 0, -\sqrt{3}/2)$. The factorization $T_3 = x(4x^2-3)$ yields the stepwise decomposition

$$\begin{aligned} & \mathbb{C}[x]/T_3 \\ \to\ & \mathbb{C}[x]/x \oplus \mathbb{C}[x]/(x^2 - \tfrac{3}{4}) && (29)\\ \to\ & \mathbb{C}[x]/x \oplus \big(\mathbb{C}[x]/(x - \tfrac{\sqrt{3}}{2}) \oplus \mathbb{C}[x]/(x + \tfrac{\sqrt{3}}{2})\big) && (30)\\ \to\ & \mathbb{C}[x]/(x - \tfrac{\sqrt{3}}{2}) \oplus \mathbb{C}[x]/x \oplus \mathbb{C}[x]/(x + \tfrac{\sqrt{3}}{2}). && (31)\end{aligned}$$

We start with the base change in (29) and choose in all three algebras a $V$-basis. The base change matrix $B$ is computed by mapping $(V_0, V_1, V_2)$ and expressing it in the basis on the right side of (29). The coordinate vectors are the columns of $B$. The first column is $(1,1,0)^T$. Because of $V_1 = 2x - 1 \equiv -1 \bmod x$, the second column is $(-1,0,1)^T$. The last column is obtained from $V_2 = 4x^2 - 2x - 1 \equiv -1 \bmod x$ and $4x^2 - 2x - 1 \equiv -2x + 2 = -V_1 + V_0 \bmod 4x^2 - 3$ as $(-1,1,-1)^T$. Step (30) requires polynomial transforms for $\mathbb{C}[x]/x$ and $\mathbb{C}[x]/(x^2 - 3/4)$ with $V$-bases, which are given by

$$[1] \quad \text{and} \quad \begin{bmatrix} V_0(\sqrt{3}/2) & V_1(\sqrt{3}/2) \\ V_0(-\sqrt{3}/2) & V_1(-\sqrt{3}/2) \end{bmatrix} = \begin{bmatrix} 1 & \sqrt{3}-1 \\ 1 & -\sqrt{3}-1 \end{bmatrix}$$

respectively. Finally we have to exchange the first two spectral components in (31). The result is

$$\overline{\text{DCT-4}}_3 = \begin{bmatrix} 0&1&0\\1&0&0\\0&0&1 \end{bmatrix} \begin{bmatrix} 1&0&0\\0&1&\sqrt{3}-1\\0&1&-\sqrt{3}-1 \end{bmatrix} \begin{bmatrix} 1&-1&-1\\1&0&1\\0&1&-1 \end{bmatrix}.$$

The corresponding algorithm for DCT-4$_3$ is obtained by scaling from left with $\operatorname{diag}(\cos(\pi/12), \cos(3\pi/12), \cos(5\pi/12))$ to get

$$\text{DCT-4}_3 = \begin{bmatrix} 0&1&0\\1&0&0\\0&0&1 \end{bmatrix} \begin{bmatrix} \sqrt{1/2} & 0 & 0 \\ 0 & \cos(\pi/12) & \sqrt{1/2} \\ 0 & \cos(5\pi/12) & -\sqrt{1/2} \end{bmatrix} \begin{bmatrix} 1&-1&-1\\1&0&1\\0&1&-1 \end{bmatrix}.$$

Similarly we get algorithms for the other DTTs of size 3 in the $T$-group. Those, which are among the best known ones, are collected in Table VIII in Section VII.

### B. $U$-Group DTT Algorithms

We use Lemma 2, ii) and iii), to derive a complete set of recursive algorithms for DTTs that are in the $U$-group, i.e., for all DTTs of type 1 and 2. As an example, we consider the DCT-2$_n$, $n = 2m$, with associated module $\mathcal{M} = \mathbb{C}[x]/(x-1)U_{n-1}(x)$ and $V$-basis $b = (V_0, \ldots, V_{n-1})$. From Table XXIII, the zeros of $(x-1)U_{n-1}(x)$ are given by $\alpha_k = \cos k\pi/n$, $0 \le k < n$. Using Lemma 2, ii) we decompose $\mathcal{M}$ in steps as

$$\begin{aligned} & \mathbb{C}[x]/(x-1)U_{n-1} \\ \to\ & \mathbb{C}[x]/(x-1)U_{m-1} \oplus \mathbb{C}[x]/T_m && (32)\\ \to\ & \bigoplus \mathbb{C}[x]/(x - \alpha_{2k}) \oplus \bigoplus \mathbb{C}[x]/(x - \alpha_{2k+1}) && (33)\\ \to\ & \bigoplus \mathbb{C}[x]/(x - \alpha_k). && (34)\end{aligned}$$

We also choose a $V$-basis $b' = (V_0, \ldots, V_{m-1})$ in both smaller algebras in (32); thus we know they are decomposed by DCT-2$_m$ and DCT-4$_m$, respectively. To determine the base change matrix $B$ for $b \to (b', b')$ we need to compute $V_i \bmod (x-1)U_{m-1}$ and $V_i \bmod T_m$ for $0 \le i < 2m$. For $0 \le i < m$ this is trivial,

$$V_i \equiv V_i \bmod (x-1)U_{m-1}, \quad V_i \equiv V_i \bmod T_m.$$

For $m \le i < 2m$ this is precisely the signal extension of the two smaller algebras in (32) (see [2]). Since the signal extension is monomial, $B$ is sparse. The equations are

$$\begin{aligned} V_{m+j} &\equiv V_{m-j-1} \bmod (x-1)U_{m-1}, \text{ and} \\ V_{m+j} &\equiv -V_{m-j-1} \bmod T_m. \end{aligned}$$

Thus, the base change matrix is given by

$$B_{2m} = \begin{bmatrix} \mathrm{I}_m & \mathrm{J}_m \\ \mathrm{I}_m & -\mathrm{J}_m \end{bmatrix} = (\mathrm{DFT}_2 \otimes \mathrm{I}_m)(\mathrm{I}_m \oplus \mathrm{J}_m). \qquad (35)$$

The two summands in (32) are decomposed recursively by DCT-2$_m$ and by DCT-4$_m$, respectively, to yield (33). Finally, we obtain (34) by the permutation matrix $\mathrm{L}^{2m}_m$ (see (12)), which interleaves the even and odd $\alpha_k$. As a result, we obtain the well-known recursive algorithm [12]:

$$\text{DCT-2}_n = \mathrm{L}^{2m}_m(\text{DCT-2}_m \oplus \text{DCT-4}_m)B_{2m}.$$

Analogous computations for all transforms in the $U$-group yield the full set of recursive algorithms due to Lemma 2, which are shown in Table VI(a). The formulas use the following building blocks. The base change matrices $B_{2m}$ in (35) and

$$B_{2m+1} = \begin{bmatrix} \mathrm{I}_m & 0 & \mathrm{J}_m \\ 0 & 1 & 0 \\ \mathrm{I}_m & 0 & -\mathrm{J}_m \end{bmatrix}. \qquad (36)$$

Further, they use the stride permutation matrices $\mathrm{L}^{2m}_m$, and the odd stride permutation matrices $\widehat{\mathrm{L}}^{2m+1}_{m+1}$ defined in (13), which reorder the one-dimensional summands into the proper order.

Note that the base change matrices $B_{2m}$ and $B_{2m+1}$ are sparse in the last $m$ columns (see (35) and (36)) because of the monomial signal extension characteristic for the DTTs. This provides another motivation for considering these extensions.

These four algorithms appeared first in the literature (to our best knowledge) in [44], [45], [12], and [46], respectively. Combining Table VI(a) with the many ways of translating DTTs into each other given by duality or base change (see Appendix III) gives a large number of different recursions, many of them, however, with suboptimal arithmetic cost. Apart from the references above, special cases have been derived in [47], [48], [49].

Table VI(a) is complemented by the decompositions in Table VI(b) which are due to Lemma 2, iii). We did not find these in the literature.

As one application, we can use Table VI(b) to obtain DTT algorithms for small sizes, where the smaller DTTs of type 5–8 are base cases. As a simple example, we get

$$\begin{aligned} \mathrm{DCT\text{-}2}_3 &= \widehat{\mathrm{L}}_2^3(\mathrm{DCT\text{-}6}_2 \oplus \mathrm{DCT\text{-}8}_1)B_3 \\ &= \begin{bmatrix} 1&0&0\\0&0&1\\0&1&0 \end{bmatrix} \left( \begin{bmatrix} 1&1\\ 1/2&-1 \end{bmatrix} \oplus \tfrac{\sqrt{3}}{2}\cdot \mathrm{I}_1 \right) \begin{bmatrix} 1&0&1\\0&1&0\\1&0&-1 \end{bmatrix}. \end{aligned} \tag{37}$$

Transposition yields a $\mathrm{DCT\text{-}3}_3$ algorithm, equivalent to the one obtained in Section V-A.

### C. $V$-Group DTT Algorithms

In this section, we derive algorithms from Lemma 2, iv), for all DTTs in the $V$-group, i.e., for all DTTs of type 7 and 8. Since the second factor in this factorization is $T_{2n+1}-1/2$, the skew DTTs (see Section II-B) introduced in [2] come into play. We use $1/2 = \cos \pi/3$. We do not give the detailed derivation, which is analogous to the one in the previous section, but only state the result in Table VI(c) using the following base change matrices and permutations.

$$B^{(C7)}_{3m+2} = \left[\begin{array}{ccc|c} & & & 1/2 \\ & \mathrm{I}_{2m+1} & & \mathrm{I}_m \\ & & & -\mathrm{J}_m \\ \hline 1 & & & -1 \\ & \mathrm{I}_m & -\mathrm{J}_m & -\mathrm{I}_m \end{array}\right]$$

$$B^{(S7)}_{3m+1} = \left[\begin{array}{ccc|c} & & & \mathrm{I}_m \\ & \mathrm{I}_{2m+1} & & \mathrm{J}_m \\ & & & 0 \cdots 0 \\ \hline \mathrm{I}_m & \mathrm{J}_m & \begin{smallmatrix}0\\ \vdots\\ 0\end{smallmatrix} & -\mathrm{I}_m \end{array}\right]$$

$$B^{(C8)}_{3m+1} = \left[\begin{array}{ccc|c} & & & \mathrm{I}_m \\ & \mathrm{I}_{2m+1} & & 0 \cdots 0 \\ & & & -\mathrm{J}_m \\ \hline \mathrm{I}_m & \begin{smallmatrix}0\\ \vdots\\ 0\end{smallmatrix} & -\mathrm{J}_m & -\mathrm{I}_m \end{array}\right]$$

$$B^{(S8)}_{3m+2} = \left[\begin{array}{ccc|c} & & & \mathrm{I}_m \\ & \mathrm{I}_{2m+1} & & 2 \\ & & & \mathrm{J}_m \\ \hline \mathrm{I}_m & & \mathrm{J}_m & -\mathrm{I}_m \\ & 1 & & -1 \end{array}\right]$$

To give the permutation, we decompose the index $i$ into the radix-3 format $i_1 + 3i_2$. Then $P^{3m+2}_m$ operates on the set $\{0, \dots, 3m+1\}$ and is given by

$$\begin{aligned} P^{3m+2}_m &= i_1 + 3i_2 \mapsto \begin{cases} 2i_2, & \text{for } i_1 = 0; \\ i_2 + 2m + 1, & \text{for } i_1 = 1; \\ 2i_2 + 1, & \text{for } i_1 = 2; \end{cases} \\ &= \widehat{L}^{3m+2}_{m+1} \begin{bmatrix} \mathrm{I}_{m+1} & & \\ & & \mathrm{I}_{m+1} \\ & \mathrm{I}_m & \end{bmatrix} (\widehat{L}^{2m+1}_2 \oplus \mathrm{I}_{m+1}). \end{aligned}$$

To give a visual impression of the structure we show $P^{11}_3$ as an example:

$$P^{11}_3 = \begin{bmatrix} 1&0&0&0&0&0&0&0&0&0&0 \\ 0&0&0&0&0&0&0&1&0&0&0 \\ 0&1&0&0&0&0&0&0&0&0&0 \\ 0&0&1&0&0&0&0&0&0&0&0 \\ 0&0&0&0&0&0&0&0&1&0&0 \\ 0&0&0&1&0&0&0&0&0&0&0 \\ 0&0&0&0&1&0&0&0&0&0&0 \\ 0&0&0&0&0&0&0&0&0&1&0 \\ 0&0&0&0&0&1&0&0&0&0&0 \\ 0&0&0&0&0&0&1&0&0&0&0 \\ 0&0&0&0&0&0&0&0&0&0&1 \end{bmatrix}.$$

The permutation $P^{3m+2}_m$ leaves the last point fixed. By restricting $P^{3m+2}_m$ to the set $\{0, \dots, 3m\}$, we obtain and define the permutation $\widehat{P}^{3m+1}_m$.

### D. $W$-Group DTT Algorithms

The corresponding $W$-group (DTTs of type 5 and 6) algorithms are given in Table VI(d) with

$$B^{(C5)}_{3m+2} = \left[\begin{array}{cc|c} 1 & & 1 \\ \mathrm{I}_m & \mathrm{J}_m & \mathrm{I}_m \\ \hline & & -1/2 \\ \mathrm{I}_{2m+1} & & -\mathrm{I}_m \\ & & -\mathrm{J}_m \end{array}\right]$$

$$B^{(S5)}_{3m+1} = \left[\begin{array}{ccc|c} \mathrm{I}_m & -\mathrm{J}_m & \begin{smallmatrix}0\\ \vdots\\ 0\end{smallmatrix} & \mathrm{I}_m \\ \hline & & & -\mathrm{I}_m \\ & \mathrm{I}_{2m+1} & & \mathrm{J}_m \\ & & & 0 \cdots 0 \end{array}\right]$$

$$B^{(C6)}_{3m+2} = \left[\begin{array}{cc|c} \mathrm{I}_m & \mathrm{J}_m & \mathrm{I}_m \\ & 1 & 1 \\ \hline & & -\mathrm{I}_m \\ \mathrm{I}_{2m+1} & & -2 \\ & & -\mathrm{J}_m \end{array}\right]$$

$$B^{(S6)}_{3m+1} = \left[\begin{array}{ccc|c} \mathrm{I}_m & \begin{smallmatrix}0\\ \vdots\\ 0\end{smallmatrix} & -\mathrm{J}_m & \mathrm{I}_m \\ \hline & & & -\mathrm{I}_m \\ & \mathrm{I}_{2m+1} & & 0 \cdots 0 \\ & & & \mathrm{J}_m \end{array}\right]$$

and $Q^{3m+2}_m$ operates on $\{0, \dots, 3m+1\}$ as

$$\begin{aligned} Q^{3m+2}_m &= i_1 + 3i_2 \mapsto \begin{cases} i_2, & \text{for } i_1 = 0; \\ 2i_2 + m + 1, & \text{for } i_1 = 1; \\ 2i_2 + m + 2, & \text{for } i_2 = 2. \end{cases} \\ &= \widehat{L}^{3m+2}_{m+1} \begin{bmatrix} & \mathrm{I}_{m+1} \\ \mathrm{I}_{2m+1} & \end{bmatrix} (\widehat{L}^{2m+1}_2 \oplus \mathrm{I}_{m+1}). \end{aligned}$$

TABLE VI

DTT ALGORITHMS BASED ON FACTORIZATION PROPERTIES OF THE CHEBYSHEV POLYNOMIALS. TRANSPOSITION YIELDS A DIFFERENT SET OF ALGORITHMS. REPLACING EACH TRANSFORM BY ITS POLYNOMIAL COUNTERPART YIELDS ALGORITHMS FOR THE POLYNOMIAL DTTS.

(a) $U$-group: Based on $U_{2n-1} = 2U_{n-1}T_n$

$$\begin{aligned} \text{DCT-1}_{2m+1} &= \widehat{\mathrm{L}}_{m+1}^{2m+1}(\text{DCT-1}_{m+1} \oplus \text{DCT-3}_m)B_{2m+1} \\ \text{DST-1}_{2m-1} &= \widehat{\mathrm{L}}_{m}^{2m-1}(\text{DST-3}_{m} \oplus \text{DST-1}_{m-1})B_{2m-1} \\ \text{DCT-2}_{2m} &= \mathrm{L}_{m}^{2m}(\text{DCT-2}_{m} \oplus \text{DCT-4}_m)B_{2m} \\ \text{DST-2}_{2m} &= \mathrm{L}_{m}^{2m}(\text{DST-4}_{m} \oplus \text{DST-2}_m)B_{2m} \end{aligned}$$

(b) $U$-group: Based on $U_{2n} = V_n W_n$

$$\begin{aligned} \text{DCT-1}_{2m} &= \mathrm{L}_{m}^{2m}(\text{DCT-5}_{m} \oplus \text{DCT-7}_m)B_{2m} \\ \text{DST-1}_{2m} &= \mathrm{L}_{m}^{2m}(\text{DST-7}_{m} \oplus \text{DST-5}_m)B_{2m} \\ \text{DCT-2}_{2m+1} &= \widehat{\mathrm{L}}_{m+1}^{2m+1}(\text{DCT-6}_{m+1} \oplus \text{DCT-8}_m)B_{2m+1} \\ \text{DST-2}_{2m+1} &= \widehat{\mathrm{L}}_{m+1}^{2m+1}(\text{DST-8}_{m+1} \oplus \text{DST-6}_m)B_{2m+1} \end{aligned}$$

(c) $V$-group: Based on $V_{3n+1} = 2(T_{2n+1} - 1/2)V_n$

$$\begin{aligned} \text{DCT-7}_{3m+2} &= P_m^{3m+2}(\text{DCT-3}_{2m+1}(\tfrac{1}{3}) \oplus \text{DCT-7}_{m+1})B_{3m+2}^{(C7)} \\ \text{DST-7}_{3m+1} &= \widehat{P}_m^{3m+1}(\text{DST-3}_{2m+1}(\tfrac{1}{3}) \oplus \text{DST-7}_{m})B_{3m+1}^{(S7)} \\ \text{DCT-8}_{3m+1} &= \widehat{P}_m^{3m+1}(\text{DCT-4}_{2m+1}(\tfrac{1}{3}) \oplus \text{DCT-8}_{m})B_{3m+1}^{(C8)} \\ \text{DST-8}_{3m+2} &= P_m^{3m+2}(\text{DST-4}_{2m+1}(\tfrac{1}{3}) \oplus \text{DST-8}_{m+1})B_{3m+2}^{(S8)} \end{aligned}$$

(d) $W$-group: Based on $W_{3n+1} = 2W_n(T_{2n+1} + 1/2)$

$$\begin{aligned} \text{DCT-5}_{3m+2} &= Q_m^{3m+2}(\text{DCT-5}_{m+1} \oplus \text{DCT-3}_{2m+1}(\tfrac{2}{3}))B_{3m+2}^{(C5)} \\ \text{DST-5}_{3m+1} &= \widehat{Q}_m^{3m+1}(\text{DST-5}_{m} \oplus \text{DST-3}_{2m+1}(\tfrac{2}{3}))B_{3m+1}^{(S5)} \\ \text{DCT-6}_{3m+2} &= Q_m^{3m+2}(\text{DCT-6}_{m+1} \oplus \text{DCT-4}_{2m+1}(\tfrac{2}{3}))B_{3m+2}^{(C6)} \\ \text{DST-6}_{3m+1} &= \widehat{Q}_m^{3m+1}(\text{DST-6}_{m} \oplus \text{DST-4}_{2m+1}(\tfrac{2}{3}))B_{3m+1}^{(S6)} \end{aligned}$$

The permutation $Q_m^{3m+2}$ leaves the first point 0 fixed. By restricting $Q_m^{3m+2}$ to the points $1 \leq i \leq 3m+1$ we obtain and define the permutation $\widehat{Q}_m^{3m+1}$. If we rename the index set into $\{0, \dots, 3m\}$, we have

$$\widehat{Q}_m^{3m+1} \;=\; i_1 \,+\, 3i_2 \;\mapsto\; \begin{cases} 2i_2 + m, & \text{for } i_1 = 0; \\ 2i_2 + m + 1, & \text{for } i_1 = 1; \\ i_2, & \text{for } i_1 = 2. \end{cases}$$

The usefulness of the above algorithms depends on the initial transform size and on the availability of algorithms for the occurring skew DTTs. These algorithms will be introduced later.

### *E. Polynomial DTTs*

Every DTT in Table VI is decomposed into two DTTs that have the same base polynomials. Thus they have the same scaling function (see Table III: $b$ and $f$ are connected), which is the reason why we see no scaling factors in the equations. As an important consequence, we get algorithms corresponding to Table VI for the polynomial transforms $\overline{\text{DTT}}$.

As an example, we derive the polynomial equivalent of (37):

$$\overline{\text{DCT-2}}_3 = \begin{bmatrix} 1&0&0\\0&0&1\\0&1&0 \end{bmatrix} \left( \begin{bmatrix} 1&1\\1&-2 \end{bmatrix} \oplus \mathrm{I}_1 \right) \begin{bmatrix} 1&0&1\\0&1&0\\1&0&-1 \end{bmatrix}, \qquad (38)$$

where $\text{DCT-2}_3 = \operatorname{diag}(1, \frac{\sqrt{3}}{2}, \frac{1}{2}) \cdot \overline{\text{DCT-2}}_3$. The algorithm requires 4 additions and 1 multiplication and is thus 1 multiplication cheaper than its non-polynomial equivalent (37).

### *F. Final Remarks*

The algorithms given in this section are based on Lemma 2, which provides factorizations of the Chebyshev polynomials $T, U, V, W$. Since all these polynomial factorizations are rational, the associated matrix factorizations are also rational. In Lemma 2, ii) and iii), the factors are again Chebyshev polynomials, and thus the smaller transforms in the decomposition are again DTTs. In Lemma 2, iv) and v), the second factor $T_{2n+1} - 1/2$ leads to skew DTTs (see Table V). The complete rational factorization of the Chebyshev polynomials $T_n, U_n$ for arbitrary $n$ is given in [50]. The rational factorization of $V_n$ and $W_n$ can be derived using [50] and Lemma 2, iii). These factorizations can be used to decompose a DTT, but the smaller transforms obtained are in general no DTTs or skew DTTs.

All algorithms in Table VI can be manipulated in numerous ways using the identities in Appendix III or transposition to obtain different algorithms.

## VI. COOLEY-TUKEY TYPE DTT ALGORITHMS (DECOMPOSITION)

In this section, we give a first overview on DTT algorithms that are based on Theorem 2, i.e., on a decomposition $p(x) = q(r(x))$ of the polynomial $p$ in the associated algebra $\mathbb{C}[x]/p$. These algorithms are, structural and in a precise mathematical sense, the equivalent of the Cooley-Tukey FFT (27), which we derived based on the decomposition of $x^n - 1 = (x^m)^k - 1$.

We will see that all 16 DTTs possess such algorithms, and that in many cases there are several reasonable variants with different characteristics to choose from. Some of these algorithms generalize the ones we introduced in Section V.

Each of these "Cooley-Tukey-like" DTT algorithms exhibits the same flexible recursion and regular and versatile structure that has been the success of the FFT. As a consequence, all FFT variants optimized for, e.g., parallel or vector computation will have counterparts for the 16 DTTs. See [32] for more details on FFT variants.

Only very few special cases of these algorithms have been found before. Again, our algebraic methods show their power: the derivation using Theorem 2 is comparatively easy, since only base changes have to be computed; in contrast, a derivation based on matrix entries becomes hopelessly complicated, and, furthermore, does not provide a guideline to which entry and how manipulations should be performed to obtain an algorithm.

**Decomposition of Chebyshev polynomials.** The DTT algorithms are based on the following lemma, which provides decomposition properties of the Chebyshev polynomials.

*Lemma 3* The Chebyshev polynomials $T, U, V, W$ have the following decomposition properties:

i) $T_{km} = T_k(T_m)$; $T_{km} - a = T_k(T_m) - a$, $a \in \mathbb{C}$.
ii) $U_{km-1} = U_{m-1} \cdot U_{k-1}(T_m)$.
iii) $V_{(k-1)/2+km} = V_m \cdot V_{(k-1)/2}(T_{2m+1})$.
iv) $W_{(k-1)/2+km} = W_m \cdot W_{(k-1)/2}(T_{2m+1})$.
v) $T_{km+m/2} = T_{m/2} \cdot V_k(T_m)$.
vi) $U_{km+m/2-1} = U_{m/2-1} \cdot W_k(T_m)$.

*Proof:* Straightforward using the closed form of $T_n$ from Table XXIII. In particular i) is well-known in the literature (e.g., [51]). ■

Inspecting the identities in Lemma 3, we observe that only i) provides a pure decomposition; the other identities are a decomposition up to a factor. Thus, in these cases, the algorithms derivation requires us to first apply Theorem 1 and then Theorem 2.

Also, we observe that Lemma 3 provides decomposition of *all four types* of Chebyshev polynomials. Thus we can expect Cooley-Tukey type algorithms for all 16 DTTs. Looking at Lemma 3, Theorem 2, and its derivation in (21)–(24), we see that the algebras in (22), will all have the form

$$\mathbb{C}[x]/(T_n - \cos r\pi),$$

and thus the decomposition (23) will require skew DTTs, which motivates their introduction in [2]. Of course, this poses the question how to further decompose the skew DTTs for non-trivial sizes. This question is answered by the second identity in Lemma 3, i): $T_n - a$ decomposes exactly as $T_n$ does, which establishes a one-to-one correspondence between algorithms for the DTTs in the $T$-group and their skew counterparts.

Fig. 2 gives an overview on the algorithms that we will derive from Lemma 3. We first organize the algorithms into the groups of DTTs (see also Table III) they apply to. In the $T$- and $U$-group, we have two types of decomposition properties. For algorithms based on $T_n = T_k(T_m)$, we have three further degrees of freedom as will be explained later. In summary, each leaf of the tree in Fig. 2 represents one class consisting of four algorithms for each of the DTTs in the respective group.

None of these algorithms is orthogonal, i.e., they do not decompose the DTTs into rotations (and butterflies). Orthogonal Cooley-Tukey type algorithms are the subject of a future paper.

## VII. $T$-Group DTT Algorithms

In this section we derive the four classes of Cooley-Tukey type algorithms for the four DTTs in the $T$-group shown in Fig. 2. We focus mainly on those algorithms based on $T_n = T_k(T_m)$.

First, we simultaneously derive the algorithms for all four DTTs to emphasize their common structure and their differences. The exact form of the algorithms, i.e., all occurring matrices, will be derived afterwards, including a discussion and cost analysis in each case.

### A. Simultaneous Derivation

We start with a fixed DTT in the $T$-group with associated algebra $\mathbb{C}[x]/T_n$ and $C$-basis $b = (C_0, \dots, C_{n-1})$, where $C \in \{T, U, V, W\}$ depends on the chosen DTT. We assume $n = km$, and use the decomposition $T_n = T_k(T_m)$. The decomposition steps (21)–(24) leading to Theorem 2 take the form

$$\begin{aligned}
\mathbb{C}[x]/T_n &\to \mathbb{C}[x]/T_k(T_m) && (39)\\
&\to \bigoplus_{0\le i<k} \mathbb{C}[x]/(T_m - \cos\tfrac{i+1/2}{k}\pi) && (40)\\
&\to \bigoplus_{0\le i<k}\bigoplus_{0\le j<m} \mathbb{C}[x]/(x - \cos r_{i,j}\pi) && (41)\\
&\to \bigoplus_{0\le i<n} \mathbb{C}[x]/(x - \cos\tfrac{i+1/2}{n}\pi), && (42)
\end{aligned}$$

where the $r_{i,j}$ are determined by Lemma 1.

In the first step (39), we change bases in $\mathbb{C}[x]/T_n = \mathbb{C}[x]/T_k(T_m)$, from the given $C$-basis $b$ to the basis $b'$ given in (25). The question arises, which basis to choose in the coarse algebra $\mathbb{C}[x]/T_k$, and which common basis to choose in the "skew" algebras $\mathbb{C}[x]/(T_m - \cos(i + 1/2)\pi/k)$. In the latter ones, we always choose the same $C$-basis as in the original algebra. For the coarse algebra, it turns out that we have two reasonable choices: a $T$-basis or a $U$-basis. We consider both cases, starting with the $U$-basis.

***U*-basis.** We choose, independently of $C$, a $U$-basis in $\mathbb{C}[x]/T_k$. Note, that the corresponding DTT is a DST-$3_m$ (Table III). The basis $b'$ in (25) is then given by

$$\begin{aligned}
b' &= (C_0U_0(T_m), \dots, C_{m-1}U_0(T_m),\\
&\qquad \dots\\
&\qquad C_0U_{k-1}(T_m), \dots, C_{m-1}U_{k-1}(T_m))\\
&= (C_jU_i(T_m) \mid 0 \le i < m,\ 0 \le j < k).
\end{aligned} \tag{43}$$

We order double indices always lexicographically $(i, j) = (0, 0), (0, 1), \dots$.

We denote the corresponding base change matrix $b \to b'$ in (39) with $\overline{B}^{(*)}_{n,k}$. Here, and in the following, the "$*$" in the superscript means that the matrix depends on the DTT. It will later be replaced by $* \in \{C3, S3, C4, S4\}$ when the precise definitions are derived.

After the base change, the decomposition is straightforward following steps (40)–(42) and Theorem 2. The coarse decomposition in step (40) is obtained with the matrix $\overline{\text{DST-3}}_k \otimes \text{I}_m$, since Theorem 2 requires us to choose a polynomial transform for the coarse decomposition. For step (41), we need a direct sum of skew DTTs: $\bigoplus_{0\le i<k} \text{DTT}_m((i + 1/2)/k)$. These are of the same type as the DTT we started with, since they have the same $C$-basis as the DTT to be decomposed.

Finally, we order the one-dimensional summands in step (42) using a permutation. This permutation does not depend on the basis, but only on the zeros of $T_k$ and $T_m$. Thus it is the same in all four cases of DTTs in the $T$-group, and, using Lemma 1, takes the form

$$\text{K}^n_m = (\text{I}_k \oplus \text{J}_k \oplus \text{I}_k \oplus \text{J}_k \oplus \dots)\,\text{L}^n_m\,.$$

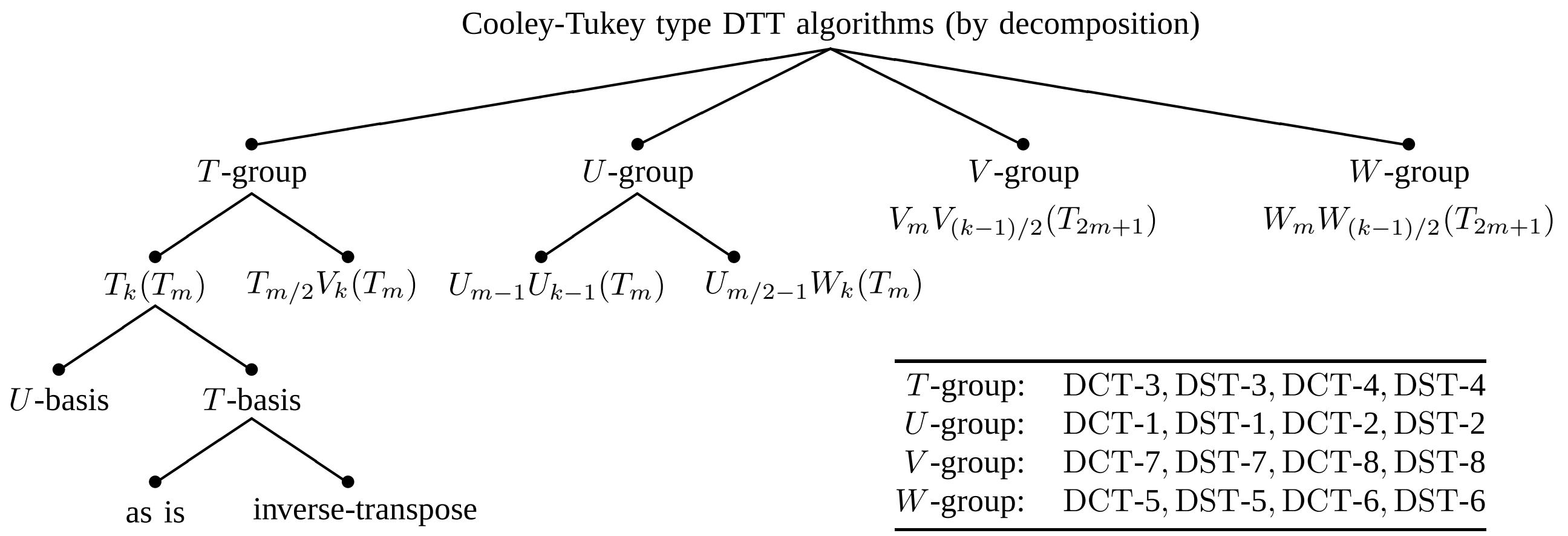

Fig. 2. Overview of Cooley-Tukey type algorithms due to decomposition properties of the Chebyshev polynomials.

This permutation is the equivalent of the stride permutation $\mathrm{L}^n_m$, occurring in the Cooley-Tukey FFT, for the DTTs in the $T$-group.

In summary, we obtain

$$\mathrm{DTT}_n = \mathrm{K}^n_m\Big(\bigoplus_{0\le i<k}\mathrm{DTT}_m(\tfrac{i+1/2}{k})\Big)(\overline{\text{DST-3}}_k\otimes \mathrm{I}_m)\overline{B}^{(*)}_{n,k}. \quad (44)$$

The question that remains is how to decompose the smaller transforms: the skew $\mathrm{DTT}_m$'s and the polynomial $\overline{\text{DST-3}}_k$. However, this poses no problem. Since for any $a\in\mathbb{C}$, $T_n - a$ decomposes exactly as $T_n$, we derive in a completely analogous way the "skew version" of (44) as

$$\mathrm{DTT}_n(r) = \mathrm{K}^n_m\Big(\bigoplus_{0\le i<k}\mathrm{DTT}_m(r_i)\Big)(\overline{\text{DST-3}}_k(r)\otimes \mathrm{I}_m)\overline{B}^{(*)}_{n,k}, \quad (45)$$

which is a generalization of (44), which arises for $r = 1/2$. The numbers $r_i$ are computed from $r$ using Lemma (1). The matrix $\mathrm{K}^n_m$ neither depends on the type of DTT, nor on $r$; the matrix $\overline{B}^{(*)}_{n,k}$ does depend on the type of DTT, but not on $r$, since the bases $b$ and $b'$ are independent of $r$.

For $k = n$, (45) translates a DTT in the $T$-group into a $\overline{\text{DST-3}}$, which is a special case of the translation by base change in Appendix III.

Further, since DTTs and skew DTTs have the same scaling function (Tables III and V), we obtain corresponding algorithms for the polynomial version of the transforms by just replacing each DTT by its polynomial counterpart:

$$\overline{\mathrm{DTT}}_n = \mathrm{K}^n_m\Big(\bigoplus_{0\le i<k}\overline{\mathrm{DTT}}_m(\tfrac{i+1/2}{k})\Big)(\overline{\text{DST-3}}_k\otimes \mathrm{I}_m)\overline{B}^{(*)}_{n,k},$$

and

$$\overline{\mathrm{DTT}}_n(r) = \mathrm{K}^n_m\Big(\bigoplus_{0\le i<k}\overline{\mathrm{DTT}}_m(r_i)\Big)(\overline{\text{DST-3}}_k(r)\otimes \mathrm{I}_m)\overline{B}^{(*)}_{n,k}.$$

The remaining task is to compute the exact form of $\overline{B}^{(*)}_{n,k}$ in the four cases. We will do this in Section VII-B and only mention at this point that in each case, $\overline{B}_{n,k}$ has a very sparse and regular structure.

Next, we derive the analogue of the above algorithms, if a $T$-basis, instead of a $U$-basis is chosen in the coarse module $\mathbb{C}[x]/T_k$.

***T*-basis.** In distinction with the above, we choose this time, independently of $C$, a $T$-basis in $\mathbb{C}[x]/T_k$. Thus, the corresponding DTT is a DCT-$3_m$. The basis $b'$ in (25) is now given by

$$\begin{aligned} b' &= (C_0T_0(T_m),\dots,C_{m-1}T_0(T_m),\\ &\qquad \dots\\ &\qquad C_0T_{k-1}(T_m),\dots,C_{m-1}T_{k-1}(T_m))\\ &= (C_{im-j}/2 + C_{im+j}/2 \mid 0\le i<m, 0\le j<k), \end{aligned} \quad (46)$$

using (104) in Appendix II. We denote the base change matrix for $b\to b'$ by $B^{(*)}_{n,k}$.

The coarse decomposition in step (40) is now performed by the matrix $\text{DCT-3}_k\otimes \mathrm{I}_m$ (note that DCT-3 is a polynomial transform). The remaining steps (41) and (42) are equal to what we had before.

As a result, we obtain

$$\mathrm{DTT}_n = \mathrm{K}^n_m\Big(\bigoplus_{0\le i<k}\mathrm{DTT}_m(\tfrac{i+1/2}{k})\Big)(\text{DCT-3}_k\otimes \mathrm{I}_m)B^{(*)}_{n,k}, \quad (47)$$

and its generalization to the skew DTTs

$$\mathrm{DTT}_n(r) = \mathrm{K}^n_m\Big(\bigoplus_{0\le i<k}\mathrm{DTT}_m(r_i)\Big)(\text{DCT-3}_k(r)\otimes \mathrm{I}_m)B^{(*)}_{n,k}. \quad (48)$$

Again, $B^{(*)}_{n,k}$ only depends on the type of DTT, and not on $r$.

The polynomial version is again given by simply replacing all DTTs by their polynomial counterparts:

$$\overline{\mathrm{DTT}}_n(r) = \mathrm{K}^n_m\Big(\bigoplus_{0\le i<k}\overline{\mathrm{DTT}}_m(r_i)\Big)(\text{DCT-3}_k(r)\otimes \mathrm{I}_m)B^{(*)}_{n,k}. \quad (49)$$

We mentioned above that choosing a $U$-basis in the coarse module $C[x]/T_k$ leads to base change matrices $B_{n,k}$ that are sparse (which will be shown in detail below). For the $T$-basis, this is somewhat different. In fact, inspecting (46) shows that the inverse base change $b' \to b$, i.e., $B_{n,k}^{-1}$ is sparse (with at most two entries in each column). For this reason, we will also consider the inverse of (47) and (48).

***T*-basis inverted.** To express the inverse, we need the inverse skew DTTs (Appendix III). The inverse of (48) will take, after minor simplifications, in each case the general form

$$\mathrm{iDTT}_n(r) = (C_{n,k}^{(*)})^{-1}(\mathrm{iDCT\text{-}3}_k(r) \otimes \mathrm{I}_m) \Big(\bigoplus_{0\le i<k} \mathrm{iDTT}_m(r_i)\Big)\mathrm{M}_k^n, \quad (50)$$

where

$$\mathrm{M}_k^n = (\mathrm{K}_m^n)^{-1} = \mathrm{L}_k^n(\mathrm{I}_k \oplus \mathrm{J}_k \oplus \mathrm{I}_k \oplus \mathrm{J}_k \oplus \dots),$$

and $C_{n,k}^{(*)}$ is closely related to $B_{n,k}^{(*)}$. (50) provides algorithms for the DTTs of type 2 and 4 (the inverses of the DTTs in the $T$-group).

**Variants.** The algorithms derived above can be further manipulated to obtain variants. We saw already an example: the inversion of (48) to obtain (50). One obvious manipulation is transposition, which turns each $T$-group DTT algorithm into an algorithm for a DCT or DST of type 2 or 4 (the transposes of the $T$-group DTTs).

More interestingly, each of the above algorithms has a corresponding "twiddle version," which is obtained by translating skew DTTs into their non-skew counterparts using (108)–(111) in Appendix III. For example, the twiddle version of (47) is given by

$$\mathrm{DTT}_n = \mathrm{K}_m^n\Big(\mathrm{I}_k \otimes \mathrm{DTT}_m\Big)D_{k,m}(\mathrm{DCT\text{-}3}_k \otimes \mathrm{I}_m)B_{n,k}, \quad (51)$$

where

$$D_{k,m} = \bigoplus_{0\le i<k} X_n^{(*)}(\tfrac{i+1/2}{k})$$

is a direct sum of the x-shaped matrices in (108)-(111) (Appendix III).

The twiddle version seems more appealing; however, we will later see that at least in the 2-power case $n = 2^k$ they incur a higher arithmetic cost. The reason is that skew and non-skew DTTs can be computed with the same cost in this case. For other sizes, the twiddle version may not incur any penalty. Most state of the art software implementations [52], [36] fuse the twiddle factors with the subsequent loop incurred by the tensor product anyway to achieve better locality.

**Base cases.** We provide the base cases for the above algorithms for size $n = 2$ in Table VII and for size $n = 3$ in Table VIII. The size 2 cases follow from the definition; most of the size 3 cases were derived in Section V-A. An exception is DCT-4, for which the algorithm was generated by AREP [42], [39], and which is a "Rader-type" algorithm (see [10]; also a future paper will discuss the algebraic origin of the Rader algorithm in detail). The DST-$4_3$ algorithm follows then by duality (105) (in Appendix III).

TABLE VII

BASE CASES FOR NORMAL AND SKEW $T$-GROUP DTTS OF SIZE 2.

$$\overline{\mathrm{DCT\text{-}3}}_2 = \mathrm{F}_2 \cdot \mathrm{diag}(1, 1/\sqrt{2})$$
$$\overline{\mathrm{DST\text{-}3}}_2 = \mathrm{F}_2 \cdot \mathrm{diag}(1, \sqrt{2})$$
$$\overline{\mathrm{DCT\text{-}4}}_2 = \mathrm{F}_2 \cdot \begin{bmatrix} 1 & -1 \\ 0 & \sqrt{2} \end{bmatrix}$$
$$\overline{\mathrm{DST\text{-}4}}_2 = \mathrm{F}_2 \cdot \begin{bmatrix} 1 & 1 \\ 0 & \sqrt{2} \end{bmatrix}$$

$$\mathrm{DCT\text{-}3}_2 = \overline{\mathrm{DCT\text{-}3}}_2$$
$$\mathrm{DST\text{-}3}_2 = \mathrm{F}_2 \cdot \mathrm{diag}(1/\sqrt{2}, 1)$$
$$\mathrm{DCT\text{-}4}_2 = \mathrm{diag}(\cos\tfrac{\pi}{8}, \sin\tfrac{\pi}{8}) \cdot \mathrm{F}_2 \cdot \begin{bmatrix} 1 & -1 \\ 0 & \sqrt{2} \end{bmatrix}$$
$$\mathrm{DST\text{-}4}_2 = \mathrm{diag}(\sin\tfrac{\pi}{8}, \cos\tfrac{\pi}{8}) \cdot \mathrm{F}_2 \cdot \begin{bmatrix} 1 & 1 \\ 0 & \sqrt{2} \end{bmatrix}$$

$$\overline{\mathrm{DCT\text{-}3}}_2(r) = \mathrm{F}_2 \cdot \mathrm{diag}(1, \cos\tfrac{r}{2}\pi)$$
$$\overline{\mathrm{DST\text{-}3}}_2(r) = \mathrm{F}_2 \cdot \mathrm{diag}(1, 2\cos\tfrac{r\pi}{2})$$
$$\overline{\mathrm{DCT\text{-}4}}_2(r) = \mathrm{F}_2 \cdot \begin{bmatrix} 1 & -1 \\ 0 & 2\cos\frac{r\pi}{2} \end{bmatrix}$$
$$\overline{\mathrm{DST\text{-}4}}_2(r) = \mathrm{F}_2 \cdot \begin{bmatrix} 1 & 1 \\ 0 & 2\cos\frac{r\pi}{2} \end{bmatrix}$$

$$\mathrm{DCT\text{-}3}_2(r) = \overline{\mathrm{DCT\text{-}3}}_2(r)$$
$$\mathrm{DST\text{-}3}_2(r) = \mathrm{F}_2 \cdot \mathrm{diag}(\sin\tfrac{r\pi}{2}, \sin r\pi)$$
$$\mathrm{DCT\text{-}4}_2(r) = \mathrm{diag}(\cos\tfrac{r\pi}{4}, \sin\tfrac{r\pi}{4}) \cdot \mathrm{F}_2 \cdot \begin{bmatrix} 1 & -1 \\ 0 & 2\cos\frac{r\pi}{2} \end{bmatrix}$$
$$\mathrm{DST\text{-}4}_2(r) = \mathrm{diag}(\sin\tfrac{r\pi}{4}, \cos\tfrac{r\pi}{4}) \cdot \mathrm{F}_2 \cdot \begin{bmatrix} 1 & 1 \\ 0 & 2\cos\frac{r\pi}{2} \end{bmatrix}$$

$$\mathrm{iDCT\text{-}3}_2(r) = \mathrm{diag}(1, \tfrac{1}{2\cos\frac{r}{2}\pi)} \cdot \mathrm{F}_2,$$
$$\mathrm{iDST\text{-}3}_2(r) = \mathrm{diag}(\tfrac{1}{2\sin\frac{r\pi}{2}}, \tfrac{1}{\sin r\pi}) \cdot \mathrm{F}_2$$
$$\mathrm{iDCT\text{-}4}_2(r) = \begin{bmatrix} 1 & 1 \\ 0 & \frac{1}{2\cos\frac{r\pi}{2}} \end{bmatrix} \cdot \mathrm{F}_2 \cdot \mathrm{diag}(\tfrac{1}{2\cos\frac{r\pi}{4}}, \tfrac{1}{2\sin\frac{r\pi}{4}})$$
$$\mathrm{iDCT\text{-}4}_2(r) = \begin{bmatrix} 1 & -1 \\ 0 & \frac{1}{2\cos\frac{r\pi}{2}} \end{bmatrix} \cdot \mathrm{F}_2 \cdot \mathrm{diag}(\tfrac{1}{2\sin\frac{r\pi}{4}}, \tfrac{1}{2\cos\frac{r\pi}{4}})$$

TABLE VIII

BASE CASES FOR NORMAL AND SKEW $T$-GROUP DTTS OF SIZE 3.

$$\overline{\mathrm{DCT\text{-}3}}_3 = \begin{bmatrix} 1 & 0 & 1 \\ 0 & 1 & 0 \\ 1 & 0 & -1 \end{bmatrix} \begin{bmatrix} 1 & 0 & 1/2 \\ 1 & 0 & -1 \\ 0 & \sqrt{3}/2 & 0 \end{bmatrix}$$
$$\overline{\mathrm{DST\text{-}3}}_3 = \begin{bmatrix} 0 & 1 & 1 \\ 1 & 0 & 0 \\ 0 & 1 & -1 \end{bmatrix} \begin{bmatrix} 1 & 0 & -1 \\ 1 & 0 & 2 \\ 0 & \sqrt{3} & 0 \end{bmatrix}$$
$$\overline{\mathrm{DCT\text{-}4}}_3 = \begin{bmatrix} 0 & 1 & \sqrt{3}-1 \\ 1 & 0 & 0 \\ 0 & 1 & -\sqrt{3}-1 \end{bmatrix} \begin{bmatrix} 1 & -1 & -1 \\ 1 & 0 & 1 \\ 0 & 1 & -1 \end{bmatrix}$$
$$\overline{\mathrm{DST\text{-}4}}_3 = \begin{bmatrix} 0 & 1 & \sqrt{3}+1 \\ 1 & 0 & 0 \\ 0 & 1 & -\sqrt{3}+1 \end{bmatrix} \begin{bmatrix} 1 & 1 & -1 \\ 1 & 0 & 1 \\ 0 & 1 & 1 \end{bmatrix}$$

$$\mathrm{DCT\text{-}3}_3 = \overline{\mathrm{DCT\text{-}3}}_3$$
$$\mathrm{DST\text{-}3}_3 = \begin{bmatrix} 1 & 0 & 1 \\ 0 & 1 & 0 \\ 1 & 0 & -1 \end{bmatrix} \begin{bmatrix} 1/2 & 0 & 1 \\ 1 & 0 & -1 \\ 0 & \sqrt{3}/2 & 0 \end{bmatrix}$$
$$\mathrm{DCT\text{-}4}_3 = \begin{bmatrix} 1 & -1 & 0 \\ 0 & 0 & 1 \\ 1 & 1 & 0 \end{bmatrix} \begin{bmatrix} 1 & 0 & 0 \\ 0 & 1 & -1 \\ 0 & -2 & -1 \end{bmatrix} \mathrm{diag}(\sqrt{\tfrac{3}{2}}, \sqrt{\tfrac{1}{8}}, \sqrt{\tfrac{1}{2}}) \begin{bmatrix} 1 & 0 & 1 \\ -1 & 0 & 1 \\ 0 & 1 & 0 \end{bmatrix}$$
$$\mathrm{DST\text{-}4}_3 = \begin{bmatrix} 1 & -1 & 0 \\ 0 & 0 & 1 \\ 1 & 1 & 0 \end{bmatrix} \begin{bmatrix} 1 & 0 & 0 \\ 0 & 1 & -1 \\ 0 & 2 & 1 \end{bmatrix} \mathrm{diag}(\sqrt{\tfrac{3}{2}}, \sqrt{\tfrac{1}{8}}, \sqrt{\tfrac{1}{2}}) \begin{bmatrix} 1 & 0 & 1 \\ 1 & 0 & -1 \\ 0 & 1 & 0 \end{bmatrix}$$

$$\overline{\mathrm{DCT\text{-}3}}_3(r) = \begin{bmatrix} 1 & 1 & 1 \\ 1 & -1 & 0 \\ 1 & 0 & -1 \end{bmatrix} \Big(\mathrm{I}_1 \oplus \begin{bmatrix} \cos(\frac{1+r}{3}\pi) & \cos(\frac{1-2r}{3}\pi) \\ \cos(\frac{1-r}{3}\pi) & \cos(\frac{1+2r}{3}\pi) \end{bmatrix}\Big)$$
$$\overline{\mathrm{DST\text{-}3}}_3(r) = \begin{bmatrix} 1 & 1 & 1 \\ 1 & -1 & 0 \\ 1 & 0 & -1 \end{bmatrix} \Big(\mathrm{I}_1 \oplus 2\begin{bmatrix} \cos(\frac{1+r}{3}\pi) & \cos(\frac{1-2r}{3}\pi) \\ \cos(\frac{1-r}{3}\pi) & \cos(\frac{1+2r}{3}\pi) \end{bmatrix}\Big) \begin{bmatrix} 1 & 0 & 1 \\ 0 & 1 & 0 \\ 0 & 0 & 1 \end{bmatrix}$$
$\overline{\mathrm{DCT\text{-}4}}_3(r) =$ by definition
$\overline{\mathrm{DST\text{-}4}}_3(r) =$ by definition

$$\mathrm{DCT\text{-}3}_3(r) = \overline{\mathrm{DCT\text{-}3}}_3(r)$$
$$\mathrm{DST\text{-}3}_3(r) = \mathrm{diag}(\sin\tfrac{r}{3}\pi, \sin\tfrac{2-r}{3}\pi, \sin\tfrac{2+r}{3}\pi)\overline{\mathrm{DST\text{-}3}}_3(r)$$
$\mathrm{DCT\text{-}4}_3(r) =$ by definition
$\mathrm{DST\text{-}4}_3(r) =$ by definition

The remaining task is to derive the exact form of the base change matrices, which are the only parts of the above algorithms that depend on the DTT. We will do this in the remainder of this section including a cost analysis for the most important cases and sizes.

### *B. Details: $T$-Group and $U$-Basis*

In this section, we compute the exact form of $\overline{B}_{k,m}^{(*)}$ for $* \in \{C3, S3, C4, S4\}$. We derive $\overline{B}_{k,m}^{(C3)}$ as an example in detail. The others are derived analogously and only the result will be presented.

**Derivation of base change matrices.** We consider DCT-3. The matrix $\overline{B}_{k,m}^{(*)} = \overline{B}_{k,m}^{(C3)}$ in (44) performs a base change in $\mathbb{C}[x]/T_n$ from a $T$-basis to the basis $b'$ in (43) with $C = T$. To compute $\overline{B}_{k,m}^{(C3)}$ we have to express every element $T_i$ in $b$ as a linear combination in $b'$. To do this, we first write $b$ as

$$b = (T_{im+j} \mid 0 \le i < m,\ 0 \le j < k).$$

We did not change $b$, but only decomposed the index into a radix-$m$ representation. The double indices are ordered as usual lexicographically: $(i, j) = (0, 0), (0, 1) \ldots.$ Similarly, we write $b'$ as a special case of (43)

$$b' = (T_j U_i(T_m) \mid 0 \le i < m,\ 0 \le j < k).$$

First, we consider the case $j = 0$. From Table XXIV, we know that $T_i = (U_i - U_{i-2})/2$ and thus

$$T_{im} = T_i(T_m) = \tfrac{1}{2}U_i(T_m) - \tfrac{1}{2}U_{i-2}(T_m) \qquad (52)$$

is the desired representation in $b'$.

Now, let $j \neq 0$, i.e., $1 \le j < m$. We claim that

$$T_{im+j} = T_j U_i(T_m) - T_{m-j} U_{i-1}(T_m). \qquad (53)$$

To prove it, we define the recursion

$$\begin{aligned} p_0 &= T_{m-j} = T_{-m+j}, \\ p_1 &= T_j, \\ p_{i+1} &= 2T_m T_i - T_{i-1}. \end{aligned}$$

First, because of (104) (Appendix III), we see that

$$p_{i+1} = T_{im+j},$$

which is the left hand side of (53). On the other hand, using (103) (Appendix II) with $T_m$ as variable, $p_{i+1}$ is precisely the right hand side of (53), as desired.

The equations (52) and (53) define the columns of the base change matrix, which is thus given by

$$\overline{B}_{k,m}^{(C3)} = \begin{bmatrix} 1 & & & -\frac{1}{2} & & & \\ & \mathrm{I}_{m-1} & -\mathrm{J}_{m-1} & & \ddots & & \\ & & \frac{1}{2} & \ddots & & -\frac{1}{2} & \\ & & & \ddots & -\mathrm{J}_{m-1} & & \\ & & & & \frac{1}{2} & & \\ & & & & \mathrm{I}_{m-1} & & -\mathrm{J}_{m-1} \\ & & & & & \frac{1}{2} & \\ & & & & & & \mathrm{I}_{m-1} \end{bmatrix} \qquad (54)$$

For example, all rows with an index that is a multiple of $m$ are determined by (52) and thus contain the numbers $1/2$.

Using

$$C_{im+j} = C_j U_i(T_m) - C_{j-m} U_{i-1}(T_m),$$

which generalizes (53), yields the base change matrices in the other three cases:

$$\overline{B}_{k,m}^{(S3)} = \begin{bmatrix} \mathrm{I}_m & \overline{Z}_m & & & \\ & \mathrm{I}_m & \overline{Z}_m & & \\ & & \ddots & \ddots & \\ & & & \mathrm{I}_m & \overline{Z}_m \\ & & & & \mathrm{I}_m \end{bmatrix}$$

$$\overline{B}_{k,m}^{(C4)} = \begin{bmatrix} \mathrm{I}_m & -\mathrm{J}_m & & & \\ & \mathrm{I}_m & -\mathrm{J}_m & & \\ & & \ddots & \ddots & \\ & & & \mathrm{I}_m & -\mathrm{J}_m \\ & & & & \mathrm{I}_m \end{bmatrix}$$

$$\overline{B}_{k,m}^{(S4)} = \begin{bmatrix} \mathrm{I}_m & \mathrm{J}_m & & & \\ & \mathrm{I}_m & \mathrm{J}_m & & \\ & & \ddots & \ddots & \\ & & & \mathrm{I}_m & \mathrm{J}_m \\ & & & & \mathrm{I}_m \end{bmatrix}.$$

**Exact forms of algorithms.** Table IX summarizes the exact form of all algorithms based on (45). Each algorithm has a polynomial counterpart obtained by replacing the DTTs by $\overline{\text{DTT}}$s.

### *C. Details: $T$-Group and $T$-Basis*

In this section, we derive the exact base change matrices for the algorithms in (47), (48), and (50).

**Derivation of base change matrices.** Again, we use the DCT-$3_n$ as detailed example. The matrix $B_{k,m}^{(*)} = B_{k,m}^{(C3)}$ in (47) performs a base change in $\mathbb{C}[x]/T_n$ from a $T$-basis to the basis

$$b' = (T_{im-j}/2 + T_{im+j}/2 \mid 0 \le i < m,\ 0 \le j < k), \qquad (59)$$

a special case of (46). It is not straightforward to determine $B_{k,m}^{(C3)}$. However, the *inverse* base change is easy to compute due to the form of $b'$: (59) already expresses the elements of $b'$ as a linear combination of the elements of $b$. Further simplification in (59) is obtained for the special cases $i = 0$ (which determines the first block of length $m$), namely

$$T_{im-j}/2 + T_{im+j}/2 = T_{-j}/2 + T_j/2 = T_j,$$

and for the special case $j = 0$ (which determines every $m$th column), namely

$$T_{im-j}/2 + T_{im+j}/2 = T_{im}.$$

Hence we get

$$\left(B_{k,m}^{(C3)}\right)^{-1} = \frac{1}{2} \cdot \begin{bmatrix} 2 \cdot \mathrm{I}_m & Z_m & & & \\ & \mathrm{I}'_m & Z_m & & \\ & & \ddots & \ddots & \\ & & & \mathrm{I}'_m & Z_m \\ & & & & \mathrm{I}'_m \end{bmatrix} \qquad (60)$$

TABLE IX

COOLEY-TUKEY TYPE ALGORITHMS FOR DTTS IN THE $T$-GROUP, BASED ON THE DECOMPOSITION $T_{km} = T_k(T_m)$ AND A $U$-BASIS CHOSEN IN THE COARSE SIGNAL MODEL. THE POLYNOMIAL VERSIONS ARE OBTAINED BY REPLACING ALL TRANSFORMS BY THEIR POLYNOMIAL COUNTERPART. TRANSPOSITION YIELDS ALGORITHMS FOR DCT AND DST TYPE 2 AND 4.

$$\text{DCT-3}_n = \text{DCT-3}_n(1/2), \qquad \text{DCT-3}_{km}(r) = \mathrm{K}^n_m\big(\bigoplus_{0\le i<k} \text{DCT-3}_m(r_i)\big)\overline{\text{DST-3}}_k(r)\otimes \mathrm{I}_m)\overline{B}^{(C3)}_{k,m} \tag{55}$$

$$\text{DST-3}_n = \text{DST-3}_n(1/2), \qquad \text{DST-3}_{km}(r) = \mathrm{K}^n_m\big(\bigoplus_{0\le i<k} \text{DST-3}_m(r_i)\big)(\overline{\text{DST-3}}_k(r)\otimes \mathrm{I}_m)\overline{B}^{(S3)}_{k,m} \tag{56}$$

$$\text{DCT-4}_n = \text{DCT-4}_n(1/2), \qquad \text{DCT-4}_{km}(r) = \mathrm{K}^n_m\big(\bigoplus_{0\le i<k} \text{DCT-4}_m(r_i)\big)\overline{\text{DST-3}}_k(r)\otimes \mathrm{I}_m)\overline{B}^{(C4)}_{k,m} \tag{57}$$

$$\text{DST-4}_n = \text{DST-4}_n(1/2), \qquad \text{DST-4}_{km}(r) = \mathrm{K}^n_m\big(\bigoplus_{0\le i<k} \text{DST-4}_m(r_i)\big)(\overline{\text{DST-3}}_k(r)\otimes \mathrm{I}_m)\overline{B}^{(S4)}_{k,m} \tag{58}$$

with

$$Z_m = \begin{bmatrix} & & & 0 \\ & & 0 & 1 \\ & \iddots & \iddots & \\ 0 & 1 & & \end{bmatrix}, \quad \mathrm{I}'_m = 2 \oplus \mathrm{I}_{m-1}\,.$$

All the multiplications in (60) can be pulled out to the right and we get

$$\big(B^{(C3)}_{k,m}\big)^{-1} = \big(C^{(C3)}_{k,m}\big)^{-1}\big(D^{(C3)}_{k,m}\big)^{-1}$$

with

$$\big(C^{(C3)}_{k,m}\big)^{-1} = \begin{bmatrix} \mathrm{I}_m & Z_m & & & \\ & \mathrm{I}_m & Z_m & & \\ & & \ddots & \ddots & \\ & & & \mathrm{I}_m & Z_m \\ & & & & \mathrm{I}_m \end{bmatrix} \tag{61}$$

and the diagonal matrix

$$\big(D^{(C3)}_{k,m}\big)^{-1} = \mathrm{I}_m \oplus (\mathrm{I}_{k-1}\otimes \operatorname{diag}(1, 1/2, \dots, 1/2)).$$

To determine $B^{(C3)}_{n,k}$, we first analyze the block structure. Investigation shows that (61) consists of $k$ blocks of size 1 at positions $jm$, $0 \le j < k$, and $m-1$ blocks of size $k$ corresponding to the index sets

$$\{0m+i, 2m\pm i, 4m\pm i, \dots (k-1)m\pm i\} \quad (k \text{ odd}),$$
$$\{0m+i, 2m\pm i, 4m\pm i, \dots km-i\} \quad (k \text{ even}),$$

for $0 < i < m$. These index sets (and thus the corresponding blocks) are obtained by starting at entry $(i,i)$, $0 < i < m$, of (61) and collecting non-zero entries in a zigzag pattern going alternately to the right and down. Each of these $m-1$ blocks has the form

$$S_k = \begin{bmatrix} 1 & 1 & & & \\ 0 & 1 & 1 & & \\ & & \cdot & \cdot & \\ & & & 1 & 1 \\ & & & & 1 \end{bmatrix}. \tag{62}$$

Using the block structure, we can now write $C^{-1}_{n,k}$ as

$$\begin{aligned}\big(C^{(C3)}_{k,m}\big)^{-1} &= (\mathrm{I}_k \oplus S_k \oplus \dots \oplus S_k)^{Q_{n,k}}, \\ &= (\mathrm{I}_k \oplus (\mathrm{I}_{m-1}\otimes S_k))^{Q_{n,k}},\end{aligned}$$

with a suitable permutation $Q_{n,k}$ (the precise form is not of importance here). Inversion yields

$$C^{(C3)}_{n,k} = (\mathrm{I}_k \oplus (\mathrm{I}_{m-1}\otimes S_k^{-1}))^{Q_{n,k}}. \tag{63}$$

Multiplication with the inverse of $S_k$, i.e., $(y_0, \dots, y_{n-1})^T = S_k^{-1}(x_0, \dots, x_{n-1})^T$ can be done with the $k-1$ recursive subtractions

$$y_{n-1} = x_{n-1},\; y_{n-2} = x_{n-2} - y_{n-1}, \dots, y_0 = x_0 - y_1,$$

i.e., the critical path of $S_k^{-1}$, and thus the one of $C^{(C3)}_{n,k}$ has length $k-1$. Hence, $k$ should be small to yield an efficient algorithm. For example, for $k = 2$,

$$\begin{aligned} B^{(C3)}_{2,m} &= (\mathrm{I}_m \oplus \operatorname{diag}(1, 1/2, \dots, 1/2))^{-1} \begin{bmatrix}\mathrm{I}_m & Z_m \\ & \mathrm{I}_m\end{bmatrix}^{-1} \\ &= (\mathrm{I}_m \oplus \operatorname{diag}(1, 2, \dots, 2)) \begin{bmatrix}\mathrm{I}_m & -Z_m \\ & \mathrm{I}_m\end{bmatrix}. \end{aligned} \tag{64}$$

On the other hand, $C^{-1}_{n,k}$ in (61) has a very short critical path of length 1. This explains the motivation to invert (47) in Section VII-A to obtain (50). Doing so for the DCT-3 considered here, it turns out that all scaling factors cancel out, and we obtain the beautifully simple form

$$\text{iDCT-3}_n(r) = (C^{(C3)}_{n,k})^{-1}(\text{iDCT-3}_k(r)\otimes \mathrm{I}_m) \Big(\bigoplus_{0\le i<k} \text{iDCT-3}_m(r_i)\Big)\mathrm{M}^n_k, \tag{65}$$

where

$$\mathrm{M}^n_k = (\mathrm{K}^n_m)^{-1} = \mathrm{L}^n_k(\mathrm{I}_k \oplus \mathrm{J}_k \oplus \mathrm{I}_k \oplus \mathrm{J}_k \oplus \dots).$$

Equation (65) gives a class of algorithms for $\text{DCT-2}_n = \text{iDCT-3}_n(1/2)$, and, by transposition, we obtain again an algorithm for $\text{DCT-3}_n$.

The base change matrices for the other three $T$-group DTTs are obtained analogously. We only give the result.

$$
(C_{k,m}^{(S3)})^{-1} = \begin{bmatrix}
\mathrm{I}_{m-1} & -\mathrm{J}_{m-1} & & & & & \\
& \frac{1}{2} & & & & -\frac{1}{2} & \\
& & \mathrm{I}_{m-1} & -\mathrm{J}_{m-1} & & \ddots & \\
& & & \frac{1}{2} & \ddots & & -\frac{1}{2} \\
& & & & \ddots & -\mathrm{J}_{m-1} & \\
& & & & \frac{1}{2} & & \\
& & & & & \mathrm{I}_{m-1} & \\
& & & & & & 1
\end{bmatrix} \quad (66)
$$

$$
(C_{k,m}^{(C4)})^{-1} = \overline{B}_{k,m}^{(S4)}
$$

$$
(C_{k,m}^{(S4)})^{-1} = \overline{B}_{k,m}^{(C4)}
$$

and

$$
\begin{aligned}
B_{2,m}^{(S3)} &= \begin{bmatrix} \mathrm{I}_m & \overline{Z}_m \\ & 2\,\mathrm{I}_m \end{bmatrix} \\
B_{2,m}^{(C4)} &= \begin{bmatrix} \mathrm{I}_m & -\mathrm{J}_m \\ & 2\,\mathrm{I}_m \end{bmatrix} \\
B_{2,m}^{(S4)} &= \begin{bmatrix} \mathrm{I}_m & \mathrm{J}_m \\ & 2\,\mathrm{I}_m \end{bmatrix}
\end{aligned} \quad (67)
$$

**Exact forms of algorithms.** Tables X and XI show the final algorithms, which are special cases of (48) and (50), respectively.

Replacing all transforms by their polynomial counterparts gives the corresponding algorithms for the polynomial DTTs. Further, each algorithm in Tables X and XI has a corresponding twiddle version as shown in (51).

**Special cases.** We briefly discuss the special case (68) for $k = 2$. (64) incurs multiplications by 2, which can be fused with the multiplications incurred by the adjacent $\text{DCT-3}_2(r)$. Namely, using $\text{DCT-3}_2(r) = \mathrm{F}_2 \cdot \operatorname{diag}(1, \cos\frac{r}{2}\pi)$ (Table VII), we can manipulate (68) to take the form

$$
\begin{aligned}
\text{DCT-3}_n(r) = \mathrm{K}_m^n(&\text{DCT-3}_m(\tfrac{r}{2}) \oplus \text{DCT-3}_m(\tfrac{2-r}{2})) \\
&(\mathrm{F}_2 \otimes \mathrm{I}_m)E_{2,m}, \quad (76)
\end{aligned}
$$

where

$$
E_{2,m} = \begin{bmatrix} \mathrm{I}_m & -Z_m \\ & \cos\frac{r}{2}\pi(\mathrm{I}_1 \oplus 2\,\mathrm{I}_{m-1}). \end{bmatrix}
$$

We also briefly consider the case $k = 3$ in (68). In this case,

$$
B_{3,m}^{(C3)} = (\mathrm{I}_m \oplus (\mathrm{I}_2 \otimes \operatorname{diag}(1, 2, \dots, 2))) \begin{bmatrix} \mathrm{I}_m & -Z_m & \mathrm{I}'_m \\ & \mathrm{I}_m & -Z_m \\ & & \mathrm{I}_m \end{bmatrix},
$$

where $\mathrm{I}'_m = \operatorname{diag}(0, 1, \dots, 1)$. Note that this matrix, as mentioned before, requires only $2(m-1)$ additions, since $m-1$ additions are duplicated (row 1, columns 2/3, and row 2, columns 2/3). However, the critical path of $B_{3,m}^{(C3)}$ has then length 2. Again, the multiplications can be fused with the adjacent $\text{DCT-3}_3(r)$. We omit the details.

Similarly, the multiplications by 2 in (67) can be fused in (69)–(71).

### D. *Alternative Decomposition*

In this section, we discuss briefly algorithms based on the decomposition

$$
T_{km+m/2} = T_{m/2} \cdot V_k(T_m).
$$

The algorithms are for DTTs in the $T$-group and it turns out that the $U$-basis is the best choice for the coarse module $\mathbb{C}[x]/V_k(x)$. Thus, a simultaneous derivation yields, for $n = km + m/2$,

$$
\begin{aligned}
\text{DTT}_n = Q_{k,m}^{(*)}(&\text{DTT}_{m/2} \oplus \Big(\bigoplus \text{DTT}_m(r)\Big) \\
&(\overline{\text{DST-7}}_k \otimes \mathrm{I}_m))B_{k,m}^{(*)}. \quad (77)
\end{aligned}
$$

Note that $\overline{\text{DST-7}}$ is the polynomial transform for $\mathbb{C}[x]/V_k(x)$ with $U$-basis (Table III).

Next, we determine the best choice of size $n$. Inspecting (77) shows that, ideally, $m = 2^s$ is a 2-power and $k = (3^t - 1)/2$ is the natural size for the $\overline{\text{DST-7}}$ (explained later in Section IX). Thus $n = 3^t 2^{s-1}$, which is a size that is well handled by the algorithms in Section VII. For this reason, we omit the exact forms of the algorithms and only note that the base change matrices $B_{k,m}^{(*)}$ have structure similar to the structures in Sections IX and X.

**DCT, type 3 and 2, size 5.** Above we established that ideally $k = (3^t - 1)/2$, the ideal size for a $\overline{\text{DST-7}}$. However, if $k$ is small enough, namely $k = 2$, the algorithm (77) is still useful. In particular, if $m = 2$, then it yields algorithms for sizes $n = 5$. We use DCT-3 as an example. It turns out (by trial and error) that in this case a $V$-basis is slightly superior in $\mathbb{C}[x]/V_2$, and, after a minor manipulation, we get the algorithm in Table XII. The cost can be read off as $(12, 6, 1)$. Transposition yields an algorithm for $\text{DCT-2}_3$ with identical cost, which is only slightly worse than the $(13, 5, 0)$ algorithm in [53].

### E. *Analysis*

In this section we analyze the algorithms in Tables IX, X, and XI with respect to arithmetic cost and other characteristics. We also review special cases that have been known in the literature.

**Cost analysis.** Each of the algorithms in Tables IX, X, and XI provides reasonable algorithms with regular structure. The cost difference between any two of the $T$-group algorithms is $O(n)$ where $n$ is the transform size. We determine the cost in greater detail for the most relevant cases only.

For a 2-power $n$, the costs in each case are independent of the chosen recursive split strategy. The best achieved costs are recorded in Table XIII. Note that the best costs for the skew (and inverse skew) and non-skew versions are equal since they have the same recursions *and* the base cases have equal cost (Table VII). This is different for other sizes; in general the skew DTTs are more expensive (see also Appendix III). Also note that the polynomial DTTs save multiplications (except for the DCT-3 = $\overline{\text{DCT-3}}$).

For a 3-power $n$, the skew DTTs are more expensive. Also, the stated costs in Table XIII in this case are not the best possible with the algorithms in this paper. For example, we can slightly improve a DCT-3 of 3-power size $n$ using the transpose of Table VI(b) to get a cost of

$$
\begin{aligned}
&(\tfrac{8}{3}n\log_3(n) - 2n + 2, \tfrac{4}{3}n\log_3(n) - \tfrac{7}{4}n + \tfrac{1}{2}\log_3(n) + \tfrac{7}{4}, \\
&\tfrac{1}{4}n + \tfrac{1}{2}\log_3(n) - \tfrac{1}{4}) = 4n\log_3(n) - \tfrac{7}{2}n + \log_3(n) + \tfrac{7}{2}
\end{aligned}
$$

TABLE X

COOLEY-TUKEY TYPE ALGORITHMS FOR DTTS IN THE $T$-GROUP, BASED ON THE DECOMPOSITION $T_{km} = T_k(T_m)$ AND A $T$-BASIS CHOSEN IN THE COARSE SIGNAL MODEL. THE POLYNOMIAL VERSIONS ARE OBTAINED BY REPLACING ALL TRANSFORMS BY THEIR POLYNOMIAL COUNTERPART. TRANSPOSITION YIELDS ALGORITHMS FOR DCT AND DST TYPE 2 AND 4.

$$\text{DCT-3}_n = \text{DCT-3}_n(1/2), \qquad \text{DCT-3}_{km}(r) = \mathrm{K}^n_m \big(\bigoplus_{0\le i<k} \text{DCT-3}_m(r_i)\big)\text{DCT-3}_k(r) \otimes \mathrm{I}_m)B^{(C3)}_{k,m} \tag{68}$$

$$\text{DST-3}_n = \text{DST-3}_n(1/2), \qquad \text{DST-3}_{km}(r) = \mathrm{K}^n_m \big(\bigoplus_{0\le i<k} \text{DST-3}_m(r_i)\big)(\text{DCT 3}_k(r) \otimes \mathrm{I}_m)B^{(S3)}_{k,m} \tag{69}$$

$$\text{DCT-4}_n = \text{DCT-4}_n(1/2), \qquad \text{DCT-4}_{km}(r) = \mathrm{K}^n_m \big(\bigoplus_{0\le i<k} \text{DCT-4}_m(r_i)\big)\,\text{DCT 3}_k(r) \otimes \mathrm{I}_m)B^{(C4)}_{k,m} \tag{70}$$

$$\text{DST-4}_n = \text{DST-4}_n(1/2), \qquad \text{DST-4}_{km}(r) = \mathrm{K}^n_m \big(\bigoplus_{0\le i<k} \text{DST-4}_m(r_i)\big)(\text{DCT 3}_k(r) \otimes \mathrm{I}_m)B^{(S4)}_{k,m} \tag{71}$$

TABLE XI

MANIPULATED INVERSE OF TABLE X: COOLEY-TUKEY TYPE ALGORITHMS FOR THE TRANSPOSES OF THE DTTS IN THE $T$-GROUP, BASED ON THE DECOMPOSITION $T_{km} = T_k(T_m)$ AND A $T$-BASIS CHOSEN IN THE COARSE SIGNAL MODEL. TRANSPOSITION YIELDS ALGORITHMS FOR THE DTTS IN THE $T$-GROUP.

$$\text{DCT-2}_n = \text{iDCT-3}_n(1/2), \qquad \text{iDCT-3}_{km}(r) = (C^{(C3)}_{k,m})^{-1}(\text{iDCT-3}_k(r) \otimes \mathrm{I}_m)\Big(\bigoplus_{0\le i<k} \text{iDCT-3}_m(r_i)\Big)\,\mathrm{M}^n_k \tag{72}$$

$$\text{DST-2}_n = \text{iDST-3}_n(1/2), \qquad \text{iDST-3}_{km}(r) = (C^{(S3)}_{k,m})^{-1}(\text{iDST-3}_k(r) \otimes \mathrm{I}_m)\Big(\bigoplus_{0\le i<k} \text{iDCT-3}_m(r_i)\Big)\,\mathrm{M}^n_k \tag{73}$$

$$\text{DCT-4}_n = \text{iDCT-4}_n(1/2), \qquad \text{iDCT-4}_{km}(r) = (C^{(C4)}_{k,m})^{-1}(\text{iDCT-4}_k(r) \otimes \mathrm{I}_m)\Big(\bigoplus_{0\le i<k} \text{iDCT-3}_m(r_i)\Big)\,\mathrm{M}^n_k \tag{74}$$

$$\text{DST-4}_n = \text{iDST-4}_n(1/2), \qquad \text{iDST-4}_{km}(r) = (C^{(S4)}_{k,m})^{-1}(\text{iDST-4}_k(r) \otimes \mathrm{I}_m)\Big(\bigoplus_{0\le i<k} \text{iDCT-3}_m(r_i)\Big)\,\mathrm{M}^n_k \tag{75}$$

TABLE XII

ALGORITHM FOR $\text{DCT-3}_5$ WITH COST (12, 6, 1). TRANSPOSITION YIELDS A $\text{DCT-2}_5$ ALGORITHM OF EQUAL COST.

$$\begin{bmatrix}0&1&0&0&0\\0&0&0&1&0\\1&0&0&0&0\\0&0&0&0&1\\0&0&1&0&0\end{bmatrix}\Big(\mathrm{I}_1 \oplus \big(\mathrm{F}_2\,\mathrm{diag}(1,\cos\tfrac{\pi}{5}) \oplus \mathrm{F}_2\,\mathrm{diag}(1,\cos\tfrac{3\pi}{5})\big)\begin{bmatrix}\mathrm{I}_2 & \mathrm{diag}(\cos\frac{\pi}{5}, 2\cos\frac{\pi}{5})\\ \mathrm{I}_2 & \mathrm{diag}(\cos\frac{3\pi}{5}, 2\cos\frac{3\pi}{5})\end{bmatrix}\Big)\begin{bmatrix}1&0&-1&0&1\\1&0&1/2&0&0\\0&1&0&0&0\\0&0&1&0&1\\0&0&0&1&0\end{bmatrix}$$

while sacrificing some regularity in structure. For example, for $n = 9$, Table XIII yields $(32, 12, 4) = 48$ and the above $(32, 11, 3) = 46$. The same costs apply to a DCT-2 by transposing the algorithms. Reference [53] provides an $(34, 8, 2) = 44$ algorithm (proven optimal w.r.t. non-rational multiplications), with no obvious structure. Using (106), (105), (106), and (107), this also yields better algorithms for skew and non-skew DCT-4 and DST-4.

For an arbitrary $p$-power $n$, we can compute $T$-group DTTs using the twiddle versions of the $T$-group algorithms (e.g., (51)). For example, a $\text{DCT-2}_{p^t}$ computed with (72) requires, independently of the split strategy $\frac{n}{p}\log_p(n)$ $\text{DCT-2}_p$'s, and

$$2(1-\frac{1}{p})n\log_p(n) - 2n + 2$$

additions and multiplications, respectively. For a given $\text{DCT-2}_p$ kernel (e.g., the transpose of Table XII or [53] for $p = 5, 7$), it is now easy to compute a cost. The other $T$-group DTTs are analogous.

Note that for a 2-power size $n$, the algorithms (56), (69) and transposed (73), for DST-3 have an $O(n)$ higher cost than a translation by duality (105) (Appendix III). For 3-power sizes $n$, all algorithms, except those for $\text{DCT-3}_n$ in Tables IX, X, and XI incur an $\frac{1}{3}n\log_3(n)$ higher cost compared to translating these DTTs into a DCT-3 using $O(n)$ operations (see Appendix III).

**Further comments.**

- The algorithms in Table X have the appealing feature that all multiplications occur in parallel with additions on the same operands. Further, they are a good choice if the output is to be pruned, i.e., only, say, the first half of the output matters. This was used in [54] for DCT-2. However, for large $k$, the critical path is potentially prohibitive.
- The cost of the $T$-group algorithms is independent of the chosen split.
- The algorithms in Table XI involve constants that are inverse cosines (from the base cases of the iDTTs in Table VII). This may cause numerical instability.
- Transposition yields algorithms for the transposed DTTs with equal cost. The reason is that all occurring matrices

have this property.

- If a non-skew DTT is decomposed using any of the $T$-group algorithms, then (the middle) one of the occurring skew DTTs in the direct sum has $r = 1/2$, i.e., is non-skew.
- Any odd-size DCT of type 2 or 3 can be translated into an RDFT without incurring operations [53].
- Again we note that the algorithms in this section are not all available ones. In particular, there are orthogonal algorithms, which are due to other algebraic principles [42].
- All the algorithms have, for a 2-power size $n$, a total cost of $2n\log_2(n) + O(n)$. This can be improved by roughly 5% with the recent paper [55] to $\frac{17}{9}n\log_2(n) + O(n)$, at the expense of some regularity.

**Literature.** Algorithm (68) for 2-power size and $k = 2$ was derived in [56] and in [33]; the latter also considered 3-powers and $k = 3$. For arbitrary $p$-powers ($p$ prime) and $k = p$, the derivation is in [34]. The above references also used Chebyshev polynomials in their derivation, but they do not use the algebraic framework, and they present the algorithms in an iterative form only, which avoids the definition of skew DTTs. For software implementations, it is crucial to have a recursive form as presented here. Further, the derivation for $p > 2$ produced suboptimal cost compared to Table XIII.

Special cases of (68) with the reverse split, i.e., $n = p^t$, $k = p^{t-1}$, are not practical because of the long critical path for computing $C_{n,k}$. Their discovery, however, is more straightforward, since they do not require large skew DCTs, which are unexpected without knowing the underlying algebra. The case $p = 2$ was reported in [57], $p = 3, 6$ in [58], the case of a general $p$ in [59] with examples $p = 3, 5, 7, 9$.

Algorithms (69), its transpose, and the transpose of (71) were found, also for 2-powers and $k = 2$ in [60]. The only special case of (70) we found in the literature was derived implicitly in [56], where the DCT-4 is called “odd DCT” and decomposed as part of a fast DCT-2 algorithm that first recurses using Table VI(a).

Architecture regular versions (i.e., the equivalent to the Pease FFT [61]) of the algorithms in Table X, again for $k = 2$, can be found in [62], [63].

The only case of (72) we found in the literature is $n = 2^t$, $m = 2$, in which case the skew DCTs become trivial [64].

All other algorithms for the $T$-group DTTs are to our best knowledge novel.

## VIII. $U$-GROUP DTT ALGORITHMS

We now derive Cooley-Tukey type algorithms for all four DTTs in the $U$-group, based on the decomposition property (Lemma 3, ii)):

$$U_{km-1} = U_{k-1}(T_m)U_{m-1}. \tag{78}$$

Since $U$ does not decompose directly, the derivation involves a first additional step to factor $U_{km-1}$ into $U_{k-1}(T_m)$ and $U_{m-1}$ using Theorem 1. In the special case of $k = 2$, the decomposition in (78) becomes trivial and (78) becomes Lemma 2, ii). Thus, the algorithms in Table VI(a) become a special case of the algorithms derived below.

The DTTs in the $U$-group have associated modules $\mathbb{C}[x]/p(x)$ with mutually distinct polynomials $p$, namely for $\mathrm{DCT\text{-}1}, \mathrm{DST\text{-}1}, \mathrm{DCT\text{-}2}, \mathrm{DST\text{-}2}$, respectively (see Table III)

$$p(x) = (x^2-1)U_{n-2},\ U_n,\ (x-1)U_{n-1},\ (x+1)U_{n-1}. \tag{79}$$

Thus, the decompositions are slightly different and cannot be stated in a precise unified way as for the $T$-group in (VII). From (79), it is clear that in order to apply (78) for $n = km$, we have to consider DCT-2 and DST-2 of size $n$, but DCT-1 of size $n + 1$ and DST-1 of size $n - 1$. This motivates the following definition, which we will use in the derivation.

$$n' = \begin{cases} n+1, & \text{for } \mathrm{DCT\text{-}1}_{n'}; \\ n-1, & \text{for } \mathrm{DST\text{-}1}_{n'}; \\ n, & \text{for } \mathrm{DCT\text{-}2}_{n'}, \mathrm{DST\text{-}2}_{n'}. \end{cases}$$

### A. Simultaneous Derivation

Let $\mathrm{DTT}_{n'}$ be one of the DTTs in the $U$-group with module $\mathcal{M} = \mathbb{C}[x]/p_{n'}$ and $C$-basis, where $C$ is one of $T, U, V, W$. We assume $n = km$. In the first step, we decompose $\mathcal{M}$ using the factorization (78) and Theorem 1 as

$$\mathbb{C}[x]/p_{n'} \to \mathbb{C}[x]/p_{m'} \oplus \mathbb{C}[x]/U_{k-1}(T_m). \tag{80}$$

In the first summand, we choose a $C$-basis (i.e., equal to the one in $\mathcal{M}$); in the second summand, we choose the basis $b'$ (see (25)) given by

$$\begin{aligned} b' \quad = \quad & (C_0U_0(T_m), \ldots, C_{m-1}U_0(T_m) \\ & \ldots \\ & C_0U_{k-2}(T_m), \ldots, C_{m-1}U_{k-2}(T_m)), \end{aligned}$$

which is required for the further decomposition of $\mathbb{C}[x]/U_{k-1}(T_m)$ using Theorem 2. This implies the choice of a $U$-basis in the coarse module $\mathbb{C}[x]/U_{k-1}$ in all four cases. Any other choice of basis would lead to a transform that is not a DTT (see Table III: only one DTT has $p = U$, namely DST-1). Also, it turns out that the base change matrices become more complicated for any other choice, and, in contrast to Section VII, the inversion of algorithms to improve their structure does not work this time.

Based on (80), we get the decomposition

$$\mathrm{DTT}_{n'} = P_{k,m}^{(*)}(A_{(k-1)m} \oplus \mathrm{DTT}_{m'})B_{k,m}^{(*)}, \tag{81}$$

where $A_{(k-1)m}$ is a Fourier transform for $\mathbb{C}[x]/U_{k-1}(T_m)$ with basis $b'$, and $(\cdot)^{(*)}$ signifies dependency on the DTT; the exact form of these matrices will be provided below. Note that we can exchange the order of the summands in (81), if we properly permute the columns and rows, respectively, of $P_{k,m}$ and $B_{k,m}$. In two of the four cases, we will do this to obtain permutations $P_{k,m}^{(*)}$ of a simpler structure.

To apply Theorem 2 for further decomposition of $A_{(k-1)m}$, we need the zeros of $U_{k-1}$, which are given by $\cos\frac{i\pi}{k}$, $0 < i < k$ (TableXXIII in Appendix II), and thus

$$\mathbb{C}[x]/U_{k-1}(T_m) \to \bigoplus_{0<i<k} \mathbb{C}[x]/(T_m - \cos\tfrac{i\pi}{k}) \tag{82}$$

TABLE XIII

ARITHMETIC COSTS ACHIEVABLE FOR THE DTTS IN THE $T$-GROUP WITH THE ALGORITHMS IN THIS PAPER FOR 2-POWER AND 3-POWER SIZE $n$. ALL THE 3-POWER SIZE COSTS CAN BE SLIGHTLY IMPROVED UPON (SEE SECTION VII-E).

| | Transform | Cost (adds, mults, 2-power mults) and total | Achieved by |
|---|---|---|---|
| **2-power $n$** | DCT-3$_n$ | $(\frac{3}{2}n\log_2(n)-n+1,\ \frac{1}{2}n\log_2(n),\ 0)$<br>**total:** $2n\log_2(n)-n+1$ | (68) (see also (76)), (72)$^T$, Table VI(a)$^T$ |
| | DST-3$_n$ | same as DCT-3 | duality (105), Table VI(a)$^T$ |
| | DCT-4$_n$ | $(\frac{3}{2}n\log_2(n),\ \frac{1}{2}n\log_2(n)+n,\ 0)$<br>**total:** $2n\log_2(n)+n$ **poly:** $-n$ | (57), (70), (74), (106), and their transposes |
| | DST-4$_n$ | same as DCT-4 | (58), (71), (75), duality (105) |
| | DCT-3$_n(r)$ | same as DCT-3 | (68) |
| | DST-3$_n(r)$ | $(\frac{3}{2}n\log_2(n)-n+1,\ \frac{1}{2}n\log_2(n)+\frac{1}{2}n,\ 0)$<br>**total:** $2n\log_2(n)-\frac{1}{2}n+1$ **poly:** $-\frac{1}{2}n$ | (56) |
| | DCT-4$_n(r)$ | same as DCT-4 | (57), (70) |
| | DST-4$_n(r)$ | same as DCT-4 | (58), (71) |
| **3-power $n$** | DCT-3$_n$ | $(\frac{8}{3}n\log_3(n)-2n+2,\ \frac{4}{3}n\log_3(n)-\frac{3}{2}n,\ \frac{1}{2}n-\frac{1}{2})$<br>**total:** $4n\log_3(n)-3n+3$ | (68), (72)$^T$, see also cheaper version in Section VII-E |
| | DST-3$_n$ | same as DCT-3 | duality (105) |
| | DCT-4$_n$ | $(\frac{8}{3}n\log_3(n)-n+1,\ \frac{4}{3}n\log_3(n)-\frac{1}{2}n,\ \frac{1}{2}n-\frac{1}{2})$<br>**total:** $4n\log_3(n)-n+2$ | (106) |
| | DST-4$_n$ | same as DCT-4 | duality (105) |
| | DCT-3$_n(r)$ | $(\frac{8}{3}n\log_3(n)-n+1,\ \frac{4}{3}n\log_3(n),\ 0)$<br>**total:** $4n\log_3(n)-n+1$ | (68) |
| | DST-3$_n(r)$ | $(\frac{8}{3}n\log_3(n)-n+1,\ \frac{4}{3}n\log_3(n)+\frac{1}{2}n+\frac{1}{2},\ \frac{1}{2}n-\frac{1}{2})$<br>**total:** $4n\log_3(n)+1$ | (107) |
| | DCT-4$_n(r)$ | $(\frac{8}{3}n\log_3(n),\ \frac{4}{3}n\log_3(n)+\frac{1}{2}n-\frac{1}{2},\ \frac{1}{2}n-\frac{1}{2})$<br>**total:** $4n\log_3(n)+n$ | (112) |
| | DST-4$_n(r)$ | same as DCT-4 | equivalent to (112) |

is decomposed by $\overline{\text{DST-1}}_{k-1}$ (note that Theorem 2 requires us to choose a polynomial transform). The smaller modules in (82) are decomposed, respectively, by skew DTTs as

$$\mathbb{C}[x]/(T_m - \cos\tfrac{i\pi}{k}) \to \textstyle\bigoplus_{0\le j<m}(x-\cos r_{i,j}\pi), \qquad (83)$$

where the $r_{i,j}$ and their order are computed by Lemma 1. The type of the skew DTT is determined by the $C$-basis. For example, for $\mathrm{DTT} = \mathrm{DCT\text{-}1}$, $C = T$ and thus (83) is decomposed by a DCT-3 (see Table III). The final factorization of $A_{(k-1)m}$ is given by

$$A_{(k-1)m} = Q^{(*)}_{k,m}\Big(\bigoplus_{0<i<k}\mathrm{DTT}'_m(\tfrac{i}{k})\Big)(\overline{\text{DST-1}}_{k-1}\otimes \mathrm{I}_m). \qquad (84)$$

In summary, we obtain the following algorithm for a $\mathrm{DTT}'_n$ in the $U$-group:

$$\begin{aligned} \mathrm{DTT}_{n'} &= P^{(*)}_{k,m}(\mathrm{DTT}_{m'}\oplus A_{(k-1)m})B^{(*)}_{k,m}, \qquad (85)\\ A_{(k-1)m} &= \Big(\bigoplus_{0<i<k}\mathrm{DTT}'_m(\tfrac{i}{k})\Big)(\overline{\text{DST-1}}_{k-1}\otimes \mathrm{I}_m), \end{aligned}$$

where we fused the permutations $P^{(*)}_{k,m}$ in (81) and $Q^{(*)}_{k,m}$ in (84) to a permutation $P^{(*)}_{k,m}$. Equation 85 remains valid when the occurring DTTs are replaced by their polynomial versions to yield

$$\begin{aligned} \overline{\mathrm{DTT}}_{n'} &= P^{(*)}_{k,m}(\overline{\mathrm{DTT}}_{m'}\oplus \overline{A}_{(k-1)m})B^{(*)}_{k,m}, \qquad (86)\\ \overline{A}_{(k-1)m} &= \Big(\bigoplus_{0<i<k}\overline{\mathrm{DTT}}'_m(\tfrac{i}{k})\Big)(\overline{\text{DST-1}}_{k-1}\otimes \mathrm{I}_m), \end{aligned}$$

In the following four sections, we will give the special structure of the matrices $P^{(*)}_{k,m}$ and $B^{(*)}_{k,m}$ in all four cases. They are shown in Table XIV. We will analyze the arithmetic cost for 2-power sizes $n = 2^t$. In all cases it turns out that, in contrast to the $T$-group algorithms derived above, the cost does depend on the chosen split, with the minimum obtained for the case $k = 2$, which is equivalent to Table VI(a). Further, the structure of the algorithm (86) shows that the polynomial version of the DTT requires a smaller number of multiplications than the DTT if this holds for the base case $n = 2$, which is easy to check. The result is that only DCT and DST of type 2 yield savings. For the occurring skew DTTs, we use the algorithms and the arithmetic cost provided in the previous sections.

### B. Details

We provide the exact form of the base change matrices $B^{(*)}_{k,m}$ and permutations $P^{(*)}_{k,m}$ using the mnemonic names $* \in \{C1, S1, C2, S2\}$ to denote the 4 DTTs in the $U$-group.

$$B^{(C1)}_{k,m} = C^{(C1)}_{k,m}\cdot D^{(C1)}_{k,m},$$

TABLE XIV

COOLEY-TUKEY TYPE ALGORITHMS FOR DTTS IN THE $U$-GROUP, BASED ON THE DECOMPOSITION $U_{km-1} = U_{m-1} \cdot U_{k-1}(T_m)$. THE POLYNOMIAL VERSIONS ARE OBTAINED BY REPLACING ALL TRANSFORMS BY THEIR POLYNOMIAL COUNTERPART.

$$\text{DCT-1}_{km+1} = P_{k,m}^{(C1)} \Big( \text{DCT-1}_{m+1} \oplus \big( \bigoplus_{0<i<k} \text{DCT-3}_m(\tfrac{i}{k}) \big) (\overline{\text{DST-1}}_{k-1} \otimes \text{I}_m) \Big) B_{k,m}^{(C1)} \qquad (87)$$

$$\text{DST-1}_{km-1} = P_{k,m}^{(S1)} \Big( \big( \bigoplus_{0<i<k} \text{DST-3}_m(\tfrac{i}{k}) \big) (\overline{\text{DST-1}}_{k-1} \otimes \text{I}_m) \oplus \text{DST-1}_{m-1} \Big) B_{k,m}^{(S1)} \qquad (88)$$

$$\text{DCT-2}_{km} = P_{k,m}^{(C2)} \Big( \text{DCT-2}_m \oplus \big( \bigoplus_{0<i<k} \text{DCT-4}_m(\tfrac{i}{k}) \big) (\overline{\text{DST-1}}_{k-1} \otimes \text{I}_m) \Big) B_{k,m}^{(C2)} \qquad (89)$$

$$\text{DST-2}_{km} = P_{k,m}^{(S2)} \Big( \big( \bigoplus_{0<i<k} \text{DST-4}_m(\tfrac{i}{k}) \big) (\overline{\text{DST-1}}_{k-1} \otimes \text{I}_m) \oplus \text{DST-2}_m \Big) B_{k,m}^{(S2)} \qquad (90)$$

where $C_{k,m}^{(C1)}$ is given by

$$\left[\begin{array}{cccccccccc}
1 & & & & 1 & & & & & \\
 & \text{I}_{m-1} & & \text{J}_{m-1} & & \text{I}_{m-1} & & \iddots & \ddots & \\
 & & 1 & & & & 1 & & & \\
\hline
1 & & & & -1 & & & & & \\
 & \text{I}_{m-1} & & -\text{J}_{m-1} & & & & & & \\
 & & 1 & & & & -1 & & & \\
 & & & \text{I}_{m-1} & & -\text{J}_{m-1} & & & \ddots & \\
 & & & & 1 & & \ddots & & -1 & \\
 & & & & & \ddots & & -\text{J}_{m-1} & & \\
 & & & & & & 1 & & & -2 \\
 & & & & & & & \text{I}_{m-1} & & -\text{J}_{m-1}
\end{array}\right],$$

and $D_{k,m}^{(C1)} = \text{I}_{m+1} \oplus (\text{I}_{k-1} \otimes (1/2 \oplus \text{I}_{m-1}))$.

Note that the first block row of $B_{k,m}^{(C1)}$ represents the signal extension of the signal model for $\text{DCT-1}_{m+1}$ (namely $T_k \bmod T_{m+1},\ k > m+1$, see [2]) . Similar statements hold for the matrices below.

$$B_{k,m}^{(S1)} = \left[\begin{array}{cccccccc}
\text{I}_{m-1} & & \text{J}_{m-1} & & & & & \\
 & 1 & & & & & & \\
 & & \text{I}_{m-1} & & \text{J}_{m-1} & & & \\
 & & & 1 & & & & \\
 & & & & \ddots & \ddots & & \\
 & & & & & \text{I}_{m-1} & & \text{J}_{m-1} \\
 & & & & & & 1 & \\
\hline
\text{I}_{m-1} & \begin{smallmatrix}0\\ \vdots\\ 0\end{smallmatrix} & -\text{J}_{m-1} & \begin{smallmatrix}0\\ \vdots\\ 0\end{smallmatrix} & \text{I}_{m-1} & \cdots & &
\end{array}\right]$$

$$B_{k,m}^{(C2)} = \left[\begin{array}{ccccc}
\text{I}_m & \text{J}_m & \text{I}_m & \text{J}_m & \cdots \\
\hline
\text{I}_m & -\text{J}_m & & & \\
 & \text{I}_m & -\text{J}_m & & \\
 & & \ddots & \ddots & \\
 & & & \text{I}_m & -\text{J}_m
\end{array}\right]$$

$$B_{k,m}^{(S2)} = \left[\begin{array}{ccccc}
\text{I}_m & \text{J}_m & & & \\
 & \text{I}_m & \text{J}_m & & \\
 & & \ddots & \ddots & \\
 & & & \text{I}_m & \text{J}_m \\
\hline
\text{I}_m & -\text{J}_m & \text{I}_m & -\text{J}_m & \cdots
\end{array}\right]$$

$$P_{k,m}^{(C1)} = \text{I}_1 \oplus ((\text{J}_{k-1} \oplus \text{I}_1) \oplus \text{I}_k \oplus (\text{J}_{k-1} \oplus \text{I}_1) \oplus \text{I}_k \dots) \widehat{\text{L}}_{k-1}^n$$

$$P_{k,m}^{(S1)} = (\text{I}_{k-1} \oplus (\text{I}_1 \oplus \text{J}_{k-1}) \oplus \text{I}_k \oplus (\text{I}_1 \oplus \text{J}_{k-1}) \oplus \dots) \widehat{\text{L}}_m^{n-1}$$

$$P_{k,m}^{(C2)} = (\text{I}_k \oplus (\text{I}_1 \oplus \text{J}_{k-1}) \oplus \text{I}_k \oplus (\text{I}_1 \oplus \text{J}_{k-1}) \oplus \dots)\, \text{L}_m^n$$

TABLE XV

BASE CASES FOR $U$-GROUP DTTS.

| | |
|---|---|
| $\text{DCT-1}_2 = \text{F}_2$ | $\overline{\text{DCT-1}}_2 = \text{DCT-1}_2$ |
| $\text{DST-1}_1 = \text{I}_1$ | $\overline{\text{DST-1}}_1 = \text{I}_1$ |
| $\text{DCT-2}_2 = \text{diag}(1, \frac{1}{\sqrt{2}}) \cdot \text{F}_2$ | $\overline{\text{DCT-2}}_2 = \text{F}_2$ |
| $\text{DST-2}_2 = \text{diag}(\frac{1}{\sqrt{2}}, 1) \cdot \text{F}_2$ | $\overline{\text{DST-2}}_2 = \text{F}_2$ |

$$P_{k,m}^{(S2)} = (\text{I}_k \oplus (\text{J}_{k-1} \oplus \text{I}_1) \oplus \text{I}_k \oplus (\text{J}_{k-1} \oplus \text{I}_1) \oplus \dots)\, \text{L}_m^n .$$

**Base cases.** For 2-power size, the recursions in Table XIV need as base cases Table XV and the skew DTTs in Table VII.

**Special cases.** The recursions in Table XIV take the simplest form for $k = 2$, in which case they coincide with Table VI(a).

### C. *Alternative decomposition*

We do not discuss algorithms based on Lemma 3, vi). Similar statements as in Section VII-D hold.

### D. *Analysis*

We analyze the algorithms in Table XIV.

**Arithmetic cost.** We only consider a 2-power size $n$. In contrast to the $T$-group algorithms in Section VII, the cost of the algorithms does depend on the split. The minimum is obtained for $k = 2$, in which case the algorithms coincide with Table VI(a). The cost in these cases is shown in Table XVI.

**Literature.** Except for the case $k = 2$ (see Section V-B), we did not find any of these algorithms in the literature.

## IX. $V$-GROUP DTT ALGORITHMS

In this section, we present algorithms for all DTTs in the $V$-group, i.e., the DCT and DST of type 7 and 8, based on Lemma 3, iii):

$$V_{(k-1)/2+km} = V_m \cdot V_{(k-1)/2}(T_{2m+1}).$$

Since the derivation is analogous to the previous sections, we will only state the results without a detailed derivation.

TABLE XVI

ARITHMETIC COSTS FOR DTTS IN THE $U$-GROUP ACHIEVABLE WITH THE RECURSIONS IN TABLE XIV. THE SIZE OF DCT-1 IS $n = 2^k + 1$, THE SIZE OF DST-1 IS $n = 2^k - 1$, AND THE SIZES OF DCT-2 AND DST-2 IS $n = 2^k$

| Transform | Cost (adds, mults, 2-power mults) and total | Achieved by |
|---|---|---|
| DCT-$1_n$ | $(\frac{3}{2}n\log_2(n-1) - 2n - \frac{1}{2}\log_2(n-1) + 6,$<br>$\frac{1}{2}n\log_2(n-1) - n - \frac{1}{2}\log_2(n-1) + 2,\ 0)$<br>**total:** $2n\log_2(n-1) - 3n - \log_2(n-1) + 8$ | Table VI(a) $=$ (87) for $k = 2$ |
| DST-$1_n$ | $(\frac{3}{2}n\log_2(n+1) - 2n + \frac{5}{2}\log_2(n+1) + 2,$<br>$\frac{1}{2}n\log_2(n+1) - n + \frac{1}{2}\log_2(n+1),\ 0)$<br>**total:** $2n\log_2(n+1) - 3n + 3\log_2(n+1) + 2$ | Table VI(a) $=$ (88) for $k = 2$ |
| DCT-$2_n$ | $(\frac{3}{2}n\log_2(n) - n + 1,\ \frac{1}{2}n\log_2(n),\ 0)$<br>**total:** $2n\log_2(n) - n + 1,\ 0)$ **poly:** $-(n-1)$ | Table VI(a) $=$ (89) for $k = 2$,<br>$(68)^T$ (see also $(76)^T$), (72) |
| DST-$2_n$ | same as DCT-2 | Table VI(a) $=$ (90) for $k = 2$, duality $(105)^T$ |

### A. Simultaneous Derivation

For the DTTs in the $V$-group (see Table III) with associated modules $\mathbb{C}[x]/p$, the polynomial $p$ takes (up to a constant) two different forms: $(x+1)V_{n-1}$ for DCT-$7_n$ and DST-$8_n$ and $V_n$ for DST-$7_n$ and DCT-$8_n$. To derive all four decompositions simultaneously, we thus define

$$n' = \begin{cases} n+1, & \text{for DCT-}7_{n'}, \text{DST-}8_{n'}; \\ n, & \text{for DST-}7_{n'}, \text{DCT-}8_{n'}. \end{cases}$$

Now we consider $\mathrm{DTT}_{n'}$ in the $V$-group with a $C$-basis, $C \in \{T, U, V, W\}$ and assume that $n = km + (k-1)/2 = (2m+1)(k-1)/2 + m$. Necessarily, $k$ is odd. Using Theorem 1, we first decompose

$$\mathbb{C}[x]/p_{n'} = \mathbb{C}[x]/V_{\frac{k-1}{2}}(T_{2m+1}) \oplus \mathbb{C}[x]/p_{m'}. \tag{91}$$

In the second summand, we choose a $C$-basis. In the first summand, we choose

$$\begin{aligned} b' \ = \ & (C_0V_0(T_m), \dots, C_{m-1}V_0(T_m) \\ & \dots \\ & C_0V_{(k-1)/2-1}(T_m), \dots, C_{m-1}V_{(k-1)/2-1}(T_m)), \end{aligned}$$

as required by Theorem 2. This implies a $V$-basis in the coarse module $\mathbb{C}[x]/V_{(k-1)/2}$ and a $C$-basis in the skew modules. We denote the base change matrix for (91) with $B^{(*)}_{k,m}$, where $* \in \{C7, S7, C8, S8\}$. The exact form will be shown below.

Next we decompose the second summand in (91) by $\mathrm{DTT}_{m'}$, and the first summand using Theorem 2. The occurring skew DTTs have the same $C$-basis as the given $\mathrm{DTT}_{n'}$, for example DCT-7 has a $T$-basis and hence the associated skew DTT is DCT-3. We denote that skew DTT with DTT$'$. The subalgebra $\mathbb{C}[x]/V_{(k-1)/2}$ with $V$-basis is, in all four cases, decomposed by the polynomial transform $\overline{\text{DST-7}}_{(k-1)/2}$. The final result is the decomposition

$$\mathrm{DTT}_{n'} = P^{(*)}_{k,m}(A_{(2m+1)(k-1)/2} \oplus \mathrm{DTT}_{m'})B^{(*)}_{k,m}, \tag{92}$$

with

$$A_{(2m+1)(k-1)/2} = \Big(\bigoplus_{0 \le i < (k-1)/2} \mathrm{DTT}'_{2m+1}(\tfrac{2i+1}{k})\Big)(\overline{\text{DST-7}}_{(k-1)/2} \otimes \mathrm{I}_{2m+1}).$$

We obtain the corresponding algorithm for the polynomial $\overline{\mathrm{DTT}}_{n'}$ by replacing all DTTs by their polynomial counterparts. Transposition of (92) yields a different set of algorithms.

### B. Details

In this section, we record the exact form of all four classes of decompositions based on (92). We need the following base change matrices.

For DCT-$7_n$ and DST-$8_n$ we require $n = km + (k+1)/2$. Then,

$$B^{(C7)}_{k,m} = D^{(C7)}_{k,m} \cdot C^{(C7)}_{k,m},$$

with

$$C^{(C7)}_{k,m} = \left[\begin{array}{ccccccccccc|c}
2 & & & & -1 & & & & & & & \\
 & \mathrm{I}_{2m} & & -\mathrm{J}_{2m} & & & & & & & & \\
 & & 1 & & & & & -1 & & & & \\
 & & & \mathrm{I}_{2m} & & & -\mathrm{J}_{2m} & & & & & \\
 & & & & \ddots & & & \ddots & & \ddots & & \\
 & & & & & 1 & & & & & -1 & \\
 & & & & & & & \mathrm{I}_{2m} & & -\mathrm{J}_{2m} & & \\
 & & & & & & & & 1 & & 1 & \\
 & & & & & & & & & \mathrm{I}_{2m} & & \begin{matrix}\mathrm{I}_m \\ -\mathrm{J}_m\end{matrix} \\
\hline
1 & & -1 & & 1 & & & & & & & \\
 & \mathrm{I}_m \ -\mathrm{J}_m & & -\mathrm{I}_m \ \mathrm{J}_m & & \cdots & & & & & &
\end{array}\right]$$

and the diagonal matrix

$$D^{(C7)}_{k,m} = (\mathrm{I}_{(k-1)/2} \otimes (1/2 \oplus \mathrm{I}_{2m})) \oplus \mathrm{I}_{m+1},$$

$$B^{(S8)}_{k,m} = \left[\begin{array}{cccccc|c}
\mathrm{I}_{2m+1} & \mathrm{J}_{2m+1} & & & & & \\
 & \mathrm{I}_{2m+1} & \mathrm{J}_{2m+1} & & & & \\
 & & \ddots & \ddots & & & \\
 & & & & \mathrm{I}_{2m+1} & \mathrm{J}_{2m+1} & \\
 & & & & & \mathrm{I}_{2m+1} & \begin{matrix}\mathrm{I}_m \\ \mathrm{J}_m\end{matrix} \ 2 \\
\hline
\mathrm{I}_m \ \ \mathrm{J}_m & -\mathrm{I}_m \ \ -\mathrm{J}_m & \ddots & \cdot^{\cdot^{\cdot}} & & & \\
1 & -1 & & & & &
\end{array}\right].$$

For DST-7 and DCT-8, we require $n = km+(k-1)/2$. Then

TABLE XVIII
BASE CASES FOR $V$-GROUP DTTS.

| | |
|---|---|
| $\text{DCT-7}_2 = \left[\begin{smallmatrix} 1 & 1/2 \\ 1 & -1 \end{smallmatrix}\right]$ | $\overline{\text{DCT-7}}_2 = \text{DCT-7}_2$ |
| $\text{DST-7}_1 = \frac{\sqrt{3}}{2} I_1$ | $\overline{\text{DST-7}}_1 = I_1$ |
| $\text{DCT-8}_1 = \frac{\sqrt{3}}{2} I_1$ | $\overline{\text{DCT-8}}_1 = I_2$ |
| $\text{DST-8}_2 = \left[\begin{smallmatrix} 1/2 & 1 \\ 1 & -1 \end{smallmatrix}\right]$ | $\overline{\text{DCT-8}}_2 = \left[\begin{smallmatrix} 1 & 2 \\ 1 & -1 \end{smallmatrix}\right]$ |

$$B^{(S7)}_{k,m} = \left[\begin{array}{cccccccccc|c}
I_{2m} & & J_{2m} & & & & & & & & \\
& 1 & & & & 1 & & & & & \\
& & I_{2m} & & J_{2m} & & & & & & \\
& & & \ddots & & \ddots & & & & & \\
& & & & 1 & & & & & 1 & \\
& & & & & I_{2m} & & J_{2m} & & & \\
& & & & & & 1 & & & & \\
& & & & & & & I_{2m} & & & \begin{smallmatrix} I_m \\ J_m \end{smallmatrix} \\
& & & & & & & & & 1 & \\ \hline
\multicolumn{11}{l}{I_m \;\; J_m \; \begin{smallmatrix} 0 \\ \vdots \\ 0 \end{smallmatrix} -I_m \;\; -J_m \; \begin{smallmatrix} 0 \\ \vdots \\ 0 \end{smallmatrix} \; I_m \;\; J_m \; \begin{smallmatrix} 0 \\ \vdots \\ 0 \end{smallmatrix} \; \ddots \;\; \cdot^{\cdot^{\cdot}}}
\end{array}\right],$$

$$B^{(C8)}_{k,m} = \left[\begin{array}{cccccc|c}
I_{2m+1} & -J_{2m+1} & & & & & \\
& I_{2m+1} & -J_{2m+1} & & & & \\
& & \ddots & \ddots & & & \\
& & & I_{2m+1} & -J_{2m+1} & & \\
& & & & I_{2m+1} & & \begin{smallmatrix} & I_m & \\ 0 & \cdots & 0 \\ & -J_m & \end{smallmatrix} \\ \hline
\multicolumn{7}{l}{I_m \; \begin{smallmatrix} 0 \\ \vdots \\ 0 \end{smallmatrix} -J_m \; -I_m \; \begin{smallmatrix} 0 \\ \vdots \\ 0 \end{smallmatrix} \; J_m \; I_m \; \begin{smallmatrix} 0 \\ \vdots \\ 0 \end{smallmatrix} -J_m \; \ddots \;\; \cdot^{\cdot^{\cdot}}}
\end{array}\right].$$

Correspondingly, we need two closely related types of permutations. Let $k, m$ be fixed. For a given $i$ to be mapped, we decompose $i$ into the radix-$k$ representation $i = i_1 + i_2 k$, with $i_1 = i \bmod k$ and $i_2 = i \operatorname{div} k$. Then $P^{(C7)}_{k,m} = P^{(S8)}_{k,m}$ is a permutation on the set $\{0, \dots, km + (k-1)/2\}$ defined by

$$P^{(C7)}_{k,m} = P^{(S8)}_{k,m} = i_1 + i_2 k \mapsto$$
$$\begin{cases} i_1(2m+1) + 2i_2, & \text{for } 0 \le i_1 < \frac{k-1}{2}; \\ i_1(2m+1) + i_2, & \text{for } i_1 = \frac{k-1}{2}; \\ (k-1-i_1)(2m+1) + 2i_2 + 1, & \text{for } \frac{k-1}{2} < i_1 < k. \end{cases}$$

This permutation leaves the last point fixed. By omitting this point, i.e., by restricting the permutation to the set $\{0, \dots, km + (k-1)/2 - 1\}$, we get the permutation $P^{(S7)}_{k,m} = P^{(C8)}_{k,m}$.

**Base cases.** The base cases for the DCTs and DSTs of type 7 and 8 are shown in Table XVIII. The sizes are motivated below in the cost analysis.

**Special cases.** The recursions in Table XVII take the simplest form for $k = 3$ (note that $k$ has to be odd), in which case they coincide with the algorithms in Table VI(c).

### C. Analysis

To analyze the arithmetic cost of the algorithms in Table XVII, the first question is which sizes are best decomposable or “natural” for these DTTs. For example, for all DTTs of type 1–4 the best decomposable size is $2^t$, with the exception of DCT-1, which has $2^t + 1$, and DST-1, which has $2^t - 1$. These sizes allow a complete decomposition into $2 \times 2$ transforms.

For $m = 1$, the decompositions in Table XVII are trivial, thus we obtain upon decomposition skew DTTs of odd size larger than 1. Hence, the best outcome is that $2m + 1$ is a 3-power, to allow at least a decomposition of the occurring skew DTTs into $3 \times 3$ transforms using Table X or IX. Further, we want to be able to further decompose the occurring $\overline{\text{DST-7}}_{(k-1)/2}$, which requires that $(k-1)/2$ has again the form $km + (k-1)/2$ (with a different $k$). Inspection shows that these conditions are satisfied for $n = (3^t - 1)/2$. Namely, for $0 < s < t$

$$(3^t - 1)/2 = 3^s(3^{t-s} - 1)/2 + (3^s - 1)/2,$$

which matches $n = (2m+1)(k-1)/2 + m$ for $m = (3^s - 1)/2$ and $k = 3^{t-s}$. In summary, the best decomposable sizes are thus $n' = n + 1 = (3^t + 1)/2$ for DCT-7 and DST-8, and $n' = n = (3^t - 1)/2$ for DST-7 and DCT-8. This also implies that the base sizes are 2 and 1, respectively.

Using the arithmetic cost of the skew DTTs of 3-power size in Table XIII, we get Table XIX. Note that the costs for DCT-8, DST-8 are not achieved using Table VI(c) but only by duality (105). The reason is that Table VI(c) yields for these DTTs the same recurrence as for DCT-7 and DST-7, respectively, but with more expensive skew base cases (Table VIII).

**Literature.** We did not find any of the algorithms in Table XVII in the literature.

## X. $W$-GROUP DTT ALGORITHMS

We present algorithms for the DTTs in the $W$-group based on the decomposition in Lemma 3, iv):

$$W_{(k-1)/2+km} = W_m \cdot W_{(k-1)/2}(T_{2m+1}).$$

The derivation and discussion is very similar to Section IX, so we will be brief.

### A. Simultaneous derivation

We define for the DTTs in the $W$-group

$$n' = \begin{cases} n+1, & \text{for DCT-5}_{n'}, \text{DCT-6}_{n'}; \\ n, & \text{for DST-5}_{n'}, \text{DST-6}_{n'}. \end{cases}$$

The definition is motivated by the associated polynomial $p$ in $\mathbb{C}[x]/p$, namely $p(x) = (x-1)W_{n-1}$ for DCT-5 and DCT-6 and $W_n$ for DST-5 and DST-6.

Now let $\text{DTT}_{n'}$ be in the $W$-group with $C$-basis. Then the first decomposition step yields

$$\mathbb{C}[x]/p_{n'} = \mathbb{C}[x]/W_{\frac{k-1}{2}}(T_{2m+1}) \oplus \mathbb{C}[x]/p_{m'}. \quad (97)$$

In the second summand we choose a $C$-basis. In the first summand we choose

$$\begin{aligned} b' \;\; = \;\; & (C_0 W_0(T_m), \dots, C_{m-1} W_0(T_m) \\ & \dots \\ & C_0 W_{(k-1)/2-1}(T_m), \dots, C_{m-1} W_{(k-1)/2-1}(T_m)), \end{aligned}$$

as required by Theorem 2. This implies a $W$-basis in the coarse module $\mathbb{C}[x]/W_{(k-1)/2}$ and a $C$-basis in the skew modules.

TABLE XVII

COOLEY-TUKEY TYPE ALGORITHMS FOR DTTS IN THE $V$-GROUP, BASED ON THE DECOMPOSITION $V_{(k-1)/2+km} = V_m \cdot V_{(k-1)/2}(T_{2m+1})$; $k$ IS ODD. THE POLYNOMIAL VERSIONS ARE OBTAINED BY REPLACING ALL TRANSFORMS BY THEIR POLYNOMIAL COUNTERPART.

$$\text{DCT-7}_{km+(k+1)/2} = P^{(C7)}_{k,m}\Big(\big(\bigoplus_{0\le i<\frac{k-1}{2}} \text{DCT-3}_{2m+1}(\tfrac{2i+1}{k})\big)(\overline{\text{DST-7}}_{\frac{k-1}{2}} \otimes \text{I}_{2m+1}) \oplus \text{DCT-7}_{m+1}\Big)B^{(C7)}_{k,m} \quad (93)$$

$$\text{DST-7}_{km+(k-1)/2} = P^{(S7)}_{k,m}\Big(\big(\bigoplus_{0\le i<\frac{k-1}{2}} \text{DST-3}_{2m+1}(\tfrac{2i+1}{k})\big)(\overline{\text{DST-7}}_{\frac{k-1}{2}} \otimes \text{I}_{2m+1}) \oplus \text{DST-7}_{m}\Big)B^{(S7)}_{k,m} \quad (94)$$

$$\text{DCT-8}_{km+(k-1)/2} = P^{(C8)}_{k,m}\Big(\big(\bigoplus_{0\le i<\frac{k-1}{2}} \text{DCT-4}_{2m+1}(\tfrac{2i+1}{k})\big)(\overline{\text{DST-7}}_{\frac{k-1}{2}} \otimes \text{I}_{2m+1}) \oplus \text{DCT-8}_{m}\Big)B^{(C8)}_{k,m} \quad (95)$$

$$\text{DST-8}_{km+(k+1)/2} = P^{(S8)}_{k,m}\Big(\big(\bigoplus_{0\le i<\frac{k-1}{2}} \text{DST-4}_{2m+1}(\tfrac{2i+1}{k})\big)(\overline{\text{DST-7}}_{\frac{k-1}{2}} \otimes \text{I}_{2m+1}) \oplus \text{DST-8}_{m+1}\Big)B^{(S8)}_{k,m} \quad (96)$$

TABLE XIX

ARITHMETIC COSTS FOR DTTS IN THE $V$-GROUP ACHIEVABLE WITH THE RECURSIONS IN TABLE XVII. THE SIZE FOR DCT-7 AND DST-8 IS $n = (3^k + 1)/2$, FOR DST-7 AND DCT-8 IS $n = (3^k - 1)/2$.

| Transform | Cost (adds, mults, 2-power mults) and total | Achieved by |
|---|---|---|
| DCT-$7_n$ | $(\frac{8}{3}n\log_3(2n-1) - 3n - \frac{1}{3}\log_3(2n-1) + 3,$ $\frac{4}{3}n\log_3(2n-1) - 2n - \frac{2}{3}\log_3(2n-1) + 2,\ \log_3(2n-1))$ | Table VI(c) $=$ (93) for $k = 3$ |
| | **total:** $4n\log_3(2n-1) - 5n + 5$ **poly:** same | |
| DST-$7_n$ | $(\frac{8}{3}n\log_3(2n+1) - 3n + \frac{1}{3}\log_3(2n+1),$ $\frac{4}{3}n\log_3(2n+1) - \frac{3}{2}n + \frac{7}{6}\log_3(2n+1),\ \frac{1}{2}n - \frac{1}{2}\log_3(2n+1))$ | Table VI(c) $=$ (94) for $k = 3$ |
| | **total:** $4n\log_3(2n+1) - 4n + \log_3(2n+1)$ | |
| DCT-$8_n$ | same as DST-7 | duality (105) |
| DST-$8_n$ | same as DCT-7 | duality (105) |

The corresponding base change matrix is $B^{(*)}_{k,m}$, where $* \in \{C5, S5, C6, S6\}$. The exact form will be shown below.

The full decomposition becomes

$$\text{DTT}_{n'} = P^{(*)}_{k,m}(A_{(2m+1)(k-1)/2} \oplus \text{DTT}_{m'})B^{(*)}_{k,m}, \quad (98)$$

with

$$A_{(2m+1)(k-1)/2} = \Big(\bigoplus_{0\le i<(k-1)/2} \text{DTT}'_{2m+1}(\tfrac{2i+1}{k})\Big)(\overline{\text{DST-5}}_{(k-1)/2} \otimes \text{I}_{2m+1}).$$

## B. Details

The base change matrices and permutations in (98) are as follows.

$$B^{(C5)}_{k,m} = C^{(C5)}_{k,m} \cdot D^{(C5)}_{k,m}$$

$$C^{(C5)}_{k,m} = \left[\begin{array}{ccccccccccc|c}
1 & & & 1 & & & 1 & & & & & \\
 & \text{I}_m\ \text{J}_m & & & \text{I}_m\ \text{J}_m & & & \cdots & & & & \\
\hline
2 & & & & & & -1 & & & & & \\
 & \text{I}_{2m} & & & -\text{J}_{2m} & & & & & & & \\
 & & & 1 & & & & & -1 & & & \\
 & & & & \text{I}_{2m} & & & -\text{J}_{2m} & & & & \\
 & & & & & & \ddots & & \ddots & \ddots & & \\
 & & & & & & & 1 & & & -1 & \\
 & & & & & & & & \text{I}_{2m} & -\text{J}_{2m} & & \\
 & & & & & & & & 1 & & -1 & \\
 & & & & & & & & & \text{I}_{2m} & & \begin{smallmatrix}-\text{I}_m\\ -\text{J}_m\end{smallmatrix}
\end{array}\right]$$

and the diagonal matrix

$$D^{(C5)}_{k,m} = \text{I}_{m+1} \oplus (\text{I}_{(k-1)/2} \otimes (1/2 \oplus \text{I}_{2m}))$$

$$B^{(S5)}_{k,m} = \left[\begin{array}{ccccccccccc|c}
\text{I}_m & -\text{J}_m & \begin{smallmatrix}0\\ \vdots\\ 0\end{smallmatrix} & \text{I}_m & -\text{J}_m & \begin{smallmatrix}0\\ \vdots\\ 0\end{smallmatrix} & \text{I}_m & -\text{J}_m & \begin{smallmatrix}0\\ \vdots\\ 0\end{smallmatrix} & \ddots & \cdot^{\cdot^{\cdot}} & \\
\hline
\text{I}_{2m} & & & \text{J}_{2m} & & & & & & & & \\
 & & 1 & & & & & & 1 & & & \\
 & & & \text{I}_{2m} & & & \text{J}_{2m} & & & & & \\
 & & & & \ddots & & & \ddots & & & & \\
 & & & & & & 1 & & & & 1 & \\
 & & & & & & & \text{I}_{2m} & & & \text{J}_{2m} & \\
 & & & & & & & & 1 & & & \\
 & & & & & & & & & & \text{I}_{2m} & \begin{smallmatrix}-\text{I}_m\\ \text{J}_m\end{smallmatrix} \\
 & & & & & & & & & & 1 &
\end{array}\right]$$

$$B^{(C6)}_{k,m} = \left[\begin{array}{ccccccc|c}
\text{I}_m\ \ \text{J}_m & \text{I}_m\ \ \text{J}_m & \ddots & \cdot^{\cdot^{\cdot}} & & & & \\
1 & 1 & & & & & & \\
\hline
\text{I}_{2m+1} & -\text{J}_{2m+1} & & & & & & \\
 & \text{I}_{2m+1} & -\text{J}_{2m+1} & & & & & \\
 & & \ddots & \ddots & & & & \\
 & & & & \text{I}_{2m+1} & -\text{J}_{2m+1} & & \\
 & & & & & \text{I}_{2m+1} & & \begin{smallmatrix}-\text{I}_m\\ -\text{J}_m\end{smallmatrix}\ -2
\end{array}\right]$$

TABLE XXI

BASE CASES FOR $W$-GROUP DTTS.

| | |
|---|---|
| $\mathrm{DCT\text{-}5}_2 = \begin{bmatrix} 1 & 1 \\ 1 & -1/2 \end{bmatrix}$ | $\overline{\mathrm{DCT\text{-}5}}_2 = \mathrm{DCT\text{-}5}_2$ |
| $\mathrm{DST\text{-}5}_1 = \frac{\sqrt{3}}{2}\,\mathrm{I}_1$ | $\overline{\mathrm{DST\text{-}5}}_1 = \mathrm{I}_1$ |
| $\mathrm{DCT\text{-}6}_2 = \begin{bmatrix} 1 & 1 \\ 1/2 & -1 \end{bmatrix}$ | $\overline{\mathrm{DCT\text{-}6}}_2 = \begin{bmatrix} 1 & 1 \\ 1 & -2 \end{bmatrix}$ |
| $\mathrm{DST\text{-}6}_1 = \frac{\sqrt{3}}{2}\,\mathrm{I}_1$ | $\overline{\mathrm{DCT\text{-}6}}_1 = \mathrm{I}_1$ |

$$B^{(S6)}_{k,m} = \left[\begin{array}{cccccc|c} \mathrm{I}_m \begin{smallmatrix}0\\ \vdots\\ 0\end{smallmatrix} -\mathrm{J}_m & \mathrm{I}_m \begin{smallmatrix}0\\ \vdots\\ 0\end{smallmatrix} -\mathrm{J}_m & \mathrm{I}_m \begin{smallmatrix}0\\ \vdots\\ 0\end{smallmatrix} -\mathrm{J}_m & \ddots & \cdot^{\cdot^{\cdot}} & & \\ \hline \mathrm{I}_{2m+1} & \mathrm{J}_{2m+1} & & & & & \\ & \mathrm{I}_{2m+1} & \mathrm{J}_{2m+1} & & & & \\ & & \ddots & \ddots & & & \\ & & & & \mathrm{I}_{2m+1} & \mathrm{J}_{2m+1} & \\ & & & & & \mathrm{I}_{2m+1} & \begin{smallmatrix} -\mathrm{I}_m \\ 0 \cdots 0 \\ \mathrm{J}_m \end{smallmatrix} \end{array}\right].$$

To state the permutations, we decompose the index $i$ to be mapped as before into $i = i_1 + i_2 k$. Then the permutations for for DCT-5 and DCT-6 operate on $i \in \{0, \dots, km+(k-1)/2\}$ and are given by

$$P^{(C5)}_{k,m} = P^{(C6)}_{k,m} = i_1 + i_2 k \mapsto$$
$$\begin{cases} i_2, & \text{for } i_1 = 0; \\ i_1(2m+1) + 2i_2 - m, & \text{for } 0 < i_1 \le \frac{k-1}{2}; \\ (k - i_1)(2m+1) + 2i_2 - m + 1, & \text{for } \frac{k-1}{2} < i_1 \le k - 1. \end{cases}$$

This permutation leaves the first point fixed. The corresponding permutation for the DSTs type 5 and 6 operates on one point less and arises from $P^{(C5)}_{k,m}$ by omitting row 0 and column 0. The definition is

$$P^{(S5)}_{k,m} = P^{(S6)}_{k,m} = i_1 + i_2 k \mapsto$$
$$\begin{cases} i_1(2m+1) + 2i_2 + m, & \text{for } i_1 < \frac{k-1}{2}; \\ (k - i_1)(2m+1) + 2i_2 - 3m - 1, & \text{for } \frac{k-1}{2} \le i_1 < k - 1; \\ i_2, & \text{for } i_1 = k - 1. \end{cases}$$

The final algorithms are shown in Table XX.

**Base cases.** The base cases are in Table XXI. The sizes are motivated in the cost analysis below.

**Special cases.** The algorithms in Table XX have the simplest form for $k = 3$ in which case they coincide with Table VI(d).

### *C. Analysis*

The natural sizes, i.e., those sizes that yield a decomposition into the smallest DTTs are $n = (3^t + 1)/2$ for DCT-5 and DCT-6 and $n = (3^t - 1)/2$ for DST-5 and DST-6. For these sizes we can achieve the cost in Table XXII.

**Literature.** We did not find any of the algorithms in Table XX in the literature.

## XI. CONCLUSIONS

We presented an algebraic approach to deriving fast transform algorithms; in particular, we identified the general principle behind the Cooley-Tukey FFT. By applying the approach to the 16 DTTs, we derived equivalent "Cooley-Tukey type" algorithms of similar structure. Thus, we could explain many existing algorithms, but discovered an even larger number of new algorithms that could not be found with previous methods.

The principle behind Cooley-Tukey type algorithms is polynomial decomposition for finite regular shift-invariant 1-D signal models (or, equivalently, polynomial algebras), or, more generally, induction as briefly discussed in Section IV-C.

We hope to have achieved several things with this paper.

The paper is a first step to obtaining a comprehensive theory of fast transform algorithms: a theory that classifies algorithms, provides insight to why they exist, illuminates their structure, and enables their concise, systematic derivation.

Second, the theory of algorithms in this paper is a natural application of the algebraic SP theory. In [1], [2] the concept of signal model (as in Definition 1) was motivated as the natural structure underlying SP. In this paper, it becomes the key to derive, discover, and understand algorithms. The algebraic approach ties together SP theory and transform algorithm theory.

Third, the paper reinforces the case for representing algorithms as structured matrices, an approach that was already successfully employed for the DFT in [35], [32] and occasionally in research papers (e.g., a early as [21] and more systematically developed and exploited in [65]).

Fourth, by summarizing many existing and deriving many new algorithms, this paper can be a reference paper on algorithms that is useful for application developers that are only interested in their application.

Future papers will derive and explain other algorithms available for trigonometric transforms, including real DFTs, orthogonal DTT algorithms, and generalized Rader type algorithms.

TABLE XX

COOLEY-TUKEY TYPE ALGORITHMS FOR DTTS IN THE $W$-GROUP, BASED ON THE DECOMPOSITION $W_{(k-1)/2+km} = W_m \cdot W_{(k-1)/2}(T_{2m+1})$; $k$ IS ODD. THE POLYNOMIAL VERSIONS ARE OBTAINED BY REPLACING ALL TRANSFORMS BY THEIR POLYNOMIAL COUNTERPART.

$$\text{DCT-5}_{km+(k+1)/2} = P^{(C5)}_{k,m}\Big(\text{DCT-5}_{m+1} \oplus \big(\bigoplus_{0\le i<\frac{k-1}{2}} \text{DCT-3}_{2m+1}(\tfrac{2i+2}{k})\big)(\overline{\text{DST-5}}_{\frac{k-1}{2}} \otimes \text{I}_{2m+1})\Big)B^{(C5)}_{k,m} \quad (99)$$

$$\text{DST-5}_{km+(k-1)/2} = P^{(S5)}_{k,m}\Big(\text{DST-5}_{m} \oplus \big(\bigoplus_{0\le i<\frac{k-1}{2}} \text{DST-3}_{2m+1}(\tfrac{2i+2}{k})\big)(\overline{\text{DST-5}}_{\frac{k-1}{2}} \otimes \text{I}_{2m+1})\Big)B^{(S5)}_{k,m} \quad (100)$$

$$\text{DCT-6}_{km+(k+1)/2} = P^{(C6)}_{k,m}\Big(\text{DCT-6}_{m+1} \oplus \big(\bigoplus_{0\le i<\frac{k-1}{2}} \text{DCT-4}_{2m+1}(\tfrac{2i+2}{k})\big)(\overline{\text{DST-5}}_{\frac{k-1}{2}} \otimes \text{I}_{2m+1})\Big)B^{(C6)}_{k,m} \quad (101)$$

$$\text{DST-6}_{km+(k-1)/2} = P^{(S6)}_{k,m}\Big(\text{DST-6}_{m} \oplus \big(\bigoplus_{0\le i<\frac{k-1}{2}} \text{DST-4}_{2m+1}(\tfrac{2i+2}{k})\big)(\overline{\text{DST-5}}_{\frac{k-1}{2}} \otimes \text{I}_{2m+1})\Big)B^{(S6)}_{k,m} \quad (102)$$

TABLE XXII

ARITHMETIC COSTS FOR DTTS IN THE $W$-GROUP ACHIEVABLE WITH THE RECURSIONS IN TABLE XX. THE SIZE FOR DCT-5 AND DCT-6 IS $n = (3^k+1)/2$, FOR DST-5 AND DST-6 IS $n = (3^k-1)/2$.

| **Transform** | **Cost (adds, mults, 2-power mults) and total** | **Achieved by** |
|---|---|---|
| DCT-$5_n$ | same as DCT-7 | Table VI(d) $=$ (99) for $k=3$ and its transpose |
| DST-$5_n$ | same as DST-7 | Table VI(d) $=$ (100) for $k=3$ and its transpose |
| DCT-$6_n$ | same as DCT-7 | duality (105), Table VI(c)$^T$ |
| DST-$6_n$ | same as DST-7 | duality (105), Table VI(c)$^T$ |

## Appendix I
## Chinese Remainder Theorem

Let $C[x]/p(x)$ be a polynomial algebra (see Section II-A) and assume that $p(x) = q(x)r(x)$ factors into two coprime polynomials, i.e., $\gcd(q, r) = 1$. Then the Chinese remainder theorem (for polynomials) states that

$$\begin{aligned} \phi:\ \mathbb{C}[x]/p(x) &\to \mathbb{C}[x]/q(x) \oplus \mathbb{C}[x]/r(x) \\ s(x) &\mapsto (s(x) \bmod q(x), s(x) \bmod r(x)) \end{aligned}$$

TABLE XXIII

Four series of Chebyshev polynomials. The range for the zeros is $0 \le k < n$. In the trigonometric closed form $\cos\theta = x$.

| | $n=0,1$ | closed form | symmetry | zeros |
|---|---|---|---|---|
| $T_n$ | $1, x$ | $\cos(n\theta)$ | $T_{-n} = T_n$ | $\cos\frac{(k+\frac{1}{2})\pi}{n}$ |
| $U_n$ | $1, 2x$ | $\frac{\sin(n+1)\theta}{\sin\theta}$ | $U_{-n} = -U_{n-2}$ | $\cos\frac{(k+1)\pi}{n+1}$ |
| $V_n$ | $1, 2x-1$ | $\frac{\cos(n+\frac{1}{2})\theta}{\cos\frac{1}{2}\theta}$ | $V_{-n} = V_{n-1}$ | $\cos\frac{(k+\frac{1}{2})\pi}{n+\frac{1}{2}}$ |
| $W_n$ | $1, 2x+1$ | $\frac{\sin(n+\frac{1}{2})\theta}{\sin\frac{1}{2}\theta}$ | $W_{-n} = -W_{n-1}$ | $\cos\frac{(k+1)\pi}{n+\frac{1}{2}}$ |

is an isomorphism of algebras. In words, $\mathbb{C}[x]/p(x)$ and $\mathbb{C}[x]/q(x) \oplus \mathbb{C}[x]/r(x)$ have the same structure. Formally,

$$\begin{aligned} \phi(s+s') &= \phi(s) + \phi(s') \\ \phi(s \cdot s') &= \phi(s) \cdot \phi(s'), \end{aligned}$$

which means informally that computing in $\mathbb{C}[x]/p(x)$ and elementwise computing in $\mathbb{C}[x]/q(x) \oplus \mathbb{C}[x]/r(x)$ is equivalent.

## Appendix II
## Chebyshev Polynomials

Chebyshev polynomials are a special class of orthogonal polynomials and play an important role in many mathematical areas. Excellent books are [66], [51], [67]. We only introduce the definitions and the properties of the polynomials we use in this paper.

Let $C_0 = 1$, $C_1(x)$ a polynomials of degree 1, and define $C_n$, $n > 1$ by the recurrence

$$C_n(x) = 2xC_{n-1} - C_{n-2}(x).$$

Running this recurrence backwards yields polynomials $C_{-n}$, $n \ge 0$. Each sequence $(C_n)_{n\in\mathbb{Z}}$ of polynomials defined this way is called a sequence of Chebyshev polynomials. It is uniquely determined by the choice of $C_1$. Four special cases are of particular importance in signal processing [2], [3] and in this paper. They are denoted by $C \in \{T, U, V, W\}$ and are called Chebyshev polynomials of the first, second, third, and fourth kind. Table XXIII gives their initial conditions, their closed form, their symmetry properties, and their zeros.

For example, $T_n(x) = \cos(n\theta)$, where $\cos\theta = x$. The closed form easily yields the zeros of $T_n$.

We will use the following properties of Chebyshev polynomials:

1) For any sequence of Chebyshev polynomials with initial conditions $C_0, C_1$, we have

$$C_n = C_1 U_{n-1} - C_0 U_{n-2}. \tag{103}$$

2) For any sequence of Chebyshev polynomials $C_n$,

$$T_k C_n = (C_{n-k} + C_{n+k})/2. \tag{104}$$

3) The identities in Table XXIV hold. They are based on trigonometric identities.

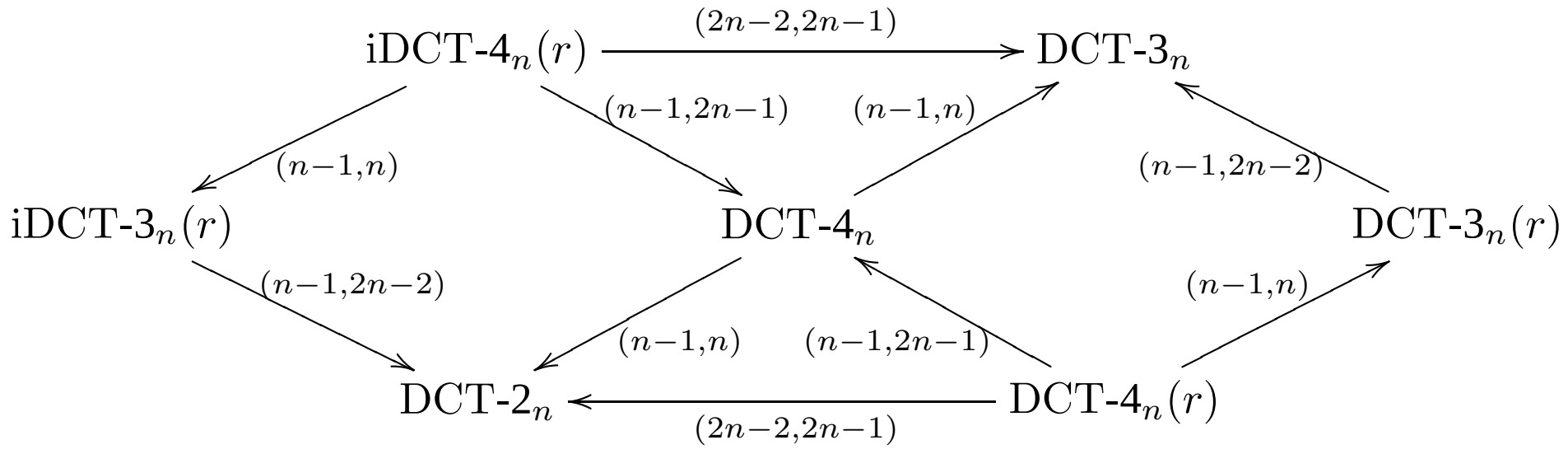

Fig. 3. Number of additions and multiplications (adds, mults) required to translate DCTs of types 2–4 into each other for odd size $n$.

TABLE XXIV
IDENTITIES AMONG THE FOUR SERIES OF CHEBYSHEV POLYNOMIALS; $C_n$ HAS TO BE REPLACED BY $T_n, U_n, V_n, W_n$ TO OBTAIN ROWS 1, 2, 3, 4, RESPECTIVELY.

| $C_n$ | $C_n - C_{n-2}$ | $C_n - C_{n-1}$ | $C_n + C_{n-1}$ |
|---|---|---|---|
| $T_n$ | $2(x^2-1)U_{n-2}$ | $(x-1)W_{n-1}$ | $(x+1)V_{n-1}$ |
| $U_n$ | $2T_n$ | $V_n$ | $W_n$ |
| $V_n$ | $2(x-1)W_{n-1}$ | $2(x-1)U_{n-1}$ | $2T_n$ |
| $W_n$ | $2(x+1)V_{n-1}$ | $2T_n$ | $2(x+1)U_{n-1}$ |

## APPENDIX III
## RELATIONSHIPS BETWEEN DTTS

We use in this paper the following relationships between DTTs. The explanation for their existence and proofs can be found in [2].

**Duality.** Two DTTs $\mathrm{DTT}_n, \mathrm{DTT}'_n$, which have flipped boundary conditions are called *dual* to each other. They are necessarily in the same group. The duality property is not visible from Table III since we omitted the boundary conditions. Thus we just state the pairs: DCT-3/DST-3, DCT-4/DST-4, the DTTs in the $U$-group are all self-dual, DCT-7/DST-8, DST-7/DCT-8, DCT-5/DCT-6, DST-5/DST-6.

The following relationship holds for dual DTTs:

$$\mathrm{diag}_{0\le k<n}((-1)^k)\cdot \mathrm{DTT}_n \cdot \mathrm{J}_n = \mathrm{DTT}'_n\,. \tag{105}$$

As a consequence any DTT algorithm can be converted into a DTT$'$ algorithm without incurring additional operations.

**Base change.** Two DTTs (or skew DTTs) in the same group (e.g., $T$-group) have (at least almost) the same associated algebra. As a consequence they can be translated into each other using a suitable base change and Table XXIV.

Examples include:

$$\begin{aligned} \text{DCT-4}_n &= S_n \cdot \text{DCT-2}_n \cdot \tfrac{1}{2} D_n(1/2)^{-1} \\ \text{iDCT-4}_n(r) &= S_n \cdot \text{iDCT-3}_n(r) \cdot \tfrac{1}{2} D_n(r)^{-1} \end{aligned} \tag{106}$$

where $S_n$ is defined in (62) and $D_n(r) = \mathrm{diag}_{0\le k<n}(\cos\frac{r_k}{2}\pi)$. The $r_k$ are computed from $r$ using Lemma 1.

**Skew and non-skew DTTs.** Every skew $\mathrm{DTT}(r)$ can be translated into its non-skew counterpart DTT:

$$\begin{aligned} \mathrm{DTT}_n(r) &= \mathrm{DTT}_n \cdot X_n^{(*)}(r), \quad \text{and} \\ \overline{\mathrm{DTT}}_n(r) &= \overline{\mathrm{DTT}}_n \cdot X_n^{(*)}(r). \end{aligned} \tag{107}$$

Here, $X_n^{(*)}(r)$ depends on the DTT and takes the following forms, indicated by $* \in \{C3, S3, C4, S4\}$.

$$X_n^{(C3)}(r) = \begin{bmatrix} 1 & 0 & \cdots & \cdots & 0 \\ 0 & c_1 & & & s_{n-1} \\ \vdots & & \ddots & \iddots & \\ \vdots & & \iddots & \ddots & \\ 0 & s_1 & & & c_{n-1} \end{bmatrix} \tag{108}$$

$$X_n^{(S3)}(r) = \begin{bmatrix} c_1 & & & -s_{n-1} & 0 \\ & \ddots & \iddots & & \vdots \\ & \iddots & \ddots & & \vdots \\ -s_1 & & & c_{n-1} & 0 \\ 0 & \cdots & \cdots & 0 & c_n \end{bmatrix} \tag{109}$$

$$X_n^{(C4)}(r) = \begin{bmatrix} c'_0 & & & s'_{n-1} \\ & \ddots & \iddots & \\ & \iddots & \ddots & \\ s'_0 & & & c'_{n-1} \end{bmatrix} \tag{110}$$

$$X_n^{(S4)}(r) = \begin{bmatrix} c'_0 & & & -s'_{n-1} \\ & \ddots & \iddots & \\ & \iddots & \ddots & \\ -s'_0 & & & c'_{n-1} \end{bmatrix} \tag{111}$$

In these equations, $c_\ell = \cos(1/2-r)\ell\pi/n$, $s_\ell = \sin(1/2-r)\ell\pi/n$, $c'_\ell = \cos(1/2-r)(2\ell+1)\pi/(2n)$, and $s'_\ell = \sin(1/2-r)(2\ell+1)\pi/(2n)$. Where the diagonals cross in (108)–(111), the elements are added.

Combining (107) with (106) gives, for example

$$\text{DCT-4}_n(r) = S_n \cdot \text{DCT-2}_n \cdot \tfrac{1}{2} D_n(1/2)^{-1} \cdot X_n^{(C4)}(r). \tag{112}$$

The diagonal matrix can be fused with the x-shaped matrix to save multiplications.

Inversion of (107) gives the corresponding identities for the $\mathrm{iDTT}(r)$'s:

$$\mathrm{iDTT}_n(r) = \big(X_n^{(*)}(r)\big)^{-1} \cdot \mathrm{DTT}_n^T\,. \tag{113}$$

The matrices $\big(X_n^{(*)}(r)\big)^{-1}$ have the same x-shaped structure and the same arithmetic complexity as $X_n^{(*)}(r)$ and can be readily computed because of their block structure. For

example:

$$\left(X_n^{(C3)}(r)\right)^{-1} =$$

$$\frac{1}{\cos(1/2-r)\pi}\begin{bmatrix} c_n & 0 & \cdots & \cdots & 0 \\ 0 & c_{n-1} & & & -s_{n-1} \\ \vdots & & \ddots & \iddots & \\ \vdots & & \iddots & \ddots & \\ 0 & -s_1 & & & c_1 \end{bmatrix}.$$

The above identities show that the complexity of the skew DTTs differ from the complexity of the DTTs by $O(n)$.

Figure 3 displays the cost, as a pair (additions, multiplications), of translating skew and non-skew DCTs of types 2–4 into each other for odd size $n$.